\documentclass{amsart}
\usepackage[english]{babel}
\usepackage{graphicx}
\usepackage[hidelinks]{hyperref}
\usepackage{amssymb}
\usepackage{amsmath}
\usepackage[foot]{amsaddr}
\usepackage{amsfonts}
\usepackage{bbm}
\usepackage{xcolor}
\usepackage{float}
\usepackage{algorithm}
\usepackage{algpseudocode}
\usepackage{comment}

\usepackage{tikz}
\usepackage[top=1in, bottom=1.25in, left=1.25in, right=1.25in]{geometry}

\graphicspath{{images/}}

\newcommand{\norm}[1]{\lVert #1 \rVert}

\newtheorem{remark}{Remark}[section]

\def\PP{{{\rm l}\kern - .15em {\rm P} }}
\def\PN2{{\PP_{N}-\PP_{N-2}}}

\newcommand{\bphi}{\boldsymbol{\varphi}}

\newcommand{\bu}{\boldsymbol{u}}

\newcommand{\bw}{\boldsymbol{w}}

\newcommand{\bx}{\boldsymbol{x}}

\newcommand{\deleted}[1]{{}}

\definecolor{maria}{HTML}{0090A0}
\newcommand{\maria}[1]{{\color{maria}\textbf{#1}}\color{black}}
\newcommand{\TI}[1]{\textcolor{black}{#1}}

\newcommand{\A}[1]{\color{black}{#1}\color{black}}

\newcommand{\C}[1]{\color{black}{#1}\color{black}}

\newcommand{\both}[1]{\color{black}{#1}\color{black}}

\begin{document}

\title[
EFR FOM-ROM Consistency
]{ 
Consistency of the full and reduced order models for
Evolve-Filter-Relax Regularization of Convection-Dominated, Marginally-Resolved Flows 
}

\author{Maria Strazzullo$^1$, Michele Girfoglio$^1$, Francesco Ballarin$^2$, Traian Iliescu$^3$ and Gianluigi Rozza$^1$}
\address{$^1$ mathLab, Mathematics Area, SISSA, via Bonomea 265, I-34136 Trieste, Italy}
\address{$^2$ Department of Mathematics and Physics, Catholic University of the Sacred Heart, via Musei 41, I-25121 Brescia, Italy}
\address{$^3$ Department of Mathematics, Virginia Tech, Blacksburg, VA 24061, USA}
\begin{abstract}

Numerical stabilization is often used to eliminate (alleviate) the spurious oscillations generally produced by full order models (FOMs) in under-resolved or marginally-resolved simulations of convection-dominated flows.
In this paper we investigate the role of numerical stabilization in reduced order models (ROMs) of convection-dominated, marginally-resolved flows.
Specifically, we investigate the FOM-ROM consistency, i.e., whether the numerical stabilization is beneficial 
both at the FOM and the ROM level.
As a numerical stabilization strategy, we focus on the evolve-filter-relax (EFR) regularization algorithm, which centers around spatial filtering.
To investigate the FOM-ROM consistency, we consider two ROM strategies:
(i) the EFR-ROM, in which the EFR stabilization is used at the FOM level, but not at the ROM level; and
(ii) the EFR-EFRROM, in which the EFR stabilization is used both at the FOM and at the ROM level.
We compare the EFR-ROM with the EFR-EFRROM in the numerical simulation of a 2D flow past a circular cylinder in the convection-dominated, marginally-resolved regime.
We also perform model reduction with respect to both time and Reynolds number.
Our numerical investigation shows that the EFR-EFRROM is more accurate than the EFR-ROM, which suggests that FOM-ROM consistency is beneficial in  convection-dominated, marginally-resolved flows.
\end{abstract}

\maketitle

\section{Introduction}



In the numerical simulation of incompressible fluid flows modeled by the Navier-Stokes equations (NSE), standard full order models (FOMs) (e.g., finite element (FE) methods, finite volume (FV) methods, finite difference (FD) methods or spectral elements (SE))  work well in the {\it resolved} regime, i.e., when the number of degrees of freedom is large enough to represent the complex flow dynamics.
However, FOMs are generally not accurate in the {\it under-resolved} regime, i.e., when the number of degrees of freedom is too low to represent the flow dynamics. 
We note that FOMs are also inaccurate in the {\it marginally-resolved} regime, when the number of degrees of freedom is just enough to capture the flow dynamics.  
However, the FOM inaccuracy is less pronounced in the marginally-resolved regime than in the under-resolved regime.
A classic illustration of under-resolved and marginally-resolved FOMs is in the {\it convection-dominated} regime (i.e., at high Reynolds numbers, when the viscosity is low and the convective term dominates the diffusion term), when the FOM mesh-size is larger than the smallest flow scales. 
In this case, the FOMs generally yield {\it spurious numerical oscillations} that significantly degrade the FOM numerical accuracy.

To eliminate or alleviate the spurious numerical oscillations of the FOMs in convection-dominated, under-resolved and marginally-resolved simulations, various {\it numerical stabilization} methods have been proposed over the years.
Many of these FOM numerical stabilization approaches are surveyed in the research monograph of Roos, Stynes, and Tobiska~\cite{roos2008robust}.
{\it Regularized} models are a popular class of numerical stabilization methods in which {\it spatial filtering} is used to regularize (smooth) various terms in the NSE and eliminate (alleviate) the spurious numerical oscillations  of the FOMs in convection-dominated, under-resolved and marginally-resolved simulations.
A plethora of regularized models are surveyed in the research monograph of Layton and Rebholz~\cite{layton2012approximate}.
A classical regularized model is the {\it Leray} model (proposed in 1934 by the great mathematician Jean Leray~\cite{leray1934sur}), which regularizes the convective field in the NSE nonlinearity.
Another classical regularized model is the {\it evolve-filter-relax (EFR)} model, which consists of three steps:
(i) in the ``evolve" step, a standard FOM is used to obtain an intermediate approximation of the velocity;
(ii) in the ``filter" step, a spatial filter is used to filter (regularize) the intermediate approximation obtained in step (i) and eliminate (alleviate) its spurious numerical oscillations 
\cite{bowers2013numerical,boyd2001chebyshev, CHQZ88, fischer2001filter, germano1986differential-b,germano1986differential,GirfoglioQuainiRozza2019,GirfoglioQuainiRozza+2021+13+28, mullen1999filtering, pasquetti2002comments};
(iii) in the ``relax" step, a more accurate velocity approximation is obtained as a convex combination between the filtered and unfiltered flow approximations~\cite{bertagna2016deconvolution, ervin2012numerical}.
The EFR model is a popular regularized model that has been used for different classical numerical methods, e.g., the FE method~\cite{bertagna2016deconvolution,xu2020backflow} and the SE method~\cite{fischer2001filter}.
The main reasons for the popularity of the EFR model are its {\it simplicity} and {\it modularity}:
given a legacy FOM code, the ``evolve" step is already implemented, the ``filter" step requires the addition of a simple subroutine, and the ``relax" step is just one line of code.

To summarize the above discussion, when FOMs are used in the convection-dominated, under-resolved (marginally-resolved) regime, regularized models (e.g., the Leray or the EFR models) can be used to eliminate (alleviate) the spurious numerical oscillations.
In general, the need for numerical stabilization in the convection-dominated, under-resolved (marginally-resolved) regime is well known and well documented in the realm of classical numerical methods:
there are hundreds (if not thousands) of papers and several research monographs on this topic, and commercial software often includes numerical stabilization strategies for the under-resolved (marginally-resolved) regime.
Our goal is to investigate this topic in a reduced order modeling context.

{\it Reduced order models (ROMs)} are relatively low-dimensional computational models that can reduce the FOM computational cost by orders of magnitude
\cite{benner2015survey, benner2017model, hesthaven2015certified, prud2002reliable, quarteroni2015reduced, RozzaHuynhPatera2008}.
The basic ROM idea is to collect solutions of the system computed for several parameter values, build a relatively low-dimensional manifold, and then perform efficient simulations on this manifold for new parameter values. 
ROM strategies have been successfully applied in several contexts, from elliptic coercive problems \cite{hesthaven2015certified, RozzaHuynhPatera2008} to Stokes flows \cite{rozza2013reduced, rozza2007stability} 
to nonlinear frameworks \cite{ballarin2015supremizer, deparis2008reduced, deparis2009reduced}.
The ultimate ROM goal is to make an impact in important applications (e.g., uncertainty quantification, shape optimization, flow control, and data assimilation), where numerical simulations need to be repeated for a large number of physical and/or geometrical parameter values. 
In these applications, ROMs could represent an efficient alternative to FOMs, whose computational cost is generally prohibitively high.
We emphasize, however, that many of these practical applications take place in the convection-dominated regime.
Since in the convection-dominated, under-resolved (marginally-resolved) regime numerical stabilization plays a central role for FOMs, a natural question is whether numerical stabilization is also beneficial in the ROM setting.
To address this issue, one could first ask the following question:\\[-0.3cm]

\textbf{(Q1)} \textit{Assuming that the FOM is run in the convection-dominated, resolved regime and the ROM is run in the convection-dominated, under-resolved  (marginally-resolved) regime, is numerical stabilization needed for the ROM?
}\\[-0.3cm]

For Galerkin projection based ROMs, it was shown in~\cite{ballarin2015supremizer, giere2015supg,gunzburger2019evolve,pacciarini2014stabilized,wang2011two,wang2012proper,wells2017evolve} 
that, in the convection-dominated regime, under-resolved (marginally-resolved) ROMs (i.e., ROMs in which the number of ROM basis functions is too low to capture all the flow scales) yield numerical oscillations even though the snapshots used to construct the ROMs were generated by FOMs used in a resolved regime (i.e., with a sufficiently large number of degrees of freedom to capture all the flow scales).
Furthermore, it was also shown that {\it regularized ROMs (Reg-ROMs)}, e.g., the Leray ROM~\cite{kaneko2020towards,sabetghadam2012alpha,wells2017evolve} and the 
EFR-ROM~\cite{wells2017evolve}, can alleviate the spurious numerical oscillations and significantly increase the standard ROM accuracy.
Finally, in~\cite{wells2017evolve}, it was shown that the 
EFR-ROM was more accurate than the Leray ROM in the numerical simulation of a three-dimensional flow past a cylinder.
These results suggest that the answer to question (Q1) is that numerical stabilization is needed for Galerkin projection based ROMs in the convection-dominated, under-resolved  (marginally-resolved) regime and that Reg-ROMs can alleviate the numerical oscillations and increase the accuracy of standard ROMs.
(See~\cite{carlberg2011efficient} for an alternative approach, based on a least-squares Petrov-Galerkin projection.)

A natural follow-up question to (Q1) is the following:

\textbf{(Q2)} \textit{Assuming that both the FOM and the ROM are run in convection-dominated, marginally-resolved regime, if numerical stabilization is used for the FOM, is numerical stabilization still needed for the ROM?}\\[-0.3cm]

To our knowledge, question (Q2) is still open.
In this paper, we take a step in answering question (Q2).
Specifically, we consider two scenarios in which both the FOM and the ROM are run in the 
marginally-resolved regime.
In this setting, we compare two types of ROMs:
\begin{itemize}
    \item[(i)] {{EFR-noEFR}}, in which we use the EFR regularization at the FOM level but not at the ROM level; and 
    \item[(ii)] {EFR-EFR}, in which we use the EFR regularization both at the FOM level and at the ROM level.
    In this paper, we call this strategy {\it ``FOM-ROM consistency."}
\end{itemize}
We compare the {{EFR-noEFR}} and the {EFR-EFR} in the numerical simulation of a 2D \A{incompressible }flow past a circular cylinder with time-dependent Reynolds number~\cite{john2004reference,schafer1996benchmark}.
The numerical results show that the {EFR-EFR} is significantly more accurate than the {{EFR-noEFR}}.
Thus, these results suggest that the answer to question (Q2) is that, in the convection-dominated,  marginally-resolved regime, numerical stabilization should be used both at a FOM level and at a ROM level, i.e., that FOM-ROM consistency is beneficial.

To our knowledge, this is the first time the FOM-ROM consistency is investigated for the EFR regularization.
The FOM-ROM consistency has been advocated only in a few other settings, e.g., for classical residual based stabilization methods~\cite{ali2020stabilized,giere2015supg, pacciarini2014stabilized} and for a variational multiscale method~\cite{stabile2019reduced}.
The FOM-ROM consistency has also been investigated for a hybrid approach in~\cite{GirfoglioQuainiRozza2020}, where which the Leray model was combined with the EF algorithm.
We emphasize that our study is different from the numerical investigation in~\cite{GirfoglioQuainiRozza2020} in several key aspects\A{.}
\begin{enumerate}
    \item \A{The } main difference between the two investigations is the algorithm used:
    we use the EFR algorithm, whereas in~\cite{GirfoglioQuainiRozza2020} the authors use a combination of Leray and EF algorithms.
    In particular, 
    in our current investigation we use the ``relax'' step, whereas the investigation in~\cite{GirfoglioQuainiRozza2020} does not.
    This is a critical difference between the two investigations, since the ``relax'' step has been shown to be essential in increasing the algorithm's accuracy~\cite{bertagna2016deconvolution,ervin2012numerical}.
    \item An important difference between the two investigations is that in the current study we perform the model reduction both in time and in the parametric space.
    Specifically, we leverage a nested proper orthogonal decomposition (POD) approach to develop ROMs that include variations with respect to the Reynolds number, which is a critical parameter in practical ROM applications.
    In contrast, the investigation in~\cite{GirfoglioQuainiRozza2020} does not consider parametric variations with respect to the Reynolds number (although it includes a standard POD reduction strategy with parametric filter radius). 
    \item Another significant difference between the two investigations is the spatial discretization at a FOM level:
    in the current study we employ the FE method, whereas the investigation in~\cite{GirfoglioQuainiRozza2020} uses the FV method. 
    The FE and FV methods are two of the most used spatial discretizations in the numerical simulations of fluid flows.
    Since the FE and FV methods yield different ROM formulations (e.g., different ROM operators \cite{lorenzi2016pod}), it is important to investigate the FOM-ROM consistency in both settings.
\end{enumerate}

{
\begin{remark}
Although enforcing FOM-ROM consistency may seem a natural choice~\cite{barone2009stable,giere2015supg,kalashnikova2014reduced,rebollo2017certified}, there exist numerous investigations that are FOM-ROM inconsistent.
For example, for turbulent flow simulations, there exist investigations that use a closure model at the FOM level, but not at the ROM level:
On page 722 in~\cite{carlberg2017galerkin}, the authors note that the FOM data is generated by using the AERO-F code, which employs a DES turbulence model based on the Spalart–Allmaras one-equation model.
Instead of using a closure model, the ROM uses a least-squares Petrov-Galerkin (LSPG) formulation.
On page 17 in~\cite{grimberg2020stability}, the authors mention that the FOM data is generated by using Vreman's LES model.
However, the ROM utilizes the LSPG formulation instead of a closure model.
On page 15 in~\cite{wang2012proper} (see also Appendix A), the authors note that the FOM data is generated by using a DNS.
However, at the ROM level, the authors use several closure models of LES type (e.g., the dynamic SGS model).
In~\cite{mou2021data}, the author employs a regularized model to generate the FOM data, and a data-driven LES closure model to construct the ROM.
On page 604 in~\cite{wells2017evolve}, the authors mention that the FOM data is generated by the same type of DNS as that used in~\cite{wang2012proper}.
However, at the ROM level, the authors employ two types of regularized ROMs, i.e., the EF-ROM and the Leray-ROM.
On page 317 in~\cite{bergmann2018zonal}, the authors note that RANS equations are used to generate the FOM data.
However, the RANS equations are not used at the ROM level. 
There are, of course, examples of FOM-ROM inconsistencies with respect to discretization choices other than closure, e.g., time discretization~\cite{carlberg2017galerkin,wang2012proper,zucatti2021calibration}, 
nonlinearity discretization~\cite{ingimarson2021full}, and stabilization. 
Examples of 
FOM-ROM inconsistency with respect to stabilization include \cite{ali2020stabilized,pacciarini2014stabilized}, where the authors employ a FOM equipped with residual-based stabilization, and discuss advantages (e.g., accuracy) and disadvantages (e.g., computational cost) of including such stabilization in the ROM. 
The FOM-ROM inconsistency with respect to stabilization
is also investigated in the case of a residual-based variational multiscale (VMS) formulation in \cite{stabile2019reduced}, where it is shown that dropping the VMS terms 
in the ROM formulation leads to a considerable deterioration of the ROM accuracy.

The reason for the relative popularity of the FOM-ROM inconsistency is most probably its practical convenience:
to build the ROM, one is not restricted by the particular choices made in the FOM numerical discretization.
Thus, the FOM-ROM inconsistency belongs to the general class of approaches that consider the FOM data and the ROM as different entities. 
In this paper, we espouse a different line of thought in which the FOM and ROM are not completely independent.
Specifically, we show numerically that building ROMs that are consistent with the FOMs with respect to the particular numerical regularization used can yield more accurate solutions.

\end{remark}
}
The rest of the paper is outlined as follows: 
In Section \ref{sec:hf}, we describe the FOM 
and the EFR algorithm. 
In Section \ref{sec:g-rom}, we focus on ROMs for time reduction, and compare the {{EFR-noEFR}} and {EFR-EFR} 
in the numerical simulation of a 2D flow past a circular cylinder.
In Section \ref{sec:rom_mu}, we compare the {{EFR-noEFR}} and {EFR-EFR} when model reduction is performed both in time and in the Reynolds number.
Finally, in Section \ref{sec:conc}, we present conclusions and future research directions.

\section{The Full Order Model and the Evolve-Filter-Relax Algorithm}
\label{sec:hf}
In this section, we present the FOM and the EFR algorithm.
As a mathematical model, we use the \A{incompressible } Navier-Stokes equations (NSE).
Given a fixed domain $\Omega \subset \mathbb R^{D}$, with $D=2,3$, we consider the motion of an incompressible fluid having velocity $\bu \doteq \bu(\bx, t) \in \mathbb U$ and pressure $p \doteq p(\bx, t) \in \mathbb Q$ represented by the NSE:
\begin{equation}
\label{eq:NSE}
\begin{cases}
 \displaystyle \frac{\partial \bu}{\partial t} + (\bu \cdot \nabla) \bu - \nu \Delta \bu + \nabla p = 0 & \text{in }  \Omega \times (t_0, T), \\
\nabla \cdot \bu = 0  & \text{in }  \Omega \times (t_0, T), \\
\bu = \bu_D & \text{on } \partial \Omega_D \times (t_0, T), \\
\displaystyle -p \ \boldsymbol n + \nu \frac{\partial \boldsymbol \bu}{\partial \boldsymbol n} = 0  & \text{on }  \partial \Omega_N \times (t_0, T), \\
\end{cases}
\end{equation}
endowed with the initial condition $\bu = \bu_0$ in $\Omega \times \{ t_0 \}$, where $\partial \Omega_D \cup \partial \Omega_N = \partial \Omega$, $\partial \Omega_D \cap \partial \Omega_N = \emptyset$, $\nu$ is the kinematic viscosity, and $\mathbb U$ and $\mathbb Q$ are suitable Hilbert function spaces. The functions $\bu_D$ and $\bu_0$ are given. 

The flow regime is defined by the Reynolds number
\begin{equation}
\label{eq:Re}
Re \doteq \frac{UL}{\nu},
\end{equation}
where $U$ and $L$ represent the characteristic velocity and length scales of the system, respectively. 
When the Reynolds number is large, the inertial forces 
dominate the viscous forces; this setting is generally referred to as the convection-dominated regime.
As explained in the introduction, it is well known that in the convection-dominated regime standard spatial discretizations yield spurious numerical oscillations in under-resolved and marginally-resolved numerical simulations.
In our numerical investigations in Sections~\ref{sec:results} and \ref{sec_pred_time} and Section~\ref{sec:rom_mu}, we consider marginally-resolved simulations.
To alleviate the spurious numerical oscillations of standard spatial discretizations, we equip the FOM with the evolve-filter-relax (EFR) algorithm.
This strategy has been exploited with standard numerical discretization techniques, ranging from FE to SE to FV methods: see, e.g., \cite{boyd2001chebyshev, CHQZ88, fischer2001filter, germano1986differential-b,germano1986differential,GirfoglioQuainiRozza2019, GirfoglioQuainiRozza+2021+13+28, mullen1999filtering, pasquetti2002comments}.
In this paper, we use the FE method and a backward differentiation formula of order 1 (BDF1) for the space and time discretization, respectively. 

In what follows, we 
denote the semi-discrete FE velocity and pressure with $\bu \in \mathbb U^{N_h^{\bu}}$ and $p \in \mathbb Q^{N_h^p}$, respectively, where ${N_h^{\bu}}$ and ${N_h^p}$ are the corresponding numbers of degrees of freedom. We 
denote the time step with $\Delta t$. 
Let $t_n = t_0 + n\Delta t$ for $n = 0, \dots, N_T$, and $T = t_0 + N_T\Delta t$. 
We denote 
with $y^n$ the approximation of a generic quantity $y$ at the time $t^n$. The EFR algorithm at the time $t^{n+1}$ yields:
\begin{eqnarray}
         &	\text{\bf (I)}& \text{\emph{ Evolve}:} \quad 
\begin{cases}
        	 \displaystyle \frac{\bw^{n + 1} - \bu^n}{\Delta t} + (\bw^{n+1} \cdot \nabla) \bw^{n+1} - \nu \Delta \bw^{n+1} + \nabla p^{n+1} = 0 & \text{in } \Omega \times \{t_{n+1}\}, \vspace{1mm}\\
\nabla \cdot \bw^{n+1} = 0 & \text{in } \Omega \times \{t_{n+1}\}, \vspace{1mm}\\
\bw^{n+1} = \bu_D^{n+1} & \text{on } \partial \Omega_D \times \{t_{n+1}\}, \vspace{1mm}\\
\displaystyle -p^{n+1} \cdot \boldsymbol n + \frac{\partial \boldsymbol \bw^{n+1}}{\partial \boldsymbol n} = 0  & \text{on } \partial \Omega_N \times \{t_{n+1}\}. \\
\end{cases}
            \label{eqn:ef-rom-1}\nonumber \\[0.3cm]
            &	\text{\bf (II)} &\text{\emph{ Filter:}} \quad
\begin{cases} 
        	 -2 \delta^2 \, \Delta \overline{\bw}^{n+1} +  \overline{\bw}^{n+1} = \bw^{n+1}& \text{in } \Omega \times \{t_{n+1}\}, \vspace{1mm}\\
\overline{\bw}^{n+1} = \bu^{n+1}_D &\text{on } \partial \Omega_D \times \{t_{n+1}\}, \vspace{1mm}\\
\displaystyle \frac{\partial \overline{\bw}^{n+1}}{\partial \boldsymbol n} = 0  & \text{on } \partial \Omega_N \times \{t_{n+1}\}.
\end{cases}
	\label{eqn:ef-rom-2} \nonumber \\[0.3cm]
            &	\text{\bf (III)} &\text{\emph{Relax:}} \qquad 
        	   \bu^{n+1}
            = (1 - \chi) \, \bw^{n+1}
            + \chi \, \overline{\bw}^{n+1} \, ,
            \label{eqn:ef-rom-3}\nonumber
\end{eqnarray}
where $\chi \in [0,1]$ is a relaxation parameter. Here $\bw$ is the evolved velocity and $\overline \bw$ 
is the filtered velocity. 
We note that using $\bw^{n+1} = \bu^{n+1}$ in step (I) (i.e., the evolve step) is equivalent to solving the NSE.
In step (II), we use a \emph{differential filter} (DF) 
with an
explicit lengthscale, $\delta$, which 
is the \emph{filtering radius} 
(i.e., the radius of the neighborhood 
from which the spatial filter extracts information). 
The success of DFs 
is due to several appealing properties \cite{BIL05}. 
For example, the DF leverages an elliptic operator and acts as a spatial filter by eliminating the small scales (i.e., high frequencies) from the input data. 
Step (III) is a relaxation step in which the EFR velocity approximation at the new time step is defined as a linear combination of the approximations in Step (I) and Step (II).
The relaxation parameter $\chi$ diminishes the magnitude of the numerical diffusion \cite{ervin2012numerical, fischer2001filter,mullen1999filtering} and increases the accuracy; see, e.g., the numerical results in \cite{bertagna2016deconvolution} and the theoretical 
results in \cite{ervin2012numerical}.
The scaling $\chi \sim \Delta t$ is commonly used \cite{ervin2012numerical}.
In \cite{bertagna2016deconvolution, GirfoglioQuainiRozza2019}, however, the authors provide heuristic formulas that advocate higher values.


\section{Model Reduction 
With Respect to Time}
    \label{sec:g-rom}
In this section, we focus on our POD-Galerkin ROM framework for model reduction with respect to time. 
In Section \ref{POD_t}, we give a brief description of the POD algorithm \cite{ballarin2015supremizer, burkardt2006pod, hesthaven2015certified}. Then, in Sections \ref{EFR-noEFR_t} and \ref{sec:differential-filter}, we describe the two different ROM algorithms proposed, i.e., EFR-noEFR and EFR-EFR.
Finally, in Section \ref{sec:results}, we report and discuss some numerical experiments.   
All the ROM computations 
are performed with RBniCS \cite{rbnics}, which is a FEniCS-based \cite{fenics} library. 

\subsection{The POD algorithm}
\label{POD_t}
The basic idea of ROMs is to build a low-dimensional framework where the problem at hand can be solved more efficiently than the FOM.
To this end, 
assume that we have two bases, $\{\bphi_j\}_{j=1}^r$ and $\{\psi_j\}_{j=1}^r$,  for the reduced velocity and pressure spaces $\mathbb U^r$ and $\mathbb Q^r$, respectively, so that 
\begin{equation}
	\bu_r \doteq {\bu}_r(\bx,t)
	= \sum_{j=1}^r a_j^{\bu}(t) \bphi_j(\bx) \quad \text{and} \quad p_r \doteq {p}_r(\bx,t)
	= \sum_{j=1}^r a_j^{p}(t) \psi_j(\bx),
	\label{eqn:g-rom-1}
\end{equation}
where $\{a_{j}^{\bu}(t)\}_{j=1}^{r}$ and $\{a_{j}^{p}(t)\}_{j=1}^{r}$  are the sought time-varying coefficients \cite{noack2011reduced}. 
The bases are linear combinations of the snapshots, 
i.e., FOM solutions computed at properly chosen time instances, 
$\{\bu_i\}_{i=1}^{N_{\bu}} \subseteq \{\bu^k\}_{k=1}^{N_T}$ and $\{p_i\}_{i=1}^{N_{p}} \subseteq \{p^k\}_{k=1}^{N_T}$, where $N_{\bu}$ and $N_p$ 
denote the number of snapshots for velocity and pressure, respectively.
We utilize the EFR at the FOM level to generate the snapshots for both the EFR-noEFR and the EFR-EFR strategies.
We employ the POD algorithm \cite{ballarin2015supremizer, burkardt2006pod, hesthaven2015certified} to compress the snapshot information and to build the reduced spaces. 

It is well known that, in a standard NSE setting, the POD may be combined with a \emph{supremizer stabilization} for the reduced velocity space in order to guarantee the well-posedness of the system. 
We emphasize that the main role of the supremizers is to avoid spurious reduced pressure modes.
To tackle the convection-dominated, marginally-resolved regime, different approaches are needed.
The supremizer stabilization proposed in \cite{rozza2007stability} relies on a supremizer operator
$S: \mathbb Q^{N_h^p} \rightarrow{{\mathbb U}}^{N_h^{\bu}}$ defined as
\begin{equation}
(S(p), \boldsymbol \tau)_{\mathbb {U}} = \biggl( p, \nabla \cdot \boldsymbol \tau \biggr), \hspace{1cm} \forall \boldsymbol \tau \in \mathbb{U}^{N_h^{\bu}}.
\end{equation}
Then, the considered reduced velocity space is
\begin{equation}
{{\mathbb U}}^{r_{\bu s}} \doteq \text{POD}(\{\bu_i\}_{i=1}^{N_{\bu}})\oplus \text{POD}(\{S(p_i)\}_{i=1}^{N_{\bu}}),
\label{eqn:definition-rom-velocity-space}
\end{equation}
where $\{p_i\}_{i=1}^{N_{\bu}}$ in~\eqref{eqn:definition-rom-velocity-space} are the pressure snapshots related to the velocity 
snapshots, i.e.,  derived from  the solution $(\bu_i, p_i)_{i=1}^{N_{\bu}}$. 
However, a standard POD procedure is applied to $\{p_i\}_{i=1}^{N_p}$: 
$$
\mathbb Q^{r_p} \doteq \text{POD}(\{p_i\}_{i=1}^{N_{p}}),
$$
where only the first $r_p$ POD eigenpairs 
are retained to build the bases.
The supremizer technique leads to a reduced velocity space of dimension $r_{\bu s} = r_{\bu} + r_{s}$. 
We 
denote the enriched reduced velocity dimension with $r_{\bu s}$, and consider $\{\bphi_j\}_{j=1}^{r_{\bu s}}$ as the enlarged velocity-supremizer basis. 
We define the pressure basis 
as $\{\psi_k\}_{k=1}^{r_p}$.\\


\subsection{EFR-noEFR}
\label{EFR-noEFR_t}
The EFR-noEFR consists of 
the Galerkin projection of the 
NSE 
on the reduced space, which
leads to the solution of the following system: at the time $t^{n+1}$, 
find the pair $(\bu_r^{n+1}, p_r^{n+1})$ 
that solves  

\begin{equation}
\begin{cases}
	\displaystyle \left( \frac{\bu_{r}^{n+1} - \bu_{r}^{n}} {\Delta t} , \bphi_i \right)
	+ \nu \, \biggl( \nabla \bu_{r}^{n+1} , \nabla \bphi_i \biggr)
	+ \biggl( (\bu_{r}^{n+1} \cdot \nabla) \bu_{r}^{n+1} , \bphi_i \biggr) - \biggl( p_{r}^{n+1} , \nabla \cdot \bphi_i \biggr)
	= 0, \\
\biggl( \nabla \cdot \bu_{r}^{n+1},  \psi_k \biggr) = 0,
\end{cases}
\label{eqn:g-rom-2-n+1}
\end{equation}
for all $ i = 1, \ldots, r_{\bu s},$ and $j = 1, \ldots, r_p$.
Algebraically, we are looking for the 
$(n+1)$-st solution of
\begin{equation}
\begin{cases}
    \displaystyle \frac{1}{\Delta t}\mathsf{M}(\mathsf u^{n+1} - \mathsf u^{n}) + 
    \nu \mathsf K \mathsf u^{n+1} + \mathsf C(\mathsf u^{n+1})\mathsf u^{n+1} - \mathsf {B^T} \mathsf p^{n+1} = 0,\\
    \mathsf B \mathsf u^{n+1} = 0,
    \end{cases}
\end{equation}
where $\mathsf u^{n+1} \in \mathbb R^{r_{\bu s}}$ and $\mathsf p^{n+1} \in \mathbb R^{r_{p}}$ are the vectors of the reduced coefficients of \eqref{eqn:g-rom-1} and represent the unknowns of the problem, $\mathsf M$ is the reduced velocity space mass matrix, $\mathsf K$ the reduced stiffness matrix, i.e., 
\begin{equation}
\label{eq:MK}
    \mathsf M_{ij} \doteq
    \biggl ( \bphi_{i} , \bphi_j \biggr)
    \quad \text{and} \quad 
    \mathsf K_{ij} \doteq \biggl( \nabla \bphi_{i},  \nabla \bphi_j \biggr),
\end{equation} while 
\begin{equation}
    \label{eq:CB}
    \mathsf C(\mathsf u^{n+1})_{ij} \doteq
    \biggl( (\mathsf u^{n+1} \cdot \nabla) \bphi_{i} , \bphi_j \biggr)
    \quad \text{and} \quad 
    \mathsf B_{ij} \doteq \biggl( \nabla \cdot \bphi_{i},  \psi_j \biggr).
\end{equation}
In Algorithm  \ref{alg:01}, we 
present the pseudocode for EFR-noEFR: the POD bases are extracted from EFR solutions, the supremizer enrichment is performed, and 
the standard NSE are projected on the reduced spaces. 

\begin{algorithm}
\caption{Pseudocode for EFR-noEFR}\label{alg:01}
\begin{algorithmic}[1]
\State{$\bu_0, \bu_{in}, N_{\bu}, N_{p}$}\Comment{Inputs needed}
\For{$n \in \{0, \dots, N_T - 1\}$}\Comment{Time loop}
\State{(I) + (II) + (III)} \Comment{EFR simulation}
\EndFor
\State{$\{\bu_i\}_{i=1}^{N_{\bu}} \subseteq \{\bu^k\}_{k=1}^{N_{T}}$ 
\quad $\{p_i\}_{i=1}^{N_{p}} \subseteq \{p^k\}_{k=1}^{N_{T}}$} \Comment{Snapshot collection}
\State{$\mathbb U^r \doteq \text{POD}(\{\bu_i\}_{i=1}^{N_{\bu}})\oplus \text{POD}(\{S(p_i)_{i=1}^{N_{\bu}})\}$ } \Comment{Supremizer enrichment for velocity space}
\State{$\mathbb Q^{r_p} \doteq \text{POD}(\{p_i\}_{i=1}^{N_p})$ } \Comment{Standard POD for pressure space}
\For{$n \in \{0, \dots, N_T - 1\}$}\Comment{Time loop}
\State{Solve system \eqref{eqn:g-rom-2-n+1}} \Comment{
Standard Galerkin projection}
\EndFor
\end{algorithmic}
\end{algorithm}
\subsection{EFR-EFR}
	\label{sec:differential-filter}
In the EFR-EFR, we apply a double stabilization. 
Specifically, we employ the EFR algorithm 
not only at the FOM level, but also at the ROM level.
Indeed, after 
the POD modes are built 
from the EFR snapshots, 
we apply the EFR steps (I), (II), and (III) in a reduced setting, as specified in Algorithm \ref{alg:02}: 

\begin{eqnarray*}
         &	\text{\bf (I)}_r& \text{\emph{}} \quad 
\begin{cases}
	\displaystyle \left( \frac{\bw_{r}^{n+1} - \bu_{r}^{n}} {\Delta t} , \bphi_i \right)
	+ \nu \, \biggl( \nabla \bw_{r}^{n+1} , \nabla \bphi_i \biggr)
	+ \biggl( (\bw_{r}^{n+1} \cdot \nabla) \bw_{r}^{n+1} , \bphi_i \biggr) - \biggl( p_{r}^{n+1} , \nabla \cdot \bphi_k \biggr)
	= 0, \\
\biggl( \nabla \cdot \bw_{r}^{n+1},  \psi_i \biggr) = 0,
\end{cases}
            \label{eqn:ef-rom-1-r}\nonumber \\[0.3cm]
            &	\text{\bf (II)}_r &\text{\emph{}} \quad
        	 2 \delta^2 \, \biggl( \nabla \overline{\bw}^{n+1}_r, \nabla \bphi_i \biggr)  +  \biggl ( \overline{\bw}^{n+1}_r, \bphi_i \biggr) = \biggl (\bw^{n+1}_r, \bphi_i \biggr),\\
	\label{eqn:ef-rom-2-r} \nonumber \\[0.3cm]
            &	\text{\bf (III)}_r &\text{\emph{}} \qquad 
        	   \bu^{n+1}_r
            = (1 - \chi) \, \bw^{n+1}_r
            + \chi \, \overline{\bw}^{n+1}_r.
            \label{eqn:ef-rom-3}\nonumber
\end{eqnarray*}
As we did in \eqref{eqn:g-rom-1}, we expand the reduced variables $\bw_r$ and $\overline \bw_r$ of $\mathbb U^{r}$ as \begin{equation}
	\bw_r \doteq {\bw}_r(\bx,t)
	= \sum_{j=1}^r a_j^{\bw}(t) \bphi_j(\bx) \quad \text{and} \quad \overline{\bw}_r \doteq {\overline{\bw}}_r(\bx,t)
	= \sum_{j=1}^r a_j^{\overline{\bw}}(t) \bphi_j(\bx).
	\label{eqn:ws}
\end{equation}
Thus, at the time instance $t^{n+1}$, we solve the following system:
\begin{equation}
\begin{cases}
    \displaystyle \frac{1}{\Delta t}\mathsf{M}(\mathsf w^{n+1} - \mathsf u^n) + 
    \nu \mathsf K \mathsf w^{n+1} + \mathsf C(\mathsf w^{n+1})\mathsf w^{n+1} - \mathsf {B^T} \mathsf p^{n+1} = 0,\\
    \mathsf B \mathsf w^{n+1} = 0, \\
    2\delta^2 \mathsf K \overline {\mathsf {w}}^{n+1} + 
    \mathsf M \overline{ \mathsf w}^{n+1} = \mathsf M \mathsf w^{n+1},\\
    \mathsf u^{n+1} = (1 -\chi) \mathsf w^{n+1} + \chi \overline{\mathsf w}^{n+1},
    \end{cases}
    \label{eqn:EFR-EFR-system}
\end{equation}
where $\mathsf w^{k+1} \in \mathbb R^{r_{\bu s}}$ and $\overline{\mathsf w}^{k+1} \in \mathbb R^{r_{\bu s}}$ are the unknown reduced coefficient vectors of the evolved and filtered velocity fields, respectively, as defined in \eqref{eqn:ws}. All the matrices in~\eqref{eqn:EFR-EFR-system} have been defined in \eqref{eq:MK} and \eqref{eq:CB}.
Although the DF 
has been widely used in a ROM framework \cite{GirfoglioQuainiRozza2020,fluids6090302,iliescu2017regularized,sabetghadam2012alpha,wells2017evolve,xie2017approximate}, to the best of our knowledge, the analysis of a \emph{Relax} step is still limited \cite{gunzburger2019evolve}.
We stress that, {at the ROM level,} the computational effort of the DF filter (II)$_r$ and the relaxation step (III)$_r$ is negligible and thus the costs of the 
EFR-noEFR and EFR-EFR will be comparable. \A{Moreover, the ROM model is consistent with respect to the choice of $\delta$ and $\chi$, 
which are the same 
as those used in the FOM model.} \\
\A{For the sake of clarity, in Table \ref{tab:acronyms}, we report 
all the acronyms that we use, 
together with the 
corresponding equations or algorithms. 
We also note that, in what follows, we make no distinction between the EFR and FOM simulation, since in the FOM numerical results 
the EFR strategy 
is always performed.}

\begin{table}[]
\caption{\A{Acronyms.}}
\label{tab:acronyms}
\centering {%
\begin{tabular}{|c|c|c| c |}
\hline
Acronym & Equation or Algorithm & Offline Stabilization & Online Stabilization \\ \hline
EFR (or FOM)     &   {(I) + (II) + (III)}    &  \checkmark      &                 \\ \hline
EFR-noEFR & Algorithm \ref{alg:01} & \checkmark &  \\ \hline
EFR-EFR &  Algorithm \ref{alg:02}     &   \checkmark       &      \checkmark      \\ \hline
EFR-noEFR and n-POD & Algorithm \ref{alg:03} & \checkmark &  \\ \hline
EFR-EFR and n-POD& Algorithm \ref{alg:04} & \checkmark & \checkmark \\ \hline
\end{tabular}%
}
\end{table}

\begin{remark}
We remark that, for the NSE, the online phase still depends on the FOM dimension and this affects the 
EFR-noEFR and EFR-EFR performances in terms of computational time. For this reason we are not presenting a comparative analysis between the FOM and the ROM solutions with respect to the computational costs. To overcome 
this issue, hyper-reduction techniques, such as the empirical interpolation method (EIM), may be employed, see, e.g. \cite{barrault2004empirical} or \cite[Chapter 5]{hesthaven2015certified}. However, this goes beyond the scope of the present work.
\end{remark}


\begin{algorithm}
\caption{Pseudocode for EFR-EFR}\label{alg:02}
\begin{algorithmic}[1]
\State{$\bu_0, \bu_{in}, N_{\bu}, N_{p}$}\Comment{Inputs needed}
\For{$n \in \{0, \dots, N_T - 1\}$}\Comment{Time loop}
\State{(I) + (II) + (III)} \Comment{EFR simulation}
\EndFor
\State{$\{\bu_i\}_{i=1}^{N_{\bu}} \subseteq \{\bu^k\}_{k=1}^{N_{T}}$ 
\quad $\{p_i\}_{i=1}^{N_{p}} \subseteq \{p^k\}_{k=1}^{N_{T}}$} \Comment{Snapshot collection}
\State{$\mathbb U^r \doteq \text{POD}(\{\bu_i\}_{i=1}^{N_{\bu}})\oplus \text{POD}(\{S(p_i)_{i=1}^{N_{\bu}})\}$ } \Comment{Supremizer enrichment for velocity space}
\State{$\mathbb Q^{r_p} \doteq \text{POD}(\{p_i\}_{i=1}^{N_p})$ } \Comment{Standard POD for pressure space}
\For{$n \in \{0, \dots, N_T - 1\}$}\Comment{Time loop}
\State{(I)$_r$ + (II)$_r$ + (III)$_r$} \Comment{EFR at the reduced level}
\EndFor
\end{algorithmic}
\end{algorithm}

\subsection{Numerical results: 
\A{reconstructive regime}}
\label{sec:results}
In this section, we 
analyze and compare the performances of {EFR-noEFR} (see Algorithm \ref{alg:01}) and {EFR-EFR} (see Algorithm \ref{alg:02}). 
The goal is to investigate the FOM-ROM consistency for the EFR stabilization algorithm.
We consider \A{an incompressible } 2D flow past a cylinder at time-dependent Reynolds number $0 \leq Re \leq 100$. This benchmark has been 
thoroughly studied at full order level \cite{john2004reference,GirfoglioQuainiRozza2020,schafer1996benchmark}. 

We consider the motion of an incompressible flow in a domain $\Omega \doteq \{[0, 2.2] \times [0, 0.41]\} \setminus \{(x, y)\in \mathbb R^2 \text{ such that } (x - 0.2)^2 + (y - 0.2)^2 - 0.05^2 = 0 \}$, which is depicted in Figure \ref{fig:domain}. 
We set $\nu = 10^{-3}$ and use no-slip boundary conditions on $\partial \Omega_D^{wall}$, representing the union of the lower and upper walls of the channel, and the cylinder wall (solid blue boundary in Figure \ref{fig:domain}), with a time varying inlet velocity profile $\bu_{in}$ on $\partial \Omega_D^{in}$ (red dashed line in Figure \ref{fig:domain}). The prescribed inlet condition is given by
\begin{equation}
\label{eq:inlet}
\bu_{in} \doteq \left ( \frac{0.6}{0.41^2}\sin(\pi t/8)y(0.41 - y), 0 \right).
\end{equation}
Furthermore, on $\partial \Omega_N$ (black line in Figure \ref{fig:domain}) we employ homogeneous Neumann conditions.
The value of the initial condition $\bu_0$ is $(0,0)$. The time-dependent inlet velocity leads to a \A{Reynolds number 
that varies in time\cite{schafer1996benchmark}, with $0 \leq Re \leq 100$. }
We perform our tests on a triangular mesh with $h_{min} = 4.46 \cdot 10^{-3}$ and $h_{max} = 4.02 \cdot 10^{-2}$. We employ the Taylor-Hood $\mathbb P^2 - \mathbb P^1$ FE pair for velocity and pressure, respectively, and this leads to a FE space of dimension $N_h \doteq N_h^{\bu} + N_h^p = 14053$. 
\begin{remark}
We 
note that the mesh 
that we use in our numerical investigation does not feature the level of refinement required by a direct numerical simulation, 
which is about $100k$ degrees of freedom \cite{john2004reference}. Therefore, although the Reynolds number is not 
high, the computational setting is still relatively challenging for the FOM and ROM simulations.
For the EFR validation at the FOM level performed by investigating this benchmark on coarse meshes, the reader 
is referred to~\cite{bertagna2016deconvolution, GirfoglioQuainiRozza2019, layton2012modular}.  
\end{remark}

\begin{figure}
\centering
\includegraphics[scale=0.4]{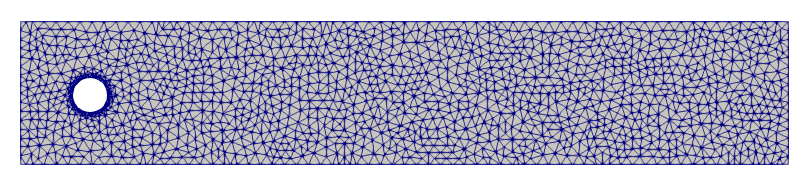}
\caption{{The 
FE mesh.}
}
\label{fig:mesh}
\end{figure}
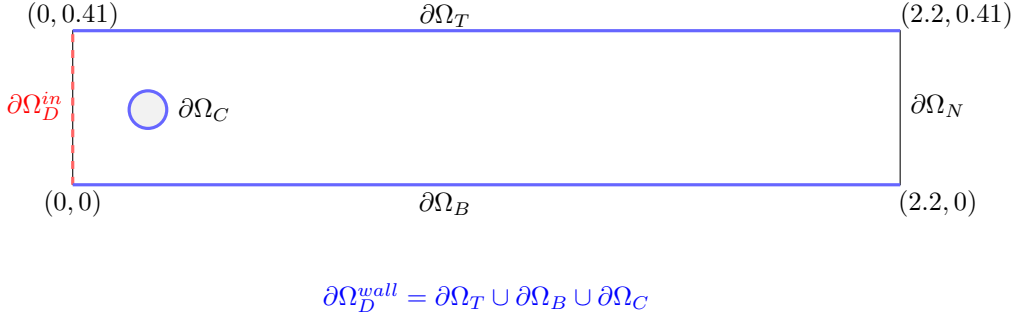
\begin{figure}
\begin{center}
\begin{tikzpicture}[scale=5.0]

\draw (0,0) -- (2.2,0) -- (2.2,0.41) -- (0,0.41) -- (0,0);
\filldraw[color=blue!60, fill=gray!10, very thick](0.2,0.2) circle (0.05);
\filldraw[color=red!60, fill=gray!10, very thick, dashed](0,0) -- (0.,0.41);
\filldraw[color=blue!60, fill=gray!10, very thick](0,0.41) -- (2.2,0.41);
\filldraw[color=blue!60, fill=gray!10, very thick](0,0.) -- (2.2,0.);
%
%
%
\node at (-.1,0.21){\color{red}{$\partial \Omega_{D}^{in}$}};
\node at (1.1,-0.3){\color{blue}{$\partial \Omega_{D}^{wall} = \partial \Omega_{T} \cup 
\partial \Omega_{B} \cup \partial \Omega_C$}};
\node at (2.3,0.21){\color{black}{$\partial \Omega_{N}$}};
\node at (0,-.05){\color{black}{$(0,0)$}};
%
\node at (0,0.45){\color{black}{$(0,0.41)$}};
\node at (2.35,0.45){\color{black}{$(2.2,0.41)$}};
\node at (2.3,-0.05){\color{black}{$(2.2,0)$}};
\node at (0.35,.2){\color{black}{$\partial \Omega_C$}};
\node at (0.99,.45){\color{black}{$\partial \Omega_T$}};
\node at (0.99,-.05){\color{black}{$\partial \Omega_B$}};
%

\end{tikzpicture}
\end{center}
\caption{The computational domain, $\Omega$. $\partial \Omega_D \doteq \partial \Omega_D^{in} \cup \partial \Omega_D^{wall}$, where the inlet boundary $\partial \Omega_D^{in}$ is represented by a dashed red line
and the no-slip boundaries by a solid blue line.
}
\label{fig:domain}
\end{figure}

For both the FOM and the ROM simulations, we
set $\Delta t = 4\cdot 10^{-4}$ and $\delta =  L \cdot Re^{-3/4} = 0.0032$ as in \cite{GirfoglioQuainiRozza2020} (i.e., we set the filter radius to the Kolmogorov scale \cite{kolmogorov1941dissipation,kolmogorov1941local}). 
\A{
\begin{remark}
\label{remark:delta}
The value of $\delta$ is still an arguable choice in CFD applications. For unstructured meshes, a common choice is to set the filtering radius as the $h_{min}$. Indeed, $\delta = h_{min}$ avoids an excessive diffusion action of system over the elements of the mesh \cite{bertagna2016deconvolution}.
However, in the following experiments, we used the Kolmogorov scale $L \cdot Re^{-\frac{3}{4}}$ 
to be consistent with the parameters of \cite{GirfoglioQuainiRozza2019}. Moreover, we underline that $L \cdot Re^{-\frac{3}{4}} \sim h_{min}$, thus the choice is reasonable.
\end{remark}}
We define the $L^2$ relative errors for the velocity and the pressure fields, respectively, as
\begin{equation}
    \label{errors}
    E_{\bu}(t) \doteq \frac{\norm{\bu(t) - \bu_r(t)}_{L^2(\Omega)}}{\norm{\bu(t)}_{L^2(\Omega)}}
\quad \text{and} \quad 
E_{p}(t) \doteq \frac{\norm{p(t) - p_r(t)}_{L^2(\Omega)}}{\norm{p(t)}_{L^2(\Omega)}}.
\end{equation}
Furthermore, we test the ROM accuracy 
by using 
the drag coefficient
\begin{equation}
\label{eq:drag}
C_D(t) \doteq \frac{2}{U^2L}\int_{\partial \Omega_C}((2\nu\nabla \bu - p\boldsymbol {I}) \cdot \boldsymbol {n}_C)\cdot \boldsymbol {t}_C \;ds,
\end{equation}
and the lift coefficient
\begin{equation}
\label{eq:lift}
C_L(t) \doteq \frac{2}{U^2 L}\int_{\partial \Omega_C}((2\nu\nabla \bu - p\boldsymbol {I}) \cdot \boldsymbol {n}_C)\cdot \boldsymbol {n}_C \;ds,
\end{equation}
where $ \boldsymbol {n}_C$ and $ \boldsymbol {t}_C$ are the normal and tangential unit vectors to the cylinder boundary $\partial \Omega_C$ (see Figure \ref{fig:domain}), respectively. 
Specifically, we compute the $L^2$-errors of the force coefficients:
\begin{equation}
\label{drag_err}
\A{
\overline{E}_{C_D}\doteq \frac{\norm{C_D - \overline{C}_D}_{L^2(t_0,T)}}{\norm{C_D}_{L^2(t_0,T)}}
\quad \text{and} \quad
\hat{E}_{C_D} \doteq \frac{\norm{C_D - \hat{C}_D}_{L^2(t_0,T)}}{\norm{C_D}_{L^2(t_0,T)}},}
\end{equation}
\begin{equation}
\A{
\label{lift_err}
\overline{E}_{C_L} \doteq \frac{\norm{C_L - \overline{C}_L}_{L^2(t_0,T)}}{\norm{C_L}_{L^2(t_0,T)}}
\quad \text{and} \quad
\hat{E}_{C_L} \doteq \frac{\norm{C_L - \hat{C}_L}_{L^2(t_0,T)}}{\norm{C_L}_{L^2(t_0,T)}}.}
\end{equation}
We 
denote the {EFR-noEFR} drag and lift coefficients 
as $\overline C_D(t)$ and $\overline C_L(t)$, respectively. 
Similarly, we denote the {EFR-EFR} drag and lift coefficients as 
$\hat C_D(t)$ and $\hat C_L(t)$. 
\\

\A{\begin{remark}
\label{remark:graddiv}
At the FOM level, 
one has to tackle the issue of preserving the incompressibility constraint when applying the DF (II). Indeed, while the 
DF 
preserves the incompressibility under periodic boundary conditions, 
it does not preserve the incompressibility under no-slip boundary conditions \cite{ervin2012numerical, layton2008numerical, van2006incompressibility}. In our specific test case, this might translate in unacceptable divergence values near the cylinder boundary, $\partial \Omega_C$. In \cite{ervin2012numerical, layton2008numerical}, a Stokes differential filter is proposed as a solution to recover the mass conservation at the 
DF level. 
However, as underlined by the authors, this is a more expensive filtering operation.
We decided to address 
this problem by exploiting another technique, easier to implement and which gave us acceptable 
divergence values: 
a div-grad stabilization that penalizes the violation of the incompressibility constraint \cite{Heavner2017,john2017divergence}. Namely, 
in the 
DF equations, we added a term of the form
\begin{equation}
\gamma \nabla(\nabla \cdot \overline{\bw}^{n+1}),
    \end{equation}
with $\gamma = 100$, as used in \cite{Heavner2017}. The reader interested in an overview 
of grad-div stabilization and 
the 
choice of $\gamma$ may refer to \cite{decaria2018determination,john2017divergence,layton2009accuracy}. 

In 
our investigation, we present numerical experiments where the relaxation is considered and others where it is not. When $\chi \neq 1$, we do not use the grad-div stabilization at the FOM level.
The rationale for our choice is the following.
When we utilize the \emph{Relax} step (III), it expresses the velocity approximation as a convex combination of the intermediate velocity approximation obtained in the \emph{Evolve} step (I) and the filtered velocity approximation  obtained in the \emph{Filter} step (II). 
Since the \emph{Evolve} step (I) enforces the incompressibility constraint in the intermediate velocity approximation, the velocity approximation in the \emph{Relax} step (III) displays low divergence values and the grad-div stabilization is no longer needed in this scenario.

We also note that other techniques 
may be employed to achieve divergence-free snapshots~\cite{john2017divergence}, such as the filters described in \cite{fluids6090302}.
\end{remark}}

\begin{figure}
\centering
        \includegraphics[scale=0.4]{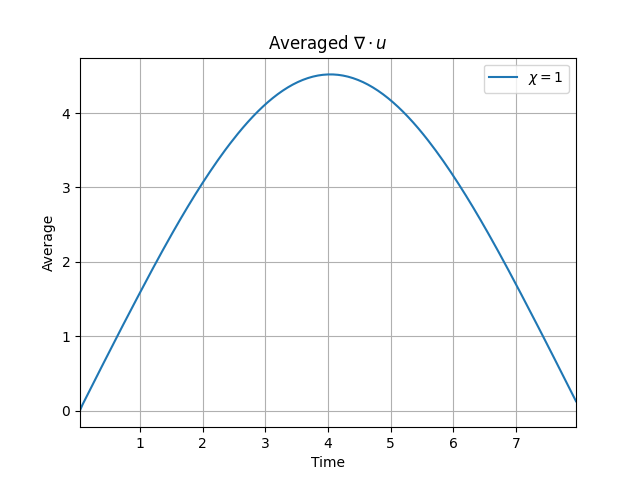}
         \includegraphics[scale=0.4]{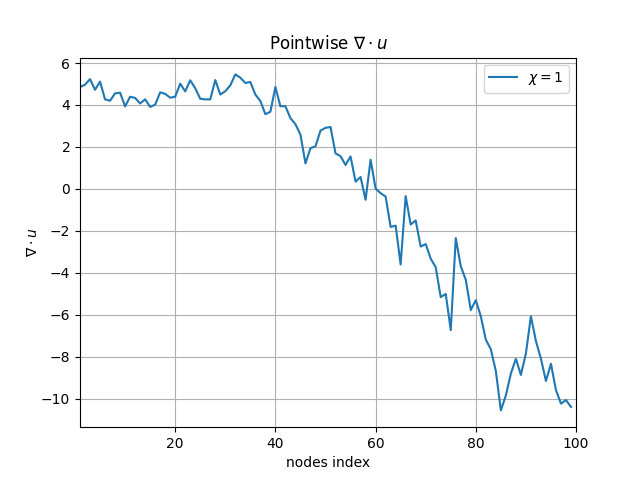} \\
        \includegraphics[scale=0.4]{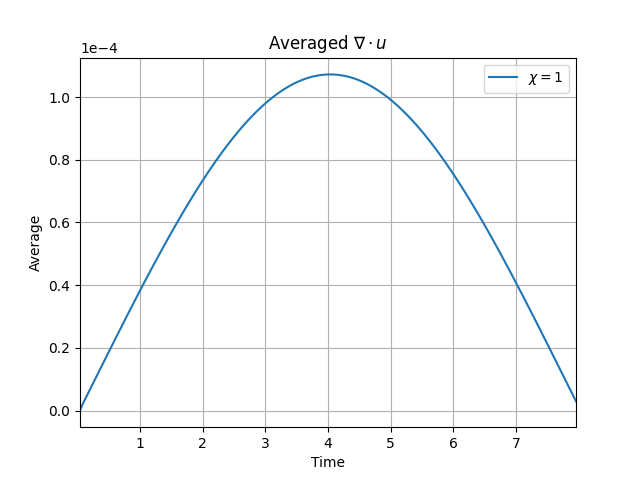}
         \includegraphics[scale=0.4]{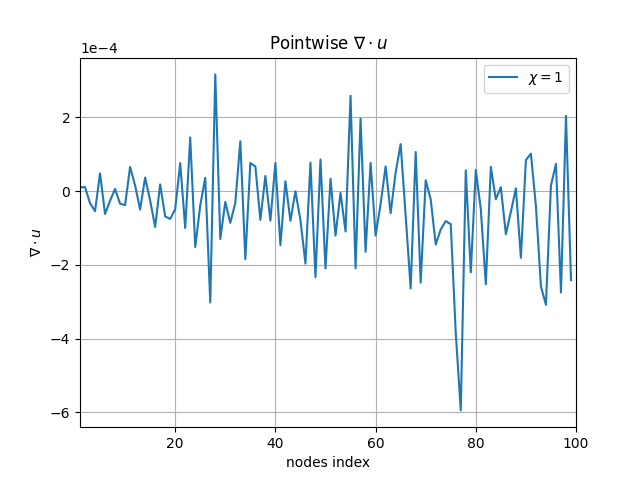}
\caption{\scriptsize{ (Experiment 1: $\delta=0.0032$, $\chi = 1$, 
full order EFR results.) \emph{Top Left and Top Right.} $Av(t)$ and divergence nodal values over the cylinder without grad-div stabilization. \emph{Bottom Left and Bottom Right.} 
$Av(t)$ and divergence nodal values over the cylinder with grad-div stabilization. 
}}
\label{fig:div}
\end{figure}

{\bf Experiment 1}. 
As a first step in the comparison of 
\A{{EFR-noEFR} } and 
\A{{EFR-EFR}}, we consider the EF stabilization strategy both at the FOM level and at the ROM level.
That is, we discard the \emph{Relax} step (i.e., we consider $\chi = 1$) both for 
\A{{EFR-noEFR} and {for EFR-EFR}}.
We note that, in a FOM (and, consequently, in a ROM) setting, the EF algorithm 
is over-diffusive and high frequency modes are completely damped (see, e.g., \cite{GirfoglioQuainiRozza2019}).
\A{As already specified in Remark \ref{remark:graddiv}, we apply the grad-div stabilization at a FOM level to 
enforce the incompressibility constraint along the cylinder boundary. }
The improvements with respect to the incompressibility of the flow are 
displayed in Figure \ref{fig:div}. 
We denote with $(x_i^c, y_i^c)_{i=1}^{N_c}$ the mesh nodal coordinates related to the cylinder boundary $\partial \Omega_C$. In our case, $N_c = 100$. We plot 
\begin{equation}
Av(t) = \frac{1}{N_c}\sum_{i}^{N_c} |\nabla \cdot \bu(x_i, y_i)|,
\end{equation}
i.e., the averaged absolute value of the divergence over the nodal coordinates at time $t$, and the nodal values $\nabla \cdot \bu(x_i, y_i)$, for $1 \leq i \leq 100$ for 
$t=4$, which 
is the time instance with the worst behaviour with respect to the incompressibility constraint violation (as can be seen 
in the left 
panel of Figure \ref{fig:div}).  
As expected, the use of the grad-div stabilization allows us to reach much smaller divergence values in time. Indeed, from values of order $O(1)$, thanks to the grad-div stabilization term, we obtain values 
of order $O(10^{-4})$. 

We note that we utilize the grad-div stabilization 
at the FOM level, but not at the ROM level.
\A{Thanks to the grad-div stabilization, the velocity snapshots 
display acceptable 
divergence values. 
Thus, the divergence values of the ROM velocity approximations are relatively low and the grad-div stabilization is 
not needed at the ROM level, i.e.\ FOM and ROM are not consistent with respect to grad-div stabilization. Moreover, in our specific case, 
using the grad-div stabilization at the ROM level with the same parameters as those used at the FOM level
leads to an over-diffusive reconstruction of the aerodynamics coefficients at the ROM level. } 
For the application of 
the grad-div stabilization at the 
ROM level, the interested reader 
is referred to \cite{caiazzo2014numerical}. 

\begin{figure}
        \centering \includegraphics[scale=0.21]{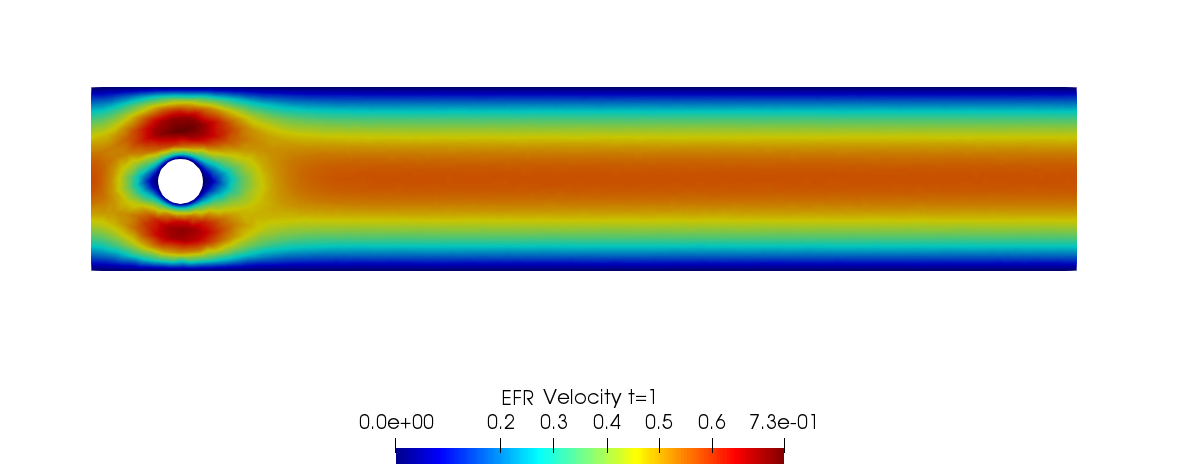}\\
\begin{minipage}{.49\textwidth}
 \centering
         \includegraphics[scale=0.21]{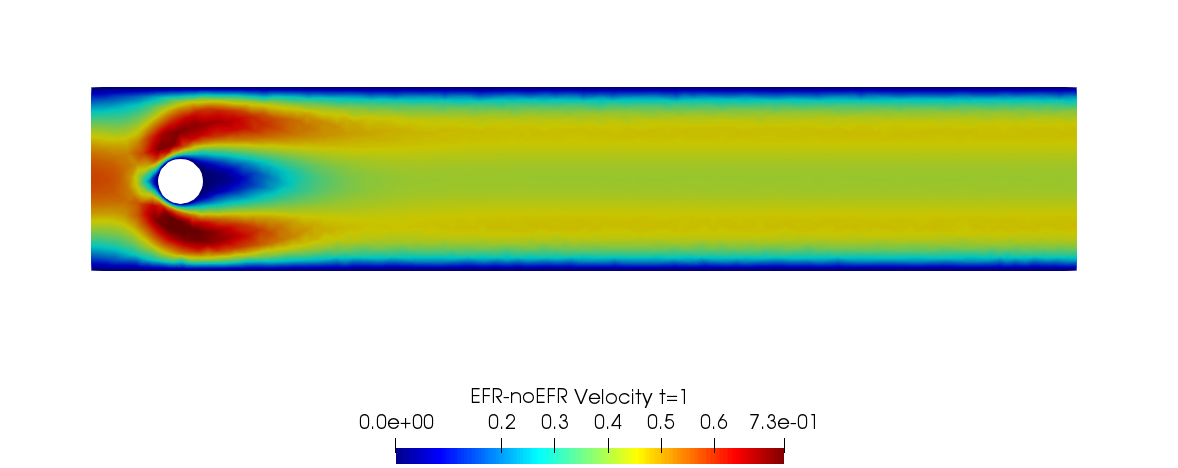}
\end{minipage}
\begin{minipage}{.49\textwidth}
 \centering
         \includegraphics[scale=0.21]{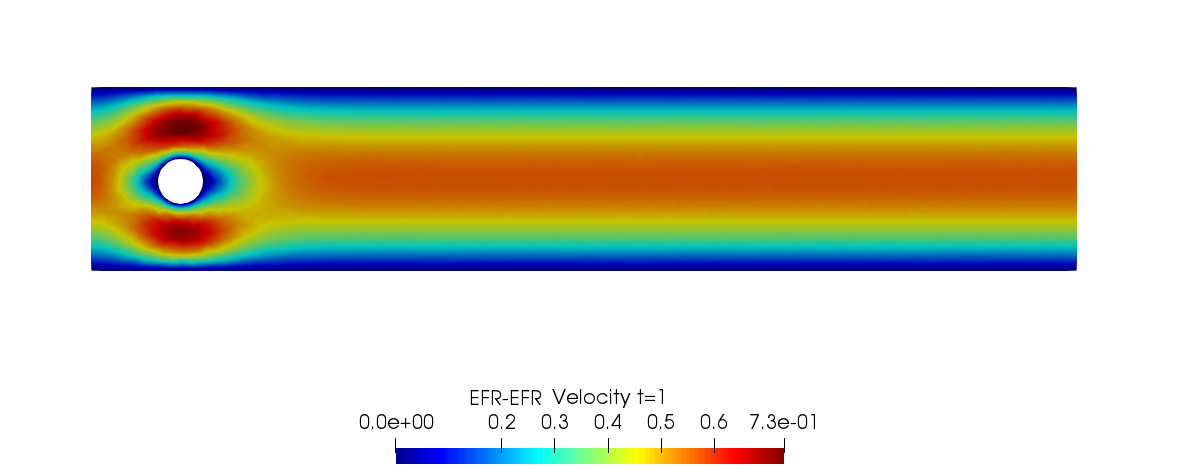}
\end{minipage}
\caption{\scriptsize{(Experiment 1: $\delta=0.0032$, $\chi = 1$, $r=2$ and $t=1$.) \emph{Top.} Full order EFR velocity magnitude. \emph{Bottom Left.} Reduced {EFR-noEFR} velocity magnitude. \emph{Bottom Right.} Reduced {EFR-EFR} velocity magnitude.}
}
\label{fig:u_chi1}
\end{figure}

\begin{figure}
        \centering \includegraphics[scale=0.21]{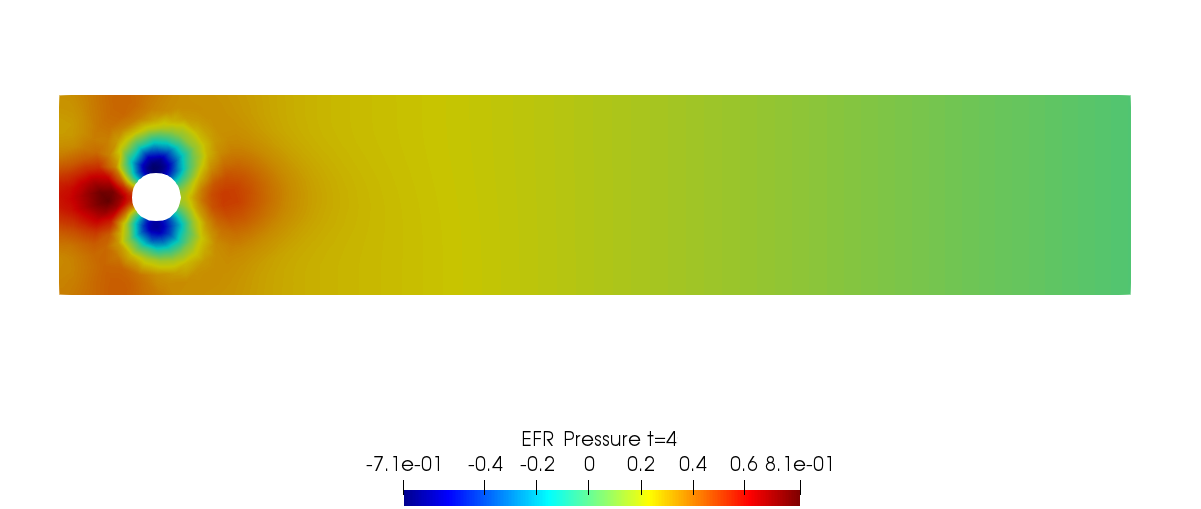}\\
\begin{minipage}{.49\textwidth}
 \centering
         \includegraphics[scale=0.21]{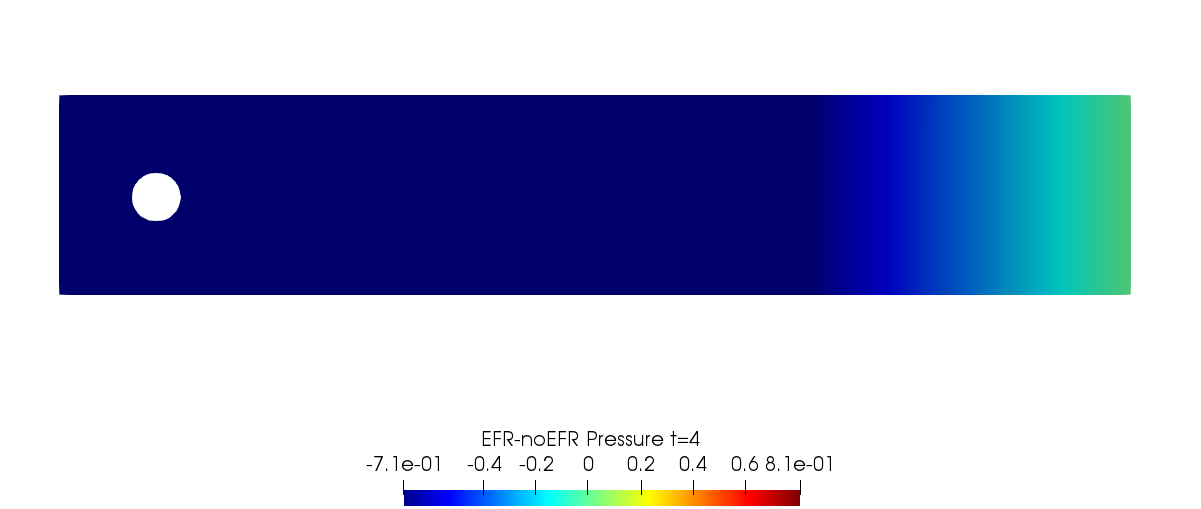}
\end{minipage}
\begin{minipage}{.49\textwidth}
 \centering
         \includegraphics[scale=0.21]{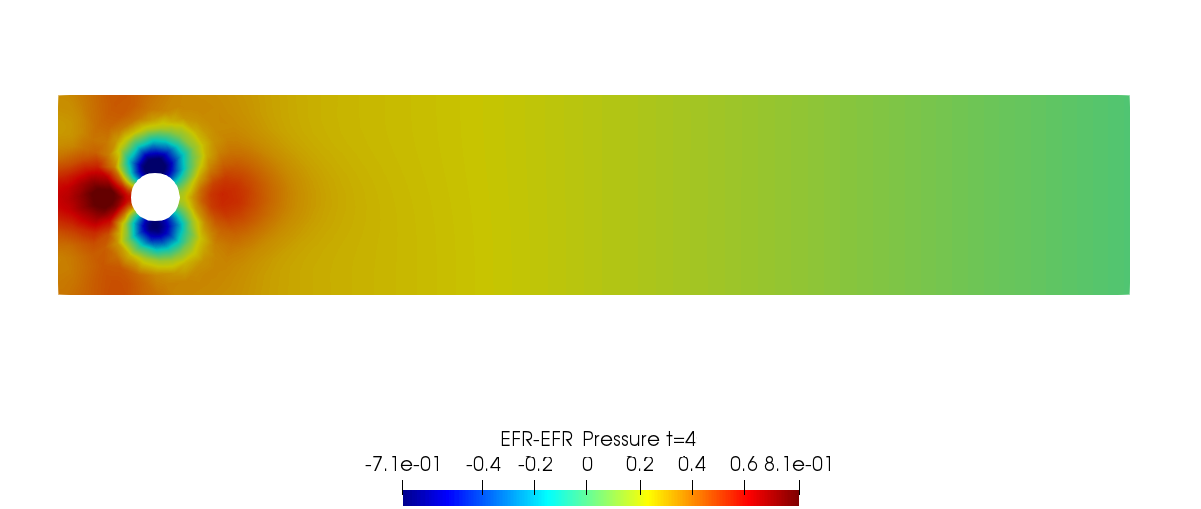}
\end{minipage}
\caption{\scriptsize{(Experiment 1: $\delta=0.0032$, $\chi = 1$, $r=2$ and $t=4$.) \emph{Top.} Full order EFR pressure field. \emph{Bottom Left.} Reduced {EFR-noEFR} pressure field. \emph{Bottom Right.} Reduced {EFR-EFR} pressure field.}}
\label{fig:p_chi1}
\end{figure}

\begin{figure}
\centering
        \includegraphics[scale=0.4]{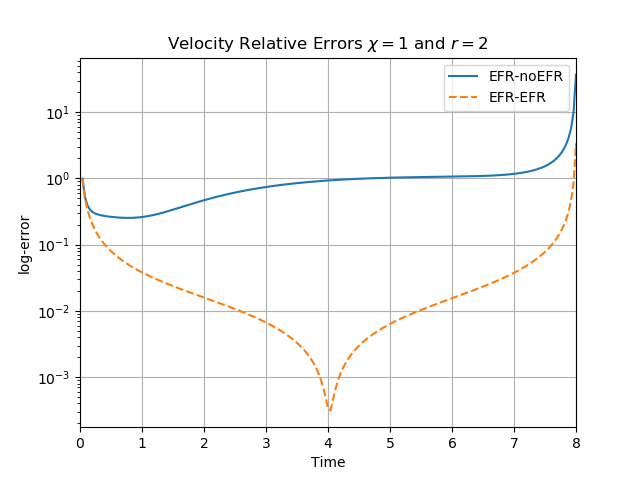}
         \includegraphics[scale=0.4]{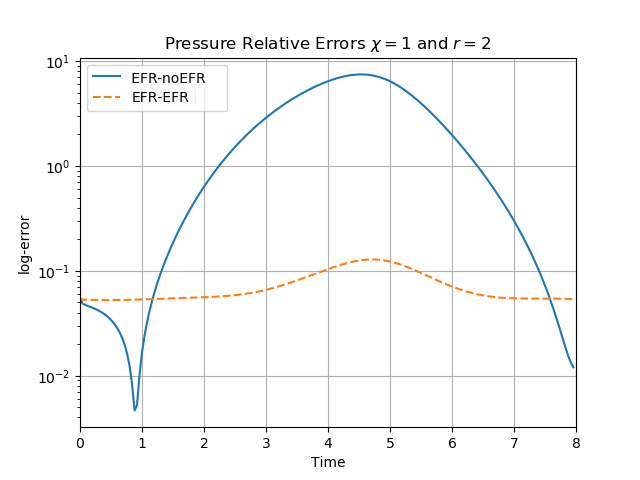}
\caption{\scriptsize{(Experiment 1: $\delta=0.0032$, $\chi = 1$ and $r=2$.) \emph{Left.}  Comparison of relative log-errors over time of the velocity profiles: EFR full order versus {EFR-noEFR} solutions and EFR full order versus {EFR-EFR} solutions, represented by solid blue  and dashed orange lines, respectively.   \emph{Right.}  Analogous representation for the relative log-errors over time of the pressure profiles.}
}
\label{fig:errs_chi1}
\end{figure}

\begin{figure}
\centering
        \includegraphics[scale=0.4]{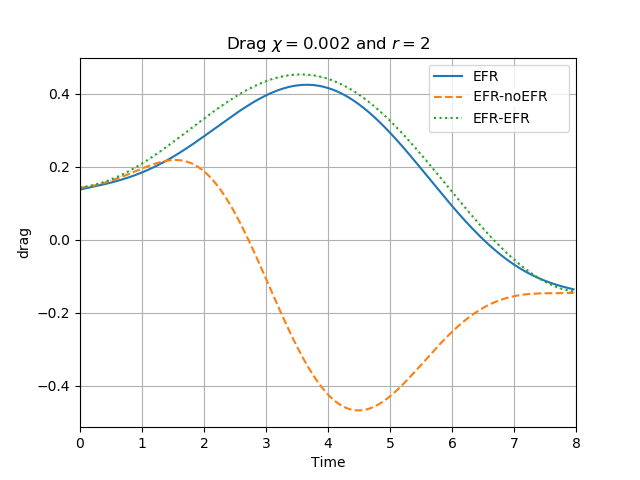}
         \includegraphics[scale=0.4]{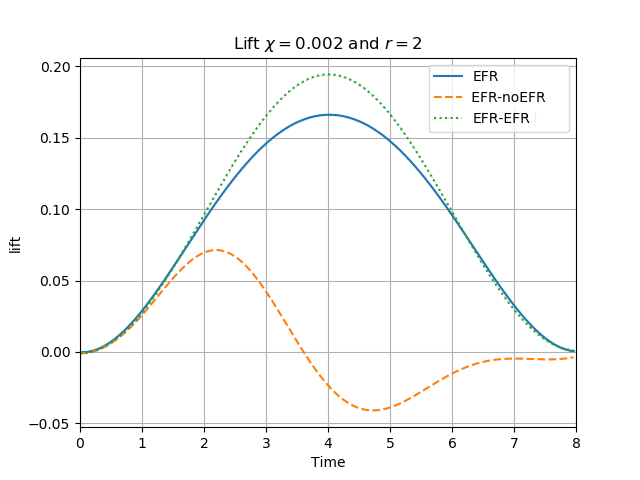}
\caption{\scriptsize{(Experiment 1: $\delta=0.0032$, $\chi = 1$ and $r=2$.) \emph{Left.}  $C_D(t)$ comparison over time. \emph{Right.}  $C_L(t)$ comparison over time.}
}
\label{fig:cd_cl}
\end{figure}
\begin{table}[]
\caption{\scriptsize{\A{(Experiment 1: $\delta=0.0032$, $\chi = 1$ and $r=2$.) Maximum, minimum, and average relative error over the considered time interval for velocity and pressure fields.}}}
\label{tab:table1}
\begin{center}
\begin{tabular}{c|ccc|ccc|}
\cline{2-7}
                                   & \multicolumn{3}{c|}{{EFR-noEFR}}                                                   & \multicolumn{3}{c|}{{EFR-EFR}}                                                \\ \cline{2-7} 
                                   & \multicolumn{1}{c|}{maximum}    & \multicolumn{1}{c|}{minimum}    & average    & \multicolumn{1}{c|}{maximum}    & \multicolumn{1}{c|}{minimum}    & average    \\ \hline
\multicolumn{1}{|c|}{$E_{\bu}(t)$} & \multicolumn{1}{c|}{$3.721e$+1} & \multicolumn{1}{c|}{$2.533e$-1} & $1.115e$+0 & \multicolumn{1}{c|}{$3.298e$+0} & \multicolumn{1}{c|}{$3.145e$-4} & $6.252$e-2 \\ \hline
\multicolumn{1}{|c|}{$E_{p}(t)$}   & \multicolumn{1}{c|}{$7.450e$+0} & \multicolumn{1}{c|}{$3.431e$-3} & $2.306e$+0 & \multicolumn{1}{c|}{$1.620e$-1} & \multicolumn{1}{c|}{$2.877e$-2} & $1.701e$-2 \\ \hline
\end{tabular}%
\end{center}
\end{table}
 We collect $N_{\bu} = N_{p} = 200$ snapshots for both 
 velocity and pressure with an 
 equally spaced grid 
 in the time interval $[t_0, T] = [0,8]$. After the POD procedure, we retain the first $2$ modes for both velocity and pressure, representing 
 $99.9\%$ of the snapshot energy. Here, for the sake of notation, we use $r \doteq r_{\bu} = r_s = r_p = 2$. We show representative solutions of velocity for $t = 1$ and pressure for $t=4$ in Figure \ref{fig:u_chi1} and Figure \ref{fig:p_chi1}, respectively. It is clear that 
 the {EFR-noEFR} 
 is not able to reconstruct the solution provided by the FOM, while the {EFR-EFR} 
 leads to very accurate results for both the fields. \A{To allow an easy comparison, we plot the velocity and pressure fields on the FOM scale, i.e., $[0, 0.7]$ and $[-0.71, 0.61]$, respectively. } The relative log-error temporal trend, displayed in Figure \ref{fig:errs_chi1}, confirms this conclusion: 
 it shows how the 
 {EFR-EFR} yields more accurate results for both 
 velocity and pressure, reaching values around $10^{-3}$ for the velocity,
 and reducing the error by two orders of magnitude with respect to the {EFR-noEFR} 
 for both 
 variables. 
The comparison between FOM and ROM aerodynamic coefficients over time is reported in Figure \ref{fig:cd_cl}. The coefficients are well recovered by the {EFR-EFR}, 
while the {EFR-noEFR} 
is not able to accurately approximate them.
Indeed, the relative errors \eqref{drag_err} and \eqref{lift_err} have the following values: $\A{\overline E_{C_D}}= 1.02$, 
$\A{\hat E_{C_D}}= 0.13$, and $\A{\overline E_{C_L}}= 1.69$, 
$\A{\hat E_{C_L}}= 0.11$. 
The advantage of using the {EFR-EFR} 
is remarkable, since we are reducing the relative $L^2$-errors of the force coefficients by an order of magnitude.
\A{Table \ref{tab:table1} 
lists the maximum, minimum, and average relative errors for the velocity and pressure fields in the 
{EFR-noEFR} and {EFR-EFR} setttings. }
\A{
Overall, the results in Table \ref{tab:table1} are consistent with the plots in Figure \ref{fig:errs_chi1}.
With respect to the velocity approximation, the {EFR-EFR} is significantly more accurate than the {EFR-noEFR}.
The maximum, minimum, and average relative errors are at least one order of magnitude lower for the {EFR-EFR} than for the {EFR-noEFR}.
With respect to the pressure approximation, the {EFR-EFR} is again more accurate than the {EFR-noEFR}, but the improvement is not as dramatic as for the velocity approximation.
}
\\


{\bf Experiment 2}. The next step in our comparison of 
\A{{EFR-noEFR} and {EFR-EFR} } is the numerical investigation of the EFR stabilization strategy both at the FOM level and at the ROM level. 
Specifically, we use $\chi=5\cdot \Delta t = 0.002$ in the \emph{Relax} step for both the {EFR-noEFR} and the {EFR-EFR}.
This choice limits the amount of dissipation introduced by the DF in the \emph{Filter} step of the EFR algorithm, and yields a more challenging test problem than Experiment 1 for both the \A{{EFR-noEFR} and the {EFR-EFR}.}
\A{ 
Specifically, Experiments 1 and 2 share the same computational setting except that for Experiment 1 no relaxation is performed.
Indeed, in Experiment 1 we use $\chi =1$, and thus
\begin{eqnarray}
\boldsymbol u^{n+1}_r
            = (1 - \chi) \, \boldsymbol w^{n+1}_r
            + \chi \, \overline{\boldsymbol w}^{n+1}_r = \overline{\boldsymbol w}^{n+1}_r.
    \label{eqn:exp1-vs-exp2}
\end{eqnarray}
The relationship in~\eqref{eqn:exp1-vs-exp2} shows that, in Experiment 1,  
the final velocity coincides with the filtered velocity. 
This setting is overly diffusive and, as a result, the vortices that appear in Experiment 2 are totally damped in Experiment 1.

\begin{figure}
        \centering \includegraphics[scale=0.21]{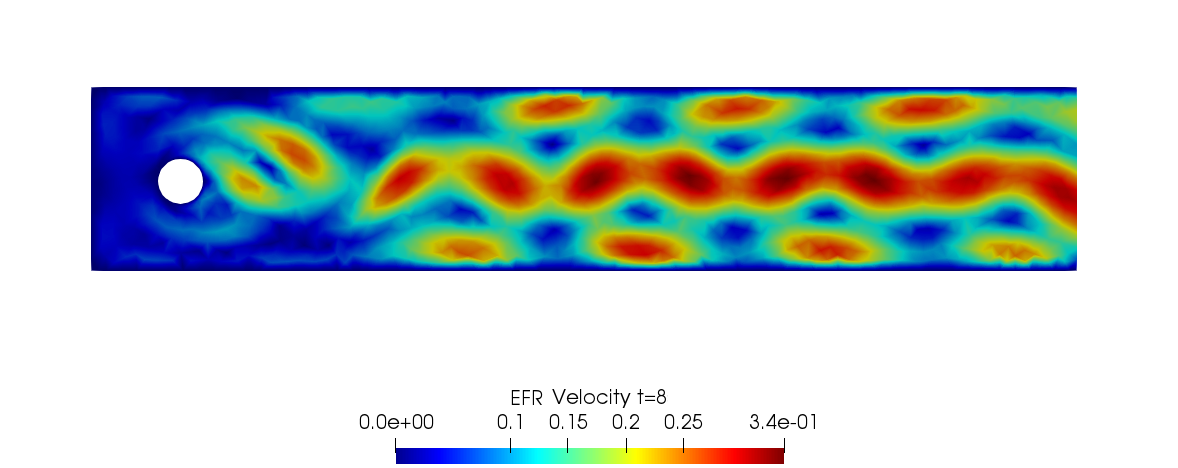}\\
\begin{minipage}{.49\textwidth}
 \centering
         \includegraphics[scale=0.21]{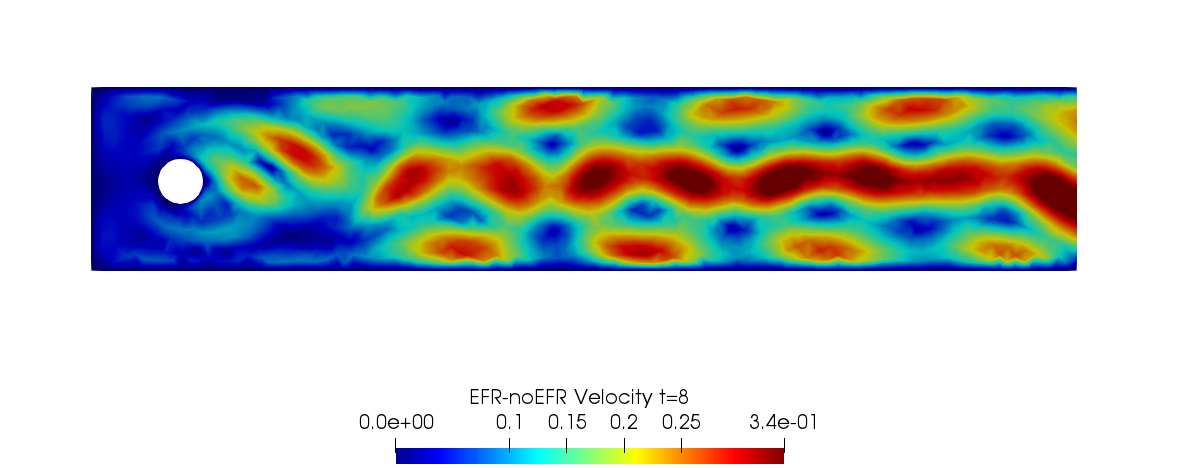}
\end{minipage}
\begin{minipage}{.49\textwidth}
 \centering
         \includegraphics[scale=0.21]{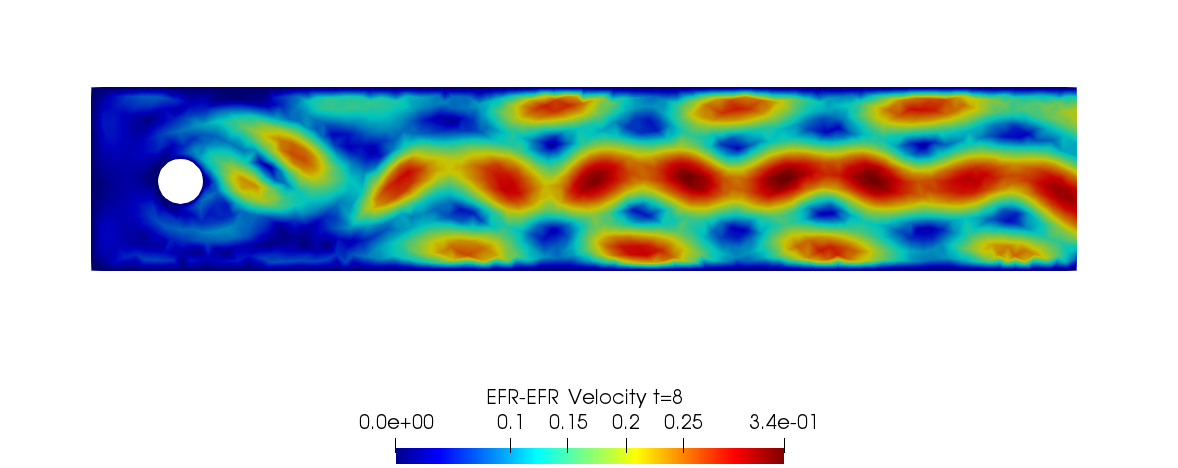}
\end{minipage}
\caption{\scriptsize{(Experiment 2: $\delta=0.0032$, $\chi = 0.002$, $r_{\bu}=43, r_p=r_s=8$ and $t=8$. Reconstruction for $t \in [4,8]$.) \emph{Top.} Full order EFR velocity magnitude. \emph{Bottom Left.} Reduced {EFR-noEFR} velocity magnitude. \emph{Bottom Right.} Reduced {EFR-EFR} velocity magnitude.}
}
\label{fig:4u_chi5dt}
\end{figure}
We also note that, in contrast with Experiment 1, in Experiment 2 we do not use the grad-div stabilization at the FOM level (see Remark \ref{remark:graddiv}). Indeed, in Experiment 2, the divergence values are 
$O(10^{-2})$ in the worst case scenario.
Thus, the grad-div stabilization is no longer needed and the FOM and ROM 
are 
consistent with respect to the grad-div stabilization. }

\begin{remark}
As already 
noted in Section \ref{sec:hf}, $\chi \sim \Delta t$ is a common choice used in the literature 
for academic benchmarks (see, e.g., \cite{ervin2012numerical}). \A{However, in \cite{bertagna2016deconvolution} the authors propose 
the scaling $\chi = c \Delta t$. Thus, we choose $c=5$, i.e., a higher 
$\chi$ value, which 
introduces 
a larger amount of dissipation. This setting can be of interest 
in more realistic applications \cite{bertagna2016deconvolution, GirfoglioQuainiRozza2019}.}
\end{remark}
\A{We construct and test the {EFR-noEFR} and {EFR-EFR} on the time interval $[4,8]$.
The rationale for our choice is that the flow dynamics is  significantly more complex on the time interval $[4,8]$ than on the time interval $[0,4]$ (see Experiment 3).}

\begin{figure} \centering
        \includegraphics[scale=0.4]{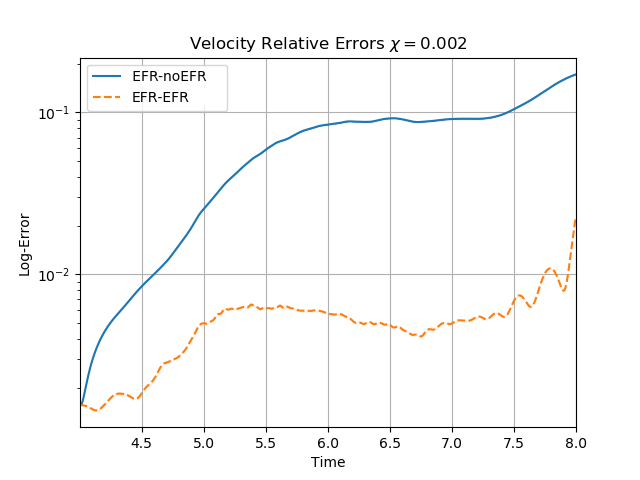}
         \includegraphics[scale=0.4]{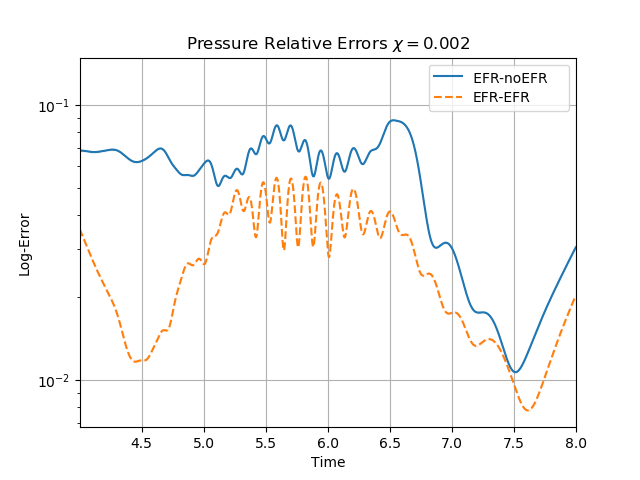}
\caption{\scriptsize{(Experiment 2: $\delta=0.0032$, $\chi = 0.002$ and $r_{\bu}=43, r_p=r_s=8$. Reconstruction for $t \in [4,8]$.) \emph{Left.}  
Comparison of relative log-errors over time for the velocity profiles: {EFR-noEFR} (solid blue line) and {EFR-EFR} (dashed orange line).
\emph{Right.}  
Comparison of relative log-errors over time for the pressure profiles: {EFR-noEFR} (solid blue line) and {EFR-EFR} (dashed orange line).
}
}
\label{fig:4err_chi5dt}
\end{figure}
\begin{figure}
       \centering \includegraphics[scale=0.21]{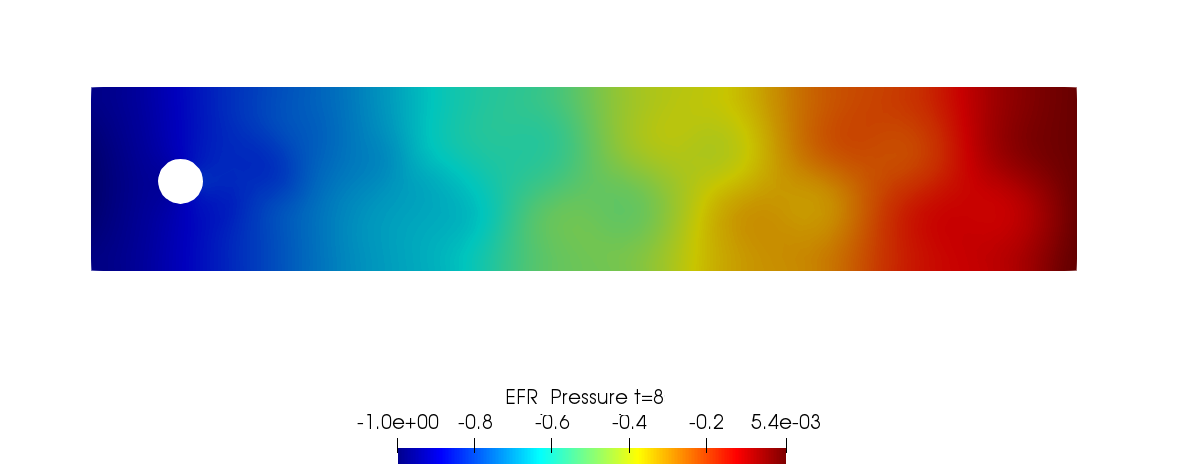}\\
\begin{minipage}{.49\textwidth}
 \centering
         \includegraphics[scale=0.21]{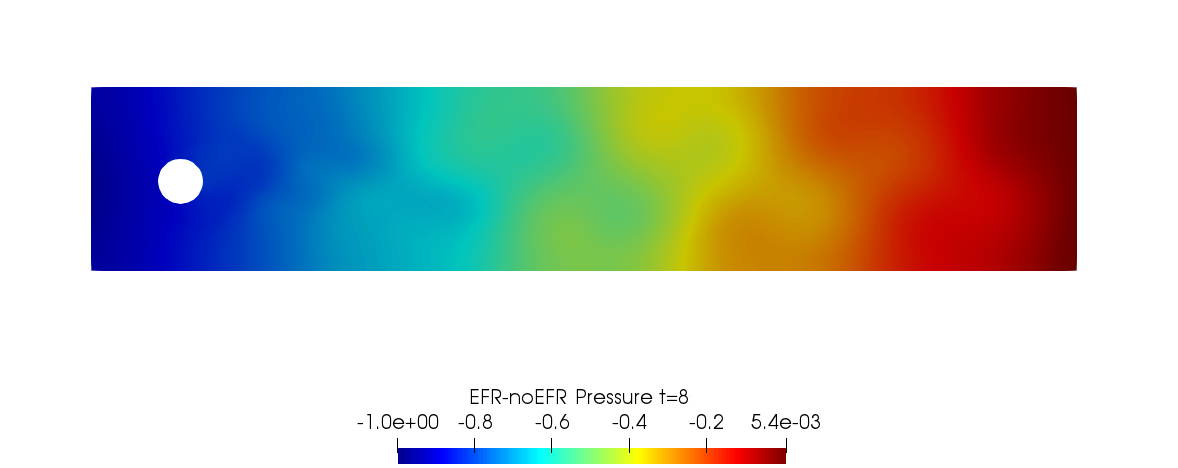}
\end{minipage}
\begin{minipage}{.49\textwidth}
 \centering
         \includegraphics[scale=0.21]{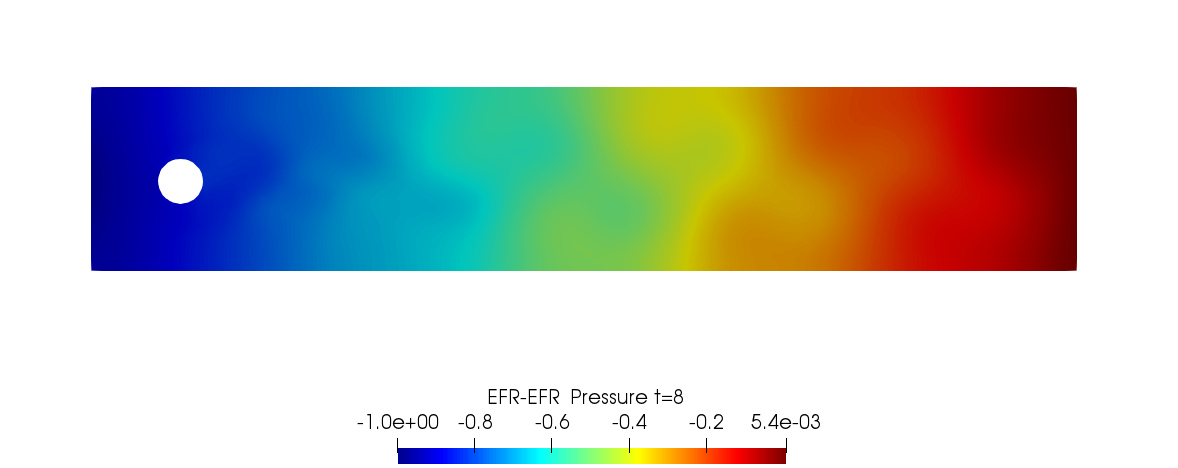}
\end{minipage}
\caption{\scriptsize{(Experiment 2: $\delta=0.0032$, $\chi = 0.002$,  $r_{\bu}=43, r_p=r_s=8$,  and $t=8$. Reconstruction for $t \in [4,8]$.) \emph{Top.} Full order EFR pressure field. \emph{Bottom Left.} Reduced {EFR-noEFR} pressure field. \emph{Bottom Right.} Reduced {EFR-EFR} pressure field.}
}
\label{fig:4p_chi5dt}
\end{figure}
\begin{figure} \centering
        \includegraphics[scale=0.4]{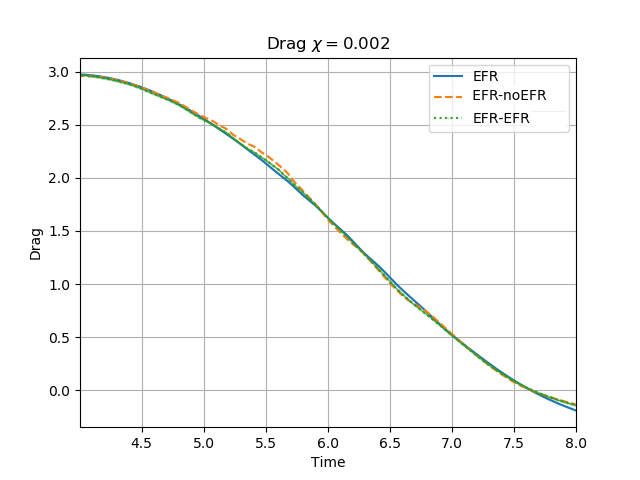}
         \includegraphics[scale=0.4]{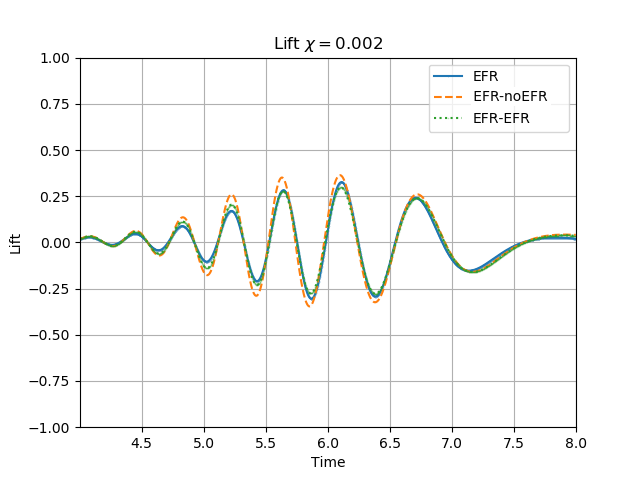}
\caption{\scriptsize{(Experiment 2: $\delta=0.0032$, $\chi = 0.002$, $r_{\bu}=43$, and $r_p=r_s=8$. Reconstruction for $t \in [4,8]$.) \emph{Left.}  $C_D(t)$ comparison over time: all the approaches almost coincide. \emph{Right.}  $C_L(t)$ comparison over time:  full order EFR and {EFR-EFR} lift coincide (solid blue and dotted green lines).}
}
\label{fig:4cd_cl_chi5dt}
\end{figure}

In order to 
approximate all the relevant features of the flow field, we increase the number of snapshots.
Indeed, to build the {EFR-noEFR} and {EFR-EFR}, we collect $N_{\bu} = N_p = 2000$ snapshots, which are equally spaced on the time interval $[4,8]$. 
To retain $99.9\%$ of the snapshots energy, we employ the following numbers of POD basis functions to build the ROMs: $r_{\bu}=43$, $r_s = r_p = 8$.
We note that, with respect to 
Experiment 1, the system shows a slower decay of the eigenvalues and therefore 
more modes need to be used to construct the ROMs.
The higher accuracy of {EFR-EFR} 
is displayed in Figure \ref{fig:4u_chi5dt}: 
the {EFR-EFR} solution perfectly matches 
the FOM solution, while the {EFR-noEFR} solution  
is slightly different from the FOM solution.
The relative log-errors in Figure \ref{fig:4err_chi5dt} (left) yield the same conclusions: 
the {EFR-EFR} velocity errors are an order of magnitude lower than the {EFR-noEFR} velocity errors.
We also note that {EFR-EFR} is more accurate than the {EFR-noEFR} in approximating the pressure field, 
especially at the beginning and at the end of the time interval
(see Figures  \ref{fig:4err_chi5dt} (right) and \ref{fig:4p_chi5dt}). 
The force coefficients $C_D(t)$ and $C_L(t)$ 
plotted in Figure \ref{fig:4cd_cl_chi5dt} show that, while both {EFR-noEFR} and {EFR-EFR} are accurate, the latter is more accurate than the former.
This is also illustrated by the $L^2$-errors of the force coefficients:
$\A{\overline E_{C_L}} = 0.26$, $\A{\hat E_{C_L}}=0.11$ and $\A{\overline E_{C_D}} = 0.018$, $\A{\hat E_{C_D}} = 0.010$.
\A{The maximum, minimum, and average error values over time for the velocity and pressure fields, which are listed in Table \ref{tab:table2}, confirm that the {EFR-EFR} is more accurate than the {EFR-noEFR}. 
The improvement in the {EFR-EFR} is also highlighted by the Pareto plot in Figure \ref{fig:pareto2}. 
Indeed, fixing $r_p=r_s=8$ and 
choosing $r_u = 30, 32, 34, 36, 38, 40, 42, 44, 46, 48, 50$ shows that, over this range of $r_u$ values, the {EFR-EFR} performs better than the {EFR-noEFR} 
with respect to both the velocity and the pressure approximations.
Indeed, the {EFR-EFR} with $r_u=30$ yields a low relative error that the {EFR-noEFR} cannot attain by increasing its $r_u$ value (and, consequently, its relative wall time).
}

\begin{table}[]
\caption{\scriptsize{\A{(Experiment 2: $\delta=0.0032$, $\chi = 0.002$,  $r_{\bu}=43, r_p=r_s=8$. Reconstruction for $t \in [4,8]$.) Maximum, minimum, and average relative error over the considered time interval for velocity and pressure fields.}}}
\label{tab:table2}
\begin{center}
\begin{tabular}{c|ccc|ccc|}
\cline{2-7}
                                   & \multicolumn{3}{c|}{{EFR-noEFR}}                                                   & \multicolumn{3}{c|}{{EFR-EFR}}                                                \\ \cline{2-7} 
                                   & \multicolumn{1}{c|}{maximum}    & \multicolumn{1}{c|}{minimum}    & average    & \multicolumn{1}{c|}{maximum}    & \multicolumn{1}{c|}{minimum}    & average    \\ \hline
\multicolumn{1}{|c|}{$E_{\bu}(t)$} & \multicolumn{1}{c|}{$1.718e$-1} & \multicolumn{1}{c|}{$1.561$e-3} & $6.781e$-2 & \multicolumn{1}{c|}{$2.267e$-2} & \multicolumn{1}{c|}{$1.446e$-3} & $5.175e$-3 \\ \hline
\multicolumn{1}{|c|}{$E_{p}(t)$}   & \multicolumn{1}{c|}{$8.815e$-2} & \multicolumn{1}{c|}{$1.068e$-2} & $5.366e$-2 & \multicolumn{1}{c|}{$5.491e$-2} & \multicolumn{1}{c|}{$7.753e$-3} & $2.688e$-2 \\ \hline
\end{tabular}%
\end{center}
\end{table}

\A{
\begin{figure}\centering 
        \includegraphics[scale=0.4]{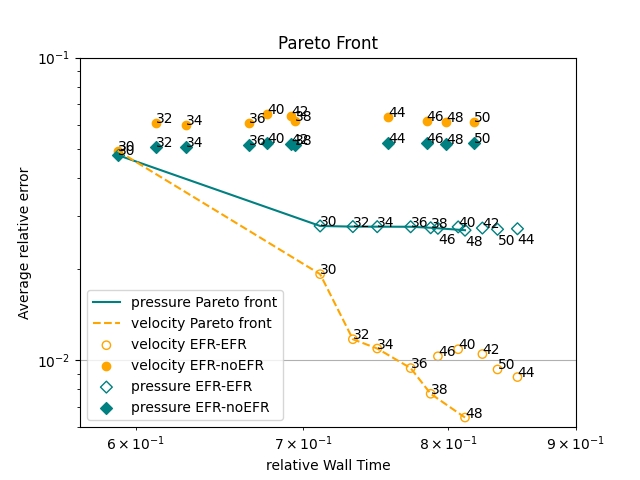}
\caption{\scriptsize{\A{(Experiment 2: $\delta=0.0032$, $\chi = 0.002$ and $r_{\bu}=\{30, 32, 34, 36, 38, 40, 42, 44, 46, 48, 50\}$, $r_p=r_s=8$.) Pareto plots for velocity (orange) and pressure (teal) fields: averaged relative error in time versus relative wall time for varying $r_{\bu}$.}}
}
\label{fig:pareto2}
\end{figure}
}
Overall, the numerical investigation in Experiment 2 yields the same conclusions as the numerical investigations in Experiment 1:
the {EFR-EFR} is more accurate than the {EFR-noEFR} with respect to all the criteria used, i.e., \A{pointwise, average, maximum, and minimum velocity and pressure errors, } 
lift and drag coefficient errors, 
\A{and Pareto front. }
Thus, these numerical results suggest that the FOM-ROM consistency is beneficial for the EFR stabilization strategy.

\A{{\bf Experiment 3}.
In this experiment, which is based on the computational setting in~\cite{john2004reference,GirfoglioQuainiRozza2020,schafer1996benchmark}, we investigate the EFR-noEFR and EFR-EFR models in a regime that is more challenging than the regime used in Experiment 2.
Specifically, we consider a regime that, while still marginally-resolved, employs a coarser resolution than Experiment 2 (i.e., relatively fewer snapshots and relatively fewer basis functions) and two different dynamical regimes (a laminar regime in the first half of the time interval, and a more complex regime in the second half). 
}
This investigation focuses on the reconstruction of the whole time interval $[0,8]$ using 
$N_{\bu} = N_p = 2000$ 
snapshots that are equally spaced in this time interval.
\A{Thus, we are using coarser resolution than the resolution used in Experiment 2, since we are employing the same number of snapshots as in Experiment 2 but consider a time interval that is twice as long as that used in Experiment 2.
Furthermore, in the first half of the time interval the flow displays laminar dynamics, whereas in the second half it displays more complex dynamics (e.g., 
vortex shedding).
We note that the mixed dynamics in Experiment 3 is more challenging to represent at the ROM level than the dynamics in Experiment 2 (i.e., in the time interval $[4,8]$).
}
In order to retain 
$99.9\%$ of the energy 
of the snapshots, we choose $r_{\bu}=47$ and $r_p=r_{s}=7$.
\A{We note that in Experiment 3 we utilize a similar number of basis functions as in Experiment 2. 
Since the dynamics in Experiment 3 is more challenging than the dynamics in Experiment 2, we conclude that the Experiment 3 ROM resolution is coarser than the Experiment 2 resolution. }

\begin{figure}
\centering
        \includegraphics[scale=0.37]{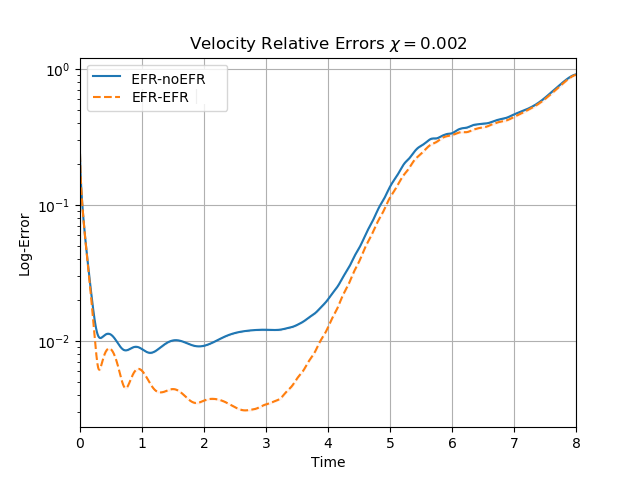}
         \includegraphics[scale=0.37]{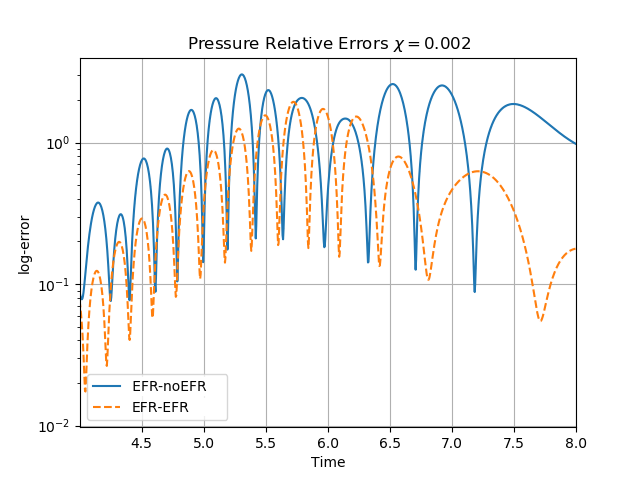}
\caption{\scriptsize{(Experiment 3:
$\delta=0.0032$, $\chi = 0.002$ and $r_{\bu}=47$, $r_p=r_s=7$.) \emph{Left.}  Comparison of relative log-errors over time for the velocity profiles: 
{EFR-noEFR} (solid blue line) and {EFR-EFR} (dashed orange line).
\emph{Right.} 
Comparison of relative log-errors over time for the pressure profiles: {EFR-noEFR} (solid blue line) and {EFR-EFR} (dashed orange line).
}
}
\label{fig:err_chi5dt}
\end{figure}

Overall, in Experiment 3, neither {EFR-noEFR} 
{nor} {EFR-EFR} give 
satisfactory results. 
Furthermore, both 
{EFR-noEFR} and {EFR-EFR} are significantly less accurate in Experiment 3 than in Experiment 2.
Indeed, the log-relative errors for the velocity field reported in Figure \ref{fig:err_chi5dt} (left) show that, even if the {EFR-EFR} 
performs better on the first half of the time interval, 
both the {EFR-noEFR} and the \A{EFR-EFR }
are inaccurate on the second half of the time interval. 
\begin{figure}
        \centering \includegraphics[scale=0.21]{exp3/EFR_v}\\
\begin{minipage}{.49\textwidth}
 \centering
         \includegraphics[scale=0.21]{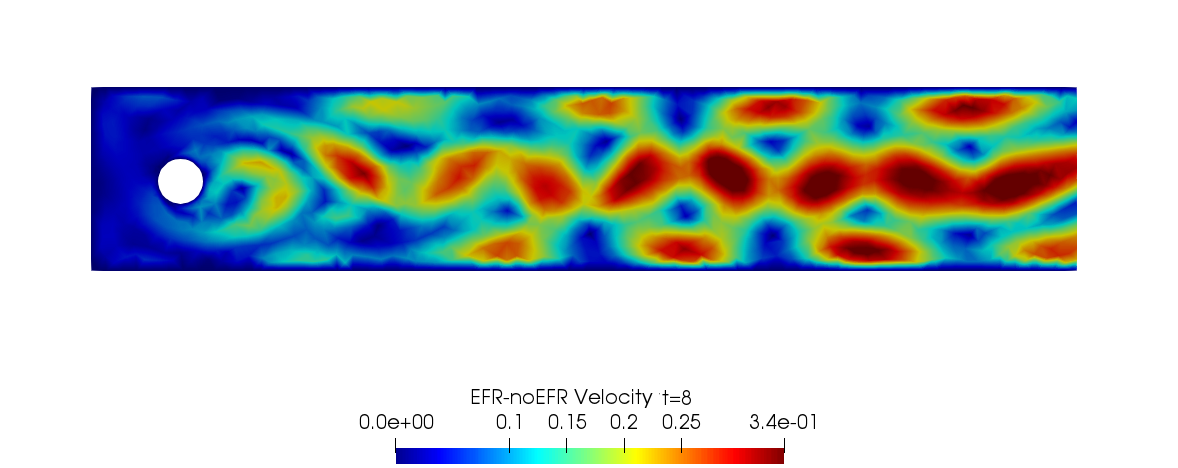}
\end{minipage}
\begin{minipage}{.49\textwidth}
 \centering
         \includegraphics[scale=0.21]{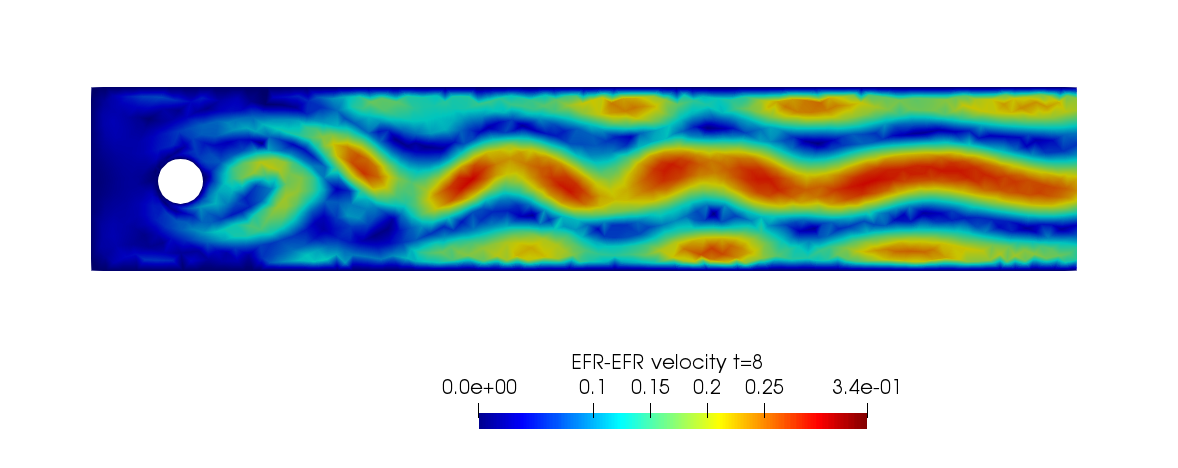}
\end{minipage}
\caption{\scriptsize{(Experiment 3: $\delta=0.0032$, $\chi = 0.002$, $r_{\bu}=47$, $r_p=r_s=7$ and $t=8$.) \emph{Top.} Full order EFR velocity magnitude. \emph{Bottom Left.} Reduced {EFR-noEFR} velocity magnitude. \emph{Bottom Right.} Reduced {EFR-EFR} velocity magnitude.}
}
\label{fig:u_chi5dt}
\end{figure}

\begin{figure}
      \centering  \includegraphics[scale=0.21]{exp3/EFR_p}\\
\begin{minipage}{.49\textwidth}
 \centering
         \includegraphics[scale=0.21]{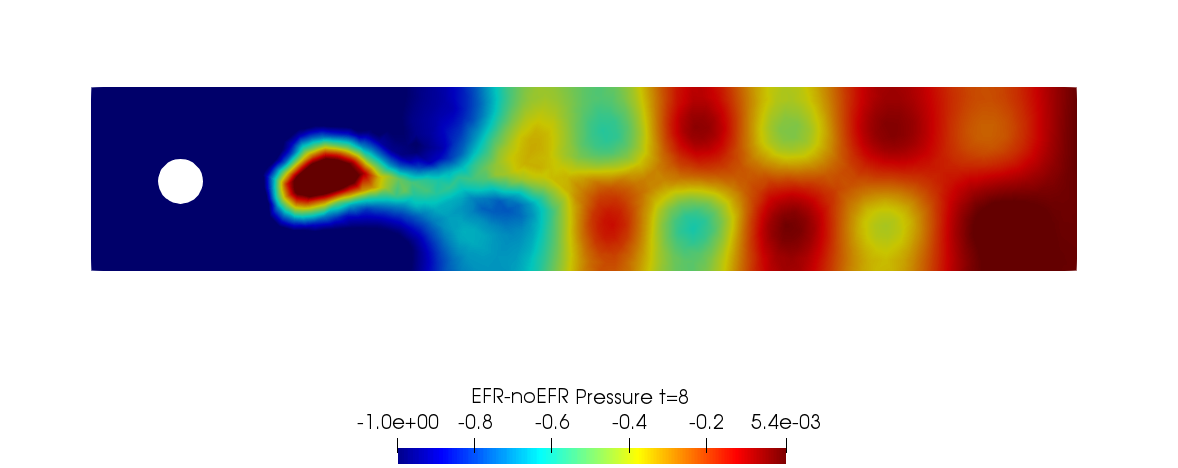}
\end{minipage}
\begin{minipage}{.49\textwidth}
 \centering
         \includegraphics[scale=0.21]{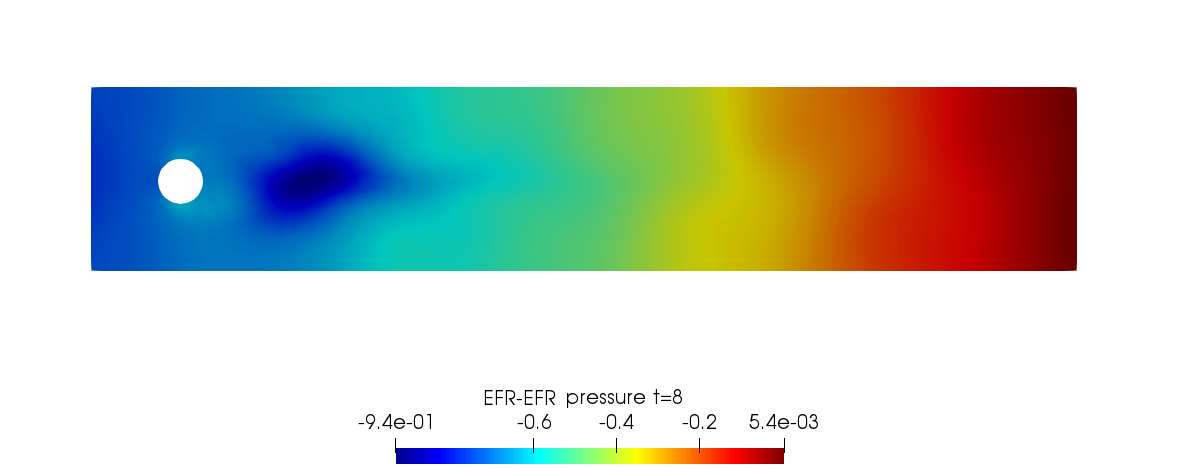}
\end{minipage}
\caption{\scriptsize{(Experiment 3: $\delta=0.0032$, $\chi = 0.002$, $r_{\bu}=47$, $r_p=r_s=7$ and $t=8$.) \emph{Top.} Full order EFR pressure field. \emph{Bottom Left.} Reduced {EFR-noEFR} pressure field. \emph{Bottom Right.} Reduced {EFR-EFR} pressure field.}
}
\label{fig:p_chi5dt}
\end{figure}

\begin{figure}\centering 
        \includegraphics[scale=0.4]{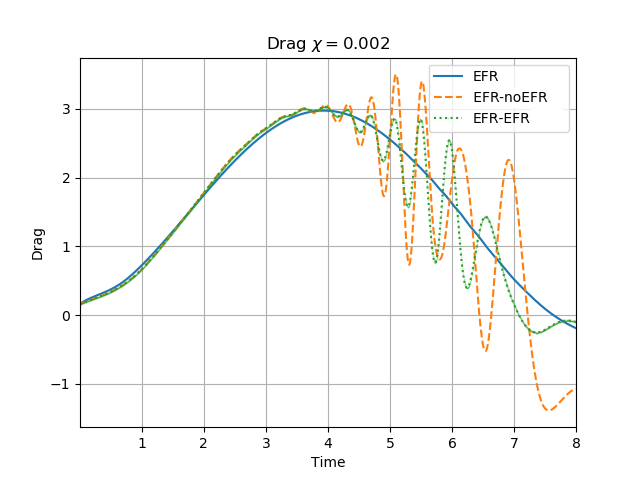}
         \includegraphics[scale=0.4]{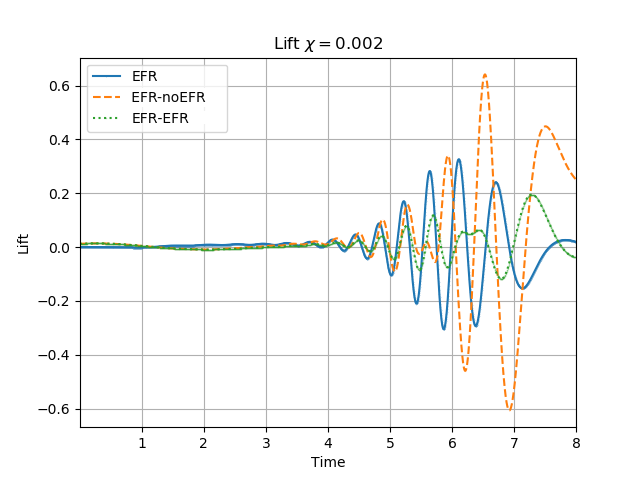}
\caption{\scriptsize{(Experiment 3: $\delta=0.0032$, $\chi = 0.002$ and $r_{\bu}=47$, $r_p=r_s=7$.) \emph{Left.}  $C_D(t)$ comparison over time: all the approaches almost coincide. \emph{Right.}  $C_L(t)$ comparison over time.}
}
\label{fig:cd_cl_chi5dt}
\end{figure}

This is confirmed by the velocity solution for $t=8$ displayed in Figure \ref{fig:u_chi5dt}. 
The plot in Figure \ref{fig:err_chi5dt} (left) shows that,
while for $t \in [0, 4]$  
{EFR-EFR} yields relative errors below $10^{-2}$ for the velocity field, this advantage is lost in the last part of the time interval, reaching unacceptable errors values (close to $1$). On the other hand, concerning the pressure field (see Figure \ref{fig:err_chi5dt} (right)), {EFR-EFR} is able to perform better than {EFR-noEFR} for almost 
the entire time window. However, the error values are high, even greater than $1$ for some time instances. \A{Moreover, the pressure field presents a checkerboard type of instability (although {EFR-noEFR} model is inf-sup stable thanks to supremizer stabilization) } and the reconstruction 
is inaccurate, as displayed in Figure \ref{fig:p_chi5dt}. 
These issues are visible also in Figure \ref{fig:cd_cl_chi5dt}: the lift coefficient $C_L(t)$ is well recovered only for the first part of the time interval, when the vortex shedding does not occur 
and the low frequency modes are dominant. 
Even the drag coefficient $C_D(t)$ is not reconstructed in a satisfactory way and shows spurious oscillations for $t>4$. 
This behavior is worse 
in the {EFR-noEFR} results, 
which exhibit larger amplitude oscillations. 
The {EFR-noEFR} and {EFR-EFR} yield the following $L^2-$errors of the force coefficients: $\A{\overline E_{C_L}} = 2.39$, $\A{\hat E_{C_L}} = 1.2$, $\A{\overline E_{C_D}} = 0.33$, and $\A{\hat E_{C_D}} = 0.15$.
Overall, although the {EFR-EFR} results are significantly more accurate than the {EFR-noEFR} results, both ROMs yield relatively inaccurate results. 
\A{
These results show that, as expected, utilizing a more challenging regime (i.e., a coarser ROM resolution and mixed dynamics) in Experiment 3 deteriorates the EFR-noEFR and EFR-EFR performance.
We emphasize, however, that even in this more challenging regime EFR-EFR performs better than EFR-noEFR.
}

\both{
\subsection{Numerical results: 
\A{predictive regime}}
\label{sec_pred_time}
This section focuses on preliminary results on 
the predictive capabilities of the  EFR-noEFR and EFR-EFR algorithms.
We stress that the offline phase and the parameters $\chi$ and $\delta$ do not change with respect to the reconstructive setting
{although this choice may be suboptimal.}
In this section, we answer the following questions: (i) Are the 
the  EFR-noEFR and EFR-EFR algorithms predictive? (ii) Which 
algorithm performs better in the predictive regime?}
\\[0.1cm]
\both{
\textbf{Experiment 1}. To study the predictability of ERF-EFR and EFR-noEFR strategies, we collect $N_{\bu} = N_{p} = 200$ equally spaced snapshots for both 
 velocity and pressure in $[0,8]$. After the POD procedure, we retain the first $2$ modes for both velocity and pressure since  they represent 
 $99.9\%$ of the snapshot energy. We recall that the filter radius is $\delta = 0.0032$ and the relaxation parameter is $\chi = 1$. We test the predictive capability of the model in the time interval [8,12]. In terms of relative velocity errors, 
 EFR-EFR performs better than EFR-noROM, reaching values around $10^{-2}$ and reducing the error by two order of magnitude, as 
 illustrated in the left plot of Figure \ref{fig:errs_chi12}. Focusing on the pressure field relative error, i.e., the right plot of Figure \ref{fig:errs_chi12}, the EFR-noEFR strategy performs better until $t=9.4$. After that value, the EFR-noEFR error increases, while  the EFR-EFR remains stable around $10^{-1}$. For the sake of completeness, we report the $L^2$-error values over the force coefficients:
$\A{\overline E_{C_L}} = 0.23$, $\A{\hat E_{C_L}}=0.08$ and $\A{\overline E_{C_D}} = 1.41$, $\A{\hat E_{C_D}} = 1.49$. These values are 
consistent with Figure \ref{fig:cd_cl12}, where the EFR-EFR lift representation is more accurate than the EFR-noEFR one, while the opposite happens for the drag coefficient. Table \ref{tab:table12} 
lists maximum, minimum, and average error values over time for the velocity and pressure fields. This table confirms that, overall,  EFR-EFR is more accurate than EFR-noEFR. 
The only exception is the minimum error for the pressure field, which is smaller for the EFR-noEFR strategy, as already pointed out in analyzing Figure \ref{fig:errs_chi12}. 
The numerical results for Experiment 1 yield the following conclusions: 
(i) both approaches are predictive in time, and (ii) EFR-EFR is, overall, more accurate than EFR-noEFR, except for the drag representation.

\begin{figure}
\centering
        \includegraphics[scale=0.4]{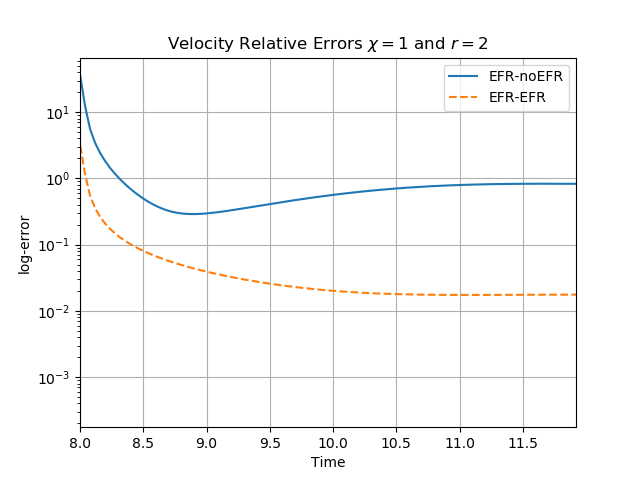}
         \includegraphics[scale=0.4]{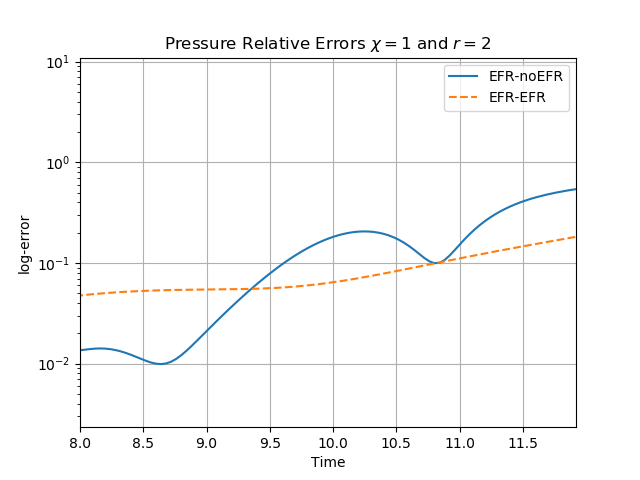}
\caption{\both{\scriptsize{(Experiment 1: $\delta=0.0032$, $\chi = 1$ and $r=2$. Prediction for $t \in [8,12]$.) \emph{Left.}  Comparison of relative log-errors over time of the velocity profiles: EFR full order versus {EFR-noEFR} solutions and EFR full order versus {EFR-EFR} solutions, represented by solid blue  and dashed orange lines, respectively.   \emph{Right.}  Analogous representation for the relative log-errors over time of the pressure profiles.}}
}
\label{fig:errs_chi12}
\end{figure}

\begin{figure}
\centering
        \includegraphics[scale=0.4]{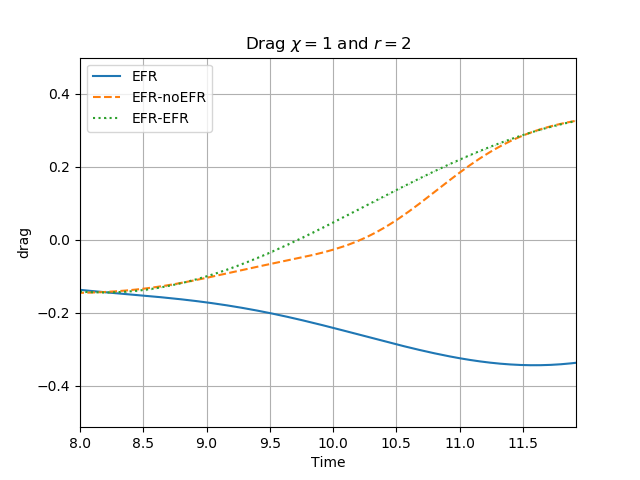}
         \includegraphics[scale=0.4]{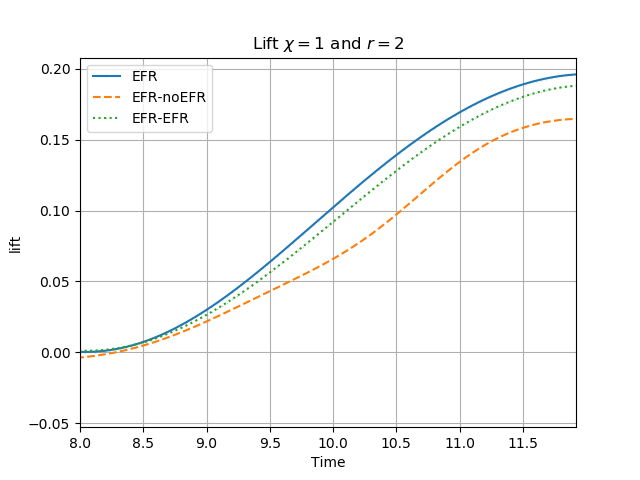}
\caption{\both{\scriptsize{(Experiment 1: $\delta=0.0032$, $\chi = 1$ and $r=2$. Prediction for $t \in [8,12]$.) \emph{Left.}  $C_D(t)$ comparison over time. \emph{Right.}  $C_L(t)$ comparison over time.}}
}
\label{fig:cd_cl12}
\end{figure}

}

\begin{table}[]
\caption{\scriptsize{\both{(Experiment 1: $\delta=0.0032$, $\chi = 1$,  $r=2$. Prediction for $t \in [8,12]$.) Maximum, minimum, and average relative error over the considered time interval for velocity and pressure fields.}}}
\label{tab:table12}
\begin{center}
\begin{tabular}{c|ccc|ccc|}
\cline{2-7}
                                   & \multicolumn{3}{c|}{{EFR-noEFR}}                                                   & \multicolumn{3}{c|}{{EFR-EFR}}                                                \\ \cline{2-7} 
                                   & \multicolumn{1}{c|}{maximum}    & \multicolumn{1}{c|}{minimum}    & average    & \multicolumn{1}{c|}{maximum}    & \multicolumn{1}{c|}{minimum}    & average    \\ \hline
\multicolumn{1}{|c|}{$E_{\bu}(t)$} & \multicolumn{1}{c|}{$1.239e$+1} & \multicolumn{1}{c|}{$2.891$e-1} & $8.466e$-1 & \multicolumn{1}{c|}{$1.157e$+0} & \multicolumn{1}{c|}{$1.739e$-2} & $5.692e$-2 \\ \hline
\multicolumn{1}{|c|}{$E_{p}(t)$}   & \multicolumn{1}{c|}{$5.508e$-1} & \multicolumn{1}{c|}{$9.898e$-3} & $1.613e$-1 & \multicolumn{1}{c|}{$1.818e$-1} & \multicolumn{1}{c|}{$4.814e$-2} & $8.480e$-2 \\ \hline
\end{tabular}%
\end{center}
\end{table}

\both{
\textbf{Experiment 2}. The next step is represented by the analysis of the predictive regime in the setting of Experiment 2. Namely, we collect $N_{\bu} = N_p = 2000$ equally spaced snapshots in the time interval $[4,8]$.  We employ
$r_{\bu}=43$, $r_s = r_p = 8$ to retain $99.9\%$ of the snapshots energy. In this case $\delta = 0.0032$ and $\chi = 0.002$.
}

\begin{figure}
\centering
        \includegraphics[scale=0.4]{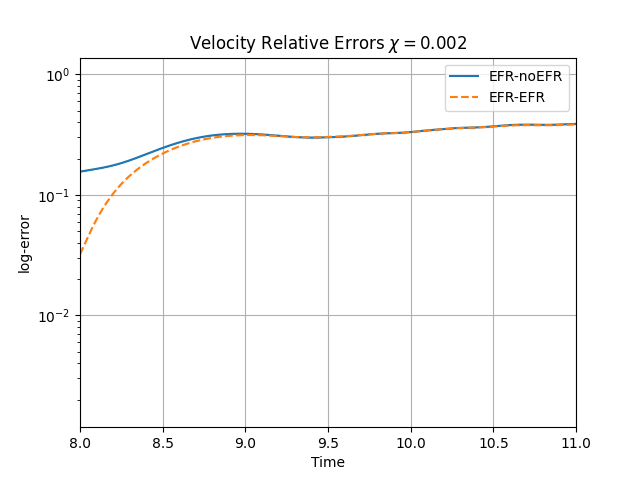}
         \includegraphics[scale=0.4]{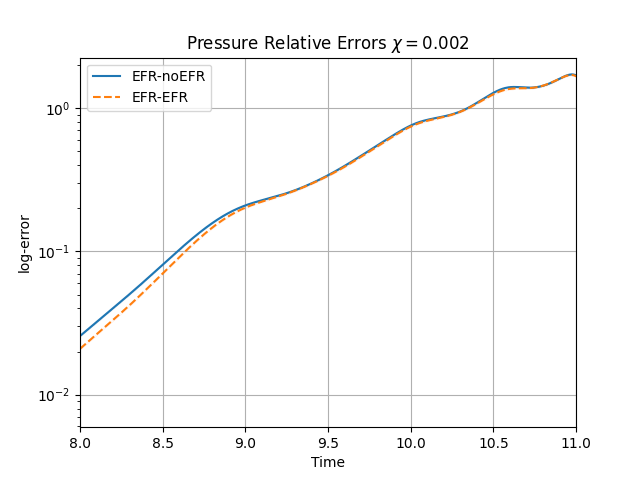}
\caption{\both{\scriptsize{(Experiment 2: $\delta=0.0032$, $\chi = 0.002$ and $r_{\bu}=43, r_p=r_s=8$. Prediction for $t \in [8,11]$.) \emph{Left.}  Comparison of relative log-errors over time of the velocity profiles: EFR full order versus {EFR-noEFR} solutions and EFR full order versus {EFR-EFR} solutions, represented by solid blue  and dashed orange lines, respectively.   \emph{Right.}  Analogous representation for the relative log-errors over time of the pressure profiles.}}
}
\label{fig:errs_chi_2_11}
\end{figure}

\begin{figure}
\centering
        \includegraphics[scale=0.4]{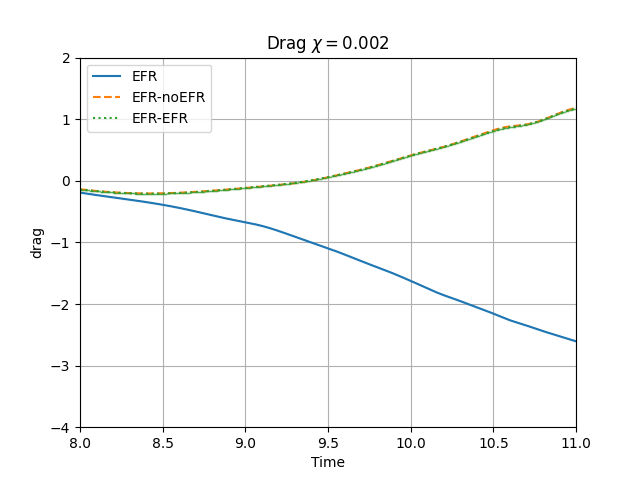}
         \includegraphics[scale=0.4]{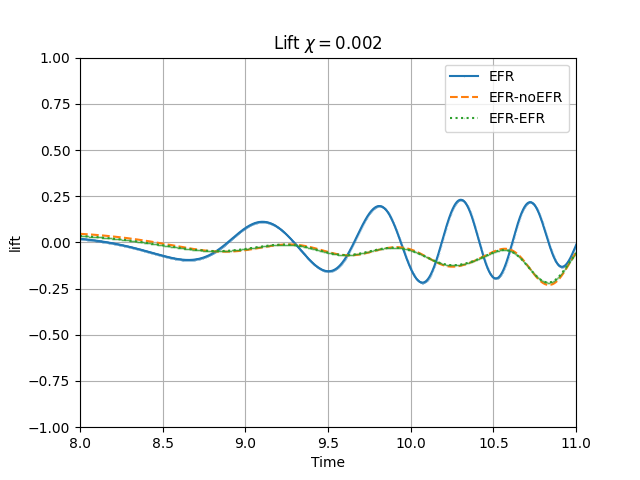}
\caption{\both{\scriptsize{(Experiment 2: $\delta=0.0032$, $\chi = 0.002$ and $r_{\bu}=43, r_p=r_s=8$. Prediction for $t \in [8,11]$.) \emph{Left.}  $C_D(t)$ comparison over time. \emph{Right.}  $C_L(t)$ comparison over time.}}
}
\label{fig:cd_cl_2_11}
\end{figure}

\both{
We analyze the predictive regime up to $T=11$. 
We do not go further in time, since for $t>11$ the Newton's solver of the FOM simulation does not converge. From the plots in Figures \ref{fig:errs_chi_2_11} and \ref{fig:cd_cl_2_11}, it is clear that EFR-noEFR and EFR-EFR are comparable. In the relative error plots of Figure \ref{fig:errs_chi_2_11}, we see how both approaches struggle to represent velocity and pressure fields for large time values. Moreover, EFR-noEFR and EFR-EFR are not capable to accurately predict the force coefficients, as illsutrated in Figure \ref{fig:cd_cl_2_11}. These results are, respectively, confirmed by Table \ref{tab:table_2_11} and by the $L^2-$error over the force coefficients: $\A{\overline E_{C_L}} = 1.00$, $\A{\hat E_{C_L}}=1.00$ and $\A{\overline E_{C_D}} = 0.99$, $\A{\hat E_{C_D}} = 0.99$. 
EFR-EFR performs slightly better than EFR-noEFR for all criteria, except for the drag coefficient.
We note that, as in Experiment 3, both the EFR-EFR and the EFR-noEFR approaches struggle since the flow we investigate displays mixed dynamics (more complex dynamics in the time interval $[4,8]$ and more laminar dynamics in $[8,11]$).} 

\both{
Overall, we conclude that (i) both the EFR-EFR and the EFR-noEFR approaches struggle in the predictive regime, and (ii) both approaches are comparable in terms of accuracy with respect to all the criteria.
Finally, we also note that we do not investigate the EFR-EFR and EFR-noEFR algorithms in the predictive regime of the more challenging Experiment 3 since the two approaches struggled in the predictive regime of Experiment 2.
}

\begin{table}[]
\caption{\scriptsize{\both{(Experiment 2: $\delta=0.0032$, $\chi = 0.002$,  $r_{\bu}=43, r_p=r_s=8$. Prediction for $t \in [8,11]$.) Maximum, minimum and average relative error over the considered time interval for velocity and pressure fields.}}}
\label{tab:table_2_11}
\begin{center}
\begin{tabular}{c|ccc|ccc|}
\cline{2-7}
                                   & \multicolumn{3}{c|}{{EFR-noEFR}}                                                   & \multicolumn{3}{c|}{{EFR-EFR}}                                                \\ \cline{2-7} 
                                   & \multicolumn{1}{c|}{maximum}    & \multicolumn{1}{c|}{minimum}    & average    & \multicolumn{1}{c|}{maximum}    & \multicolumn{1}{c|}{minimum}    & average    \\ \hline
\multicolumn{1}{|c|}{$E_{\bu}(t)$} & \multicolumn{1}{c|}{$1.000e$+0} & \multicolumn{1}{c|}{$1.558e$-1} & $4.549e$-1 & \multicolumn{1}{c|}{$1.000e$+0} & \multicolumn{1}{c|}{$3.211e$-2} & $4.437e$-1 \\ \hline
\multicolumn{1}{|c|}{$E_{p}(t)$}   & \multicolumn{1}{c|}{$1.721e$+0} & \multicolumn{1}{c|}{$2.569e$-2} & $6.856e$-1 & \multicolumn{1}{c|}{$1.704e$+0} & \multicolumn{1}{c|}{$2.040e$-2} & $6.779e$-1 \\ \hline
\end{tabular}%
\end{center}
\end{table}

\newpage

\section{Model Reduction: With Respect to Time and the Reynolds Number}
\label{sec:rom_mu}

\A{In Section~\ref{sec:g-rom}, we showed that the {{EFR-EFR}} is more accurate than the {EFR-noEFR} when model reduction is performed in the time domain.
In this section, we perform a numerical investigation of the {{EFR-EFR}} and {EFR-noEFR} when model reduction is performed not only in the time domain (as we did in Section~\ref{sec:g-rom}), but also in the parameter domain (i.e., with respect to $\nu$). }
To this end, we consider the NSE~\eqref{eq:NSE} with a variable kinematic viscosity: $\nu \in [\nu_{min}, \nu_{max}] \subset \mathbb R^+$.
From definition \eqref{eq:Re}, it is clear that 
changing 
$\nu$ will 
change the Reynolds number, which will vary in the interval $[0, Re_{max}] \subset \mathbb R^+$. 

To perform the model reduction both in the time domain and in the parameter domain, a standard POD approach that performs a simultaneous compression in time and in the parametric space would require a significant computational effort.
Thus, to avoid the high computational cost of this brute force POD approach, in our numerical investigation we use a nested-POD (n-POD) algorithm.
This compression algorithm is very popular (and goes by different names) in 
the ROM community: see, e.g., \cite{audouze2009reduced,ballarin2016fast,brands2016reduced, himpe2018hierarchical}. 
In the n-POD algorithm, the compression is performed in two different stages, which operate first in time and then in the parametric space.
Namely, first each time trajectory related to the parameter set explored is compressed, and then a  POD reduction is performed on the already reduced parametric solutions.
By decoupling the model reduction in time from the model reduction in the parameter domain, the n-POD algorithm achieves significant reductions in computational time and storage with respect to the monolithic POD algorithm.
In our numerical investigation, we use the n-POD algorithm presented in \cite{kadeethum2021nonintrusive}.  
The n-POD is based on two decoupled levels\A{.}
\begin{enumerate}
    \item \emph{\A{A } first compression of the time trajectories}. In this first phase, a training set over the parameter space is chosen: 
    $\{\nu_{i}\}_{i=1}^{N_{\nu}}$. For each $\nu_i$, a standard POD in time is applied, retaining the first $N_{\bu}^{t}, N_{p}^{t}, N_{s}^{t}$ modes for the velocity, pressure, and supremizer variables, respectively. 
    We denote these modes scaled by their singular values with $ m_{\bu(\nu_i)}^{j}, m_{p(\nu_i)}^{j}$ and $m_{S(p(\nu_i))}^{j}$, for $j = 1, \dots, N^t \doteq N_{\bu}^{t} = N_{p}^{t} = N_{s}^{t}$. For the sake of simplicity, in our setting we choose the same number of modes for all the variables and for all the parameters in the training set.  However, in principle, one can choose the number of modes for each parametric snapshot through energy criteria, and the number can be different for each variable. 
    \item \emph{A global compression of the scaled modes}. 
    Using the POD procedure presented in Section \ref{sec:g-rom}, the following reduced space for velocity
    $$
{{\mathbb U}}^{r_{\bu s}} \doteq \text{POD}(\{ m_{\bu(\nu_1)}^{j} \}_{j=1}^{N^t}, \cdots, \{ m_{\bu(\nu_{N_{\nu}})}^{j} \}_{j=1}^{N^t})\oplus \text{POD}(\{m_{S(p(\nu_1))}^{j} \}_{j=1}^{N^t}, \cdots, \{ m_{S(p(\nu_{N_{\nu}}))}^{j} \}_{j=1}^{N^t}),
$$
and the following space for pressure
    $$
{{\mathbb Q}}^{r_{p}} \doteq \text{POD}(\{m_{p(\nu_1)}^{j} \}_{j=1}^{N^t}, \cdots, \{ m_{p(\nu_{N_{\nu}})}^{j} \}_{j=1}^{N^t})
$$
are obtained.
\end{enumerate}
For the sake of readability, we introduce the following notation
$$
{{\mathbb U}}^{r} \doteq \text{n-POD}(\{\bu(\nu_i)\}_{i=1}^{N_{\nu}}; N^t) \oplus 
\text{n-POD}(\{S(p(\nu_i))\}_{i=1}^{N_{\nu}}; N^t)
$$
and
$$
{{\mathbb Q}}^{r_p} \doteq \text{n-POD}(\{p(\nu_i)\}_{i=1}^{N_{\nu}}; N^t)
$$
for the ROM velocity and pressure fields, respectively. Here, 
$N^{t}$ 
denotes the first phase of the time evolution compression, while $r$ and $r_p$ are the final reduced space dimensions for velocity and pressure, after supremizer stabilization.

Pseudocodes that describe the {EFR-noEFR} and {{EFR-EFR}} approaches coupled with the n-POD algorithm are reported in Algorithm \ref{alg:03} and Algorithm \ref{alg:04}, respectively. 

\begin{algorithm}
\caption{Pseudocode for {EFR-noEFR} with n-POD}\label{alg:03}
\begin{algorithmic}[1]
\State{$\bu_0, \bu_{in}, N_{\bu}, N_{p}, N_{\nu}, N^t$}\Comment{Inputs needed}
\For{$i \in \{1, \dots, N_{\nu}\}$}\Comment{Parameter loop}
\For{$n \in \{0, \dots, N_T - 1\}$}\Comment{Time loop}
\State{(I) + (II) + (III)} \Comment{EFR simulation}
\EndFor
\State{$\{\bu(\nu_i)_j\}_{j=1}^{N_{\bu}} \subseteq \{\bu^k(\nu_i)\}_{k=1}^{N_{T}}$ 
\quad $\{p(\nu_i)_j\}_{j=1}^{N_{p}} \subseteq \{p^k(\nu_i)\}_{k=1}^{N_{T}}$} \Comment{Snapshot collection}
\EndFor
\State{$\mathbb U^r \doteq \text{n-POD}(\{\bu(\nu_i)\}_{i=1}^{N_{\nu}}; N^{t})\oplus \text{n-POD}(\{S(p(\nu_i))_{i=1}^{N_{\nu}})\}$ } \Comment{Supremizer enrichment for velociy}
\State{$\mathbb Q^{r_p} \doteq \text{n-POD}(\{p(\nu_i)\}_{i=1}^{N_{\nu}}; N^t)$ } \Comment{Standard n-POD for pressure space} 
\For{$n \in \{0, \dots, N_T - 1\}$}\Comment{Time loop}
\State{Solve system \eqref{eqn:g-rom-2-n+1}} 
\Comment{
Standard Galerkin projection}
\EndFor
\end{algorithmic}
\end{algorithm}

\begin{algorithm}
\caption{Pseudocode for {{EFR-EFR}} with n-POD}\label{alg:04}
\begin{algorithmic}[1]
\State{$\bu_0, \bu_{in}, N_{\bu}, N_{p}, N_{\nu}, N^t$}\Comment{Inputs needed}
\For{$i \in \{1, \dots, N_{\nu}\}$}\Comment{Parameter loop}
\For{$n \in \{0, \dots, N_T - 1\}$}\Comment{Time loop}
\State{(I) + (II) + (III)} \Comment{EFR simulation}
\EndFor
\State{$\{\bu(\nu_i)_j\}_{j=1}^{N_{\bu}} \subseteq \{\bu^k(\nu_i)\}_{k=1}^{N_{T}}$ 
\quad $\{p(\nu_i)_j\}_{j=1}^{N_{p}} \subseteq \{p^k(\nu_i)\}_{k=1}^{N_{T}}$} \Comment{Snapshot collection}
\EndFor
\State{$\mathbb U^r \doteq \text{n-POD}(\{\bu(\nu_i)\}_{i=1}^{N_{\nu}}; N^{t})\oplus \text{n-POD}(\{S(p(\nu_i))_{i=1}^{N_{\nu}})\}$ } \Comment{Supremizer enrichment for velocity} 
\State{$\mathbb Q^{r_p} \doteq \text{n-POD}(\{p(\nu_i)\}_{i=1}^{N_{\nu}}, N^t)$ } \Comment{Standard n-POD for pressure space}
\For{$n \in \{0, \dots, N_T - 1\}$}\Comment{Time loop}
\State{(I)$_r$ + (II)$_r$ + (III)$_r$} \Comment{EFR at the reduced level}
\EndFor
\end{algorithmic}
\end{algorithm}

{\bf Experiment 4}. 
In this experiment, we use the same test problem 
as the one used in Section \ref{sec:results} generalized to a parametric kinematic viscosity.

In our numerical investigation, we consider the parametric domain $\nu \in 
[10^{-3},1.575 
\times 10^{-3}]$, 
which yields $Re_{max} \in [65, 100]$.
To explain the rationale for choosing this parametric domain, we 
define the averaged Reynolds number as
\begin{equation}
    \overline{Re} \doteq \frac{\overline U L}{\nu},
\end{equation}
where $\overline U$ is the time averaged magnitude of the inflow velocity, $\bu_{in}$. 
For our parametric domain, we
choose $\nu=1.575 
\times 10^{-3}$ 
since this value yields $\overline{Re}=40$, which is the lower bound of the kinematic viscosity that 
achieves a vortex shedding behavior in the case of a steady inlet condition (see, e.g., \cite{lienhard1966synopsis}). 
We note that we choose a parametric domain that ensures only one type of flow dynamics (i.e., a vortex shedding regime).
Choosing a parametric domain that spans various flow dynamics would be a more challenging test for the proposed ROMs, as noted in Experiment 3.

In the numerical investigation in Section~\ref{sec:g-rom}, 
we used the Kolmogorov scale \cite{kolmogorov1941dissipation,kolmogorov1941local} as a filtering radius. 
However, the Kolmogorov scale changes with respect to the choice of the kinematic viscosity, $\nu$. 
\A{Thus, in Experiment 4, we 
use $\delta=h_{\text{min}}$}, i.e., a classical choice for nonuniform meshes, as specified in Remark \ref{remark:delta}.
We also fix $\chi=0.002$.
To build the ROM basis, we  collect $N_{\bu} = N_p = 2000$ equally spaced snapshots in the time interval $[4,8]$, which 
is the setting used in Experiment 2.
In the n-POD algorithm, we perform the
first compression 
choosing $N^t=60$ for each sampled parametric instance. We pick $N_{\nu}=20$ parameters using a log-equispaced distribution. 
We choose the log-uniform distribution since it allows us to collect
more snapshots around the value $\nu=10^{-3}$, where we 
believe more information is needed because of higher vortex shedding frequency. 
We choose the $N^t$ value heuristically, seeking an accurate approximation of the time evolution for the various parametric snapshots.
We pick $N_{\nu}=20$ 
seeking to minimize the computational costs of the building phase. Of course, a more detailed investigation 
would probably find better parameters, i.e., parameters that yield more accurate approximations. 
In the second stage of compression of the n-POD algorithm, we retain 
$99.9\%$ of the system energy employing $r_{\bu}=33$, $r_s = r_p = 3$.

\A{
\begin{figure}\centering 
        \includegraphics[scale=0.4]{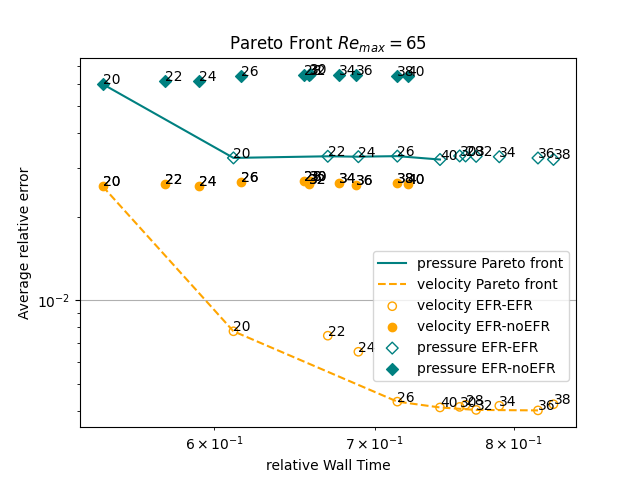} \includegraphics[scale=0.4]{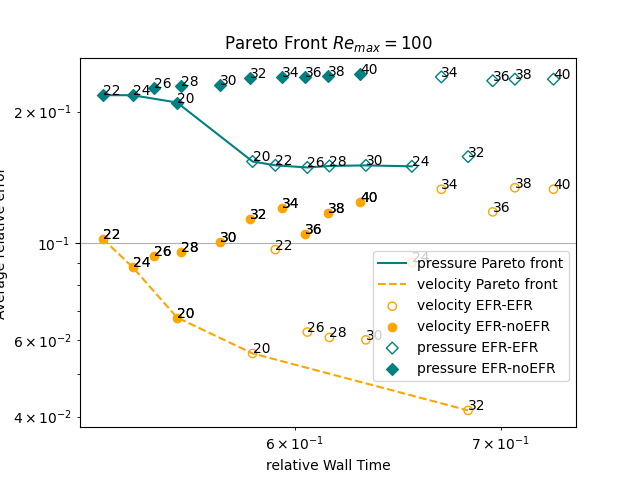} 
\caption{\scriptsize{\A{(Experiment 4: $\delta=h_{min}$, $\chi = 0.002$, 
$r_{\bu}=\{20, 22, 24, 26, 28, 30, 32, 34, 36, 38, 40\}$, and  $r_p=r_s=3$.) \emph{Left.} Pareto plots for velocity (orange) and pressure (teal) fields and $Re_{max}=65$: averaged relative error in time versus relative wall time for varying $r_{\bu}$. \emph{Right}. Analogous plot for $Re_{max}=100$.}}
}
\label{fig:pareto4}
\end{figure}
}

In Section~\ref{sec:re65}, we compare the {EFR-noEFR} with the {{EFR-EFR}} for $Re_{max}=65$ (i.e., $\nu =1.575
\times 10^{-3}$).
In Section~\ref{sec:re100}, we compare the {EFR-noEFR} with the {{EFR-EFR}} for $Re_{max}=100$ (i.e., $\nu = 10^{-3}$).

\subsection{Reynolds number $Re_{max}=65$: \A{reconstructive regime}}
    \label{sec:re65}
    
\begin{figure} \centering
        \includegraphics[scale=0.4]{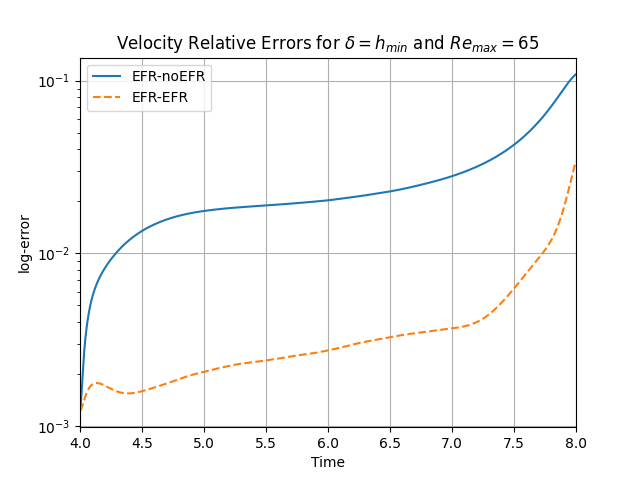}
         \includegraphics[scale=0.4]{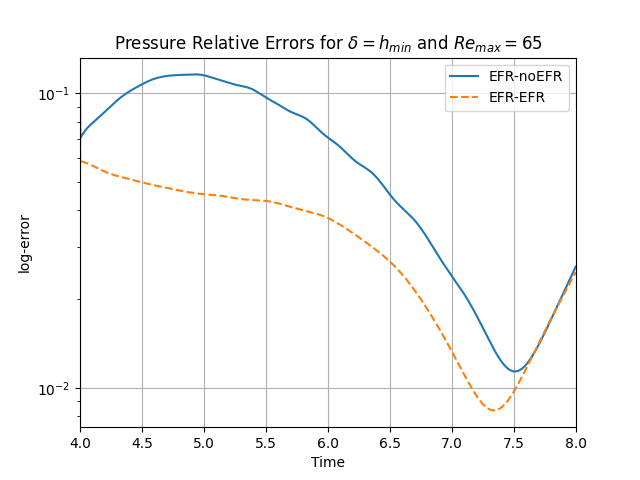}
\caption{\scriptsize{(Experiment 4: $\delta=h_{\text{min}}$, $\chi = 0.002$ and $r_{\bu}=33, r_p=r_s=3$, $Re_{max}=65$. Reconstruction for $t \in [4,8]$.) \emph{Left.}  Comparison of relative log-errors 
of the velocity profiles: 
{EFR-noEFR} (solid blue line) and {{EFR-EFR}} (dashed orange line).
\emph{Right.}  
Comparison of relative log-errors of the pressure profiles: 
{EFR-noEFR} (solid blue line) and {{EFR-EFR}} (dashed orange line).
}
}
\label{fig:4err_65}
\end{figure}

\begin{figure}
       \centering \includegraphics[scale=0.21]{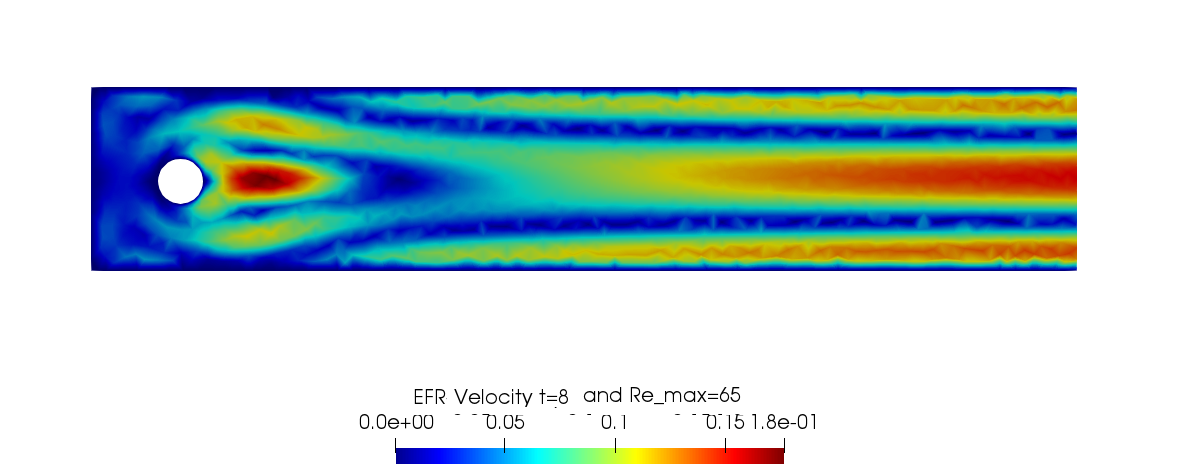}\\
\begin{minipage}{.49\textwidth}
 \centering
         \includegraphics[scale=0.25]{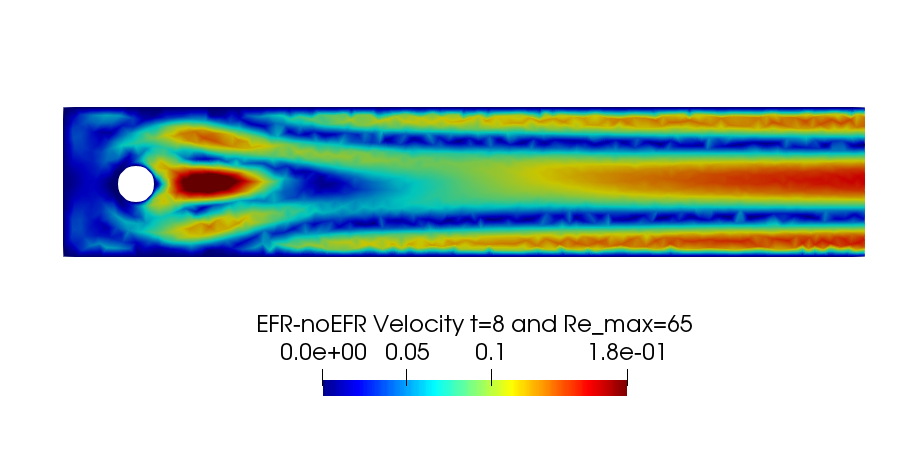}
\end{minipage}
\begin{minipage}{.49\textwidth}
 \centering
         \includegraphics[scale=0.21]{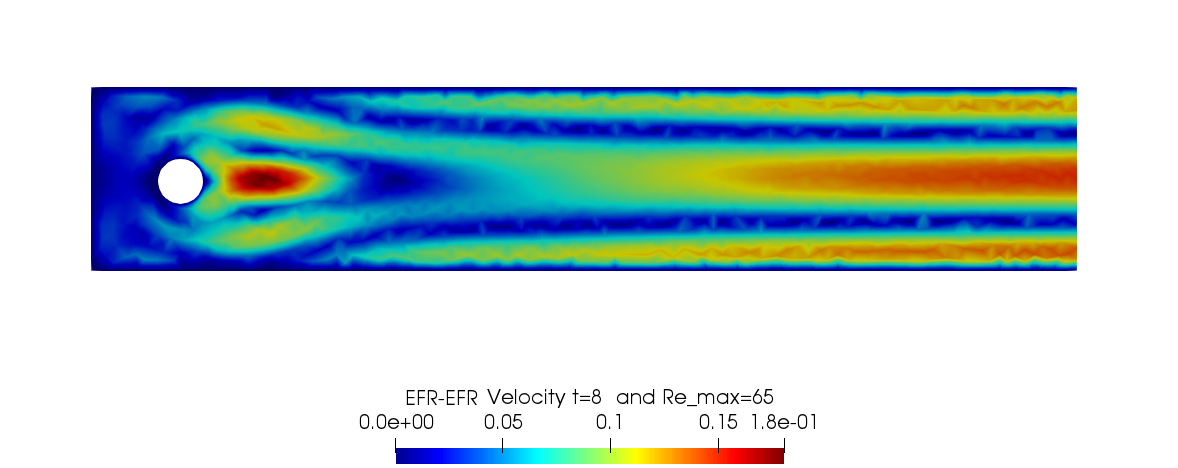}
\end{minipage}
\caption{\scriptsize{(Experiment 4: $\delta=h_{min}$, $\chi = 0.002$ and $r_{\bu}=33, r_p=r_s=3$, $Re_{max}=65$, and $t=8$.) \emph{Top.} Full order EFR velocity magnitude. \emph{Bottom Left.} Reduced {EFR-noEFR} velocity magnitude. \emph{Bottom Right.} Reduced {{EFR-EFR}} velocity magnitude.}
}
\label{fig:4u_65}
\end{figure}

\begin{figure}
       \centering \includegraphics[scale=0.25]{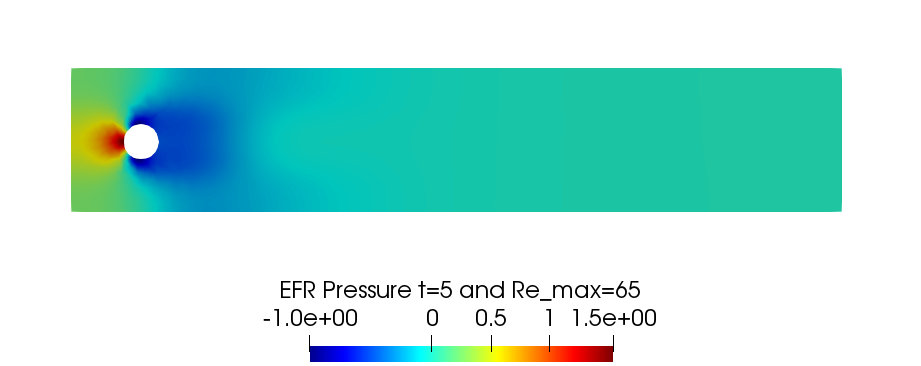}\\
\begin{minipage}{.49\textwidth}
 \centering
         \includegraphics[scale=0.25]{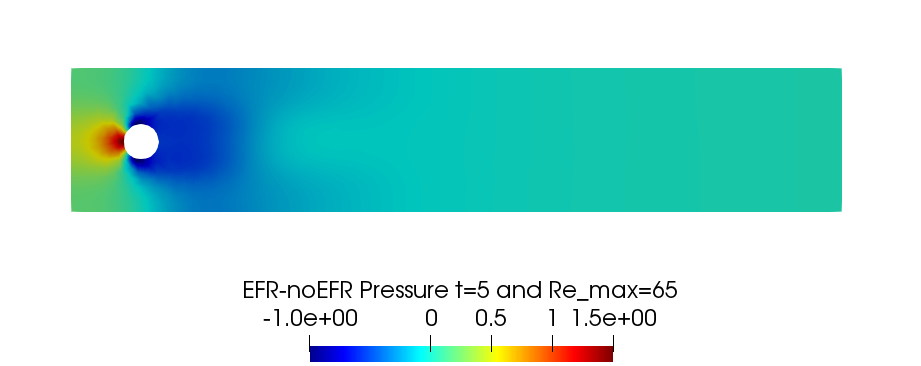}
\end{minipage}
\begin{minipage}{.49\textwidth}
 \centering
         \includegraphics[scale=0.25]{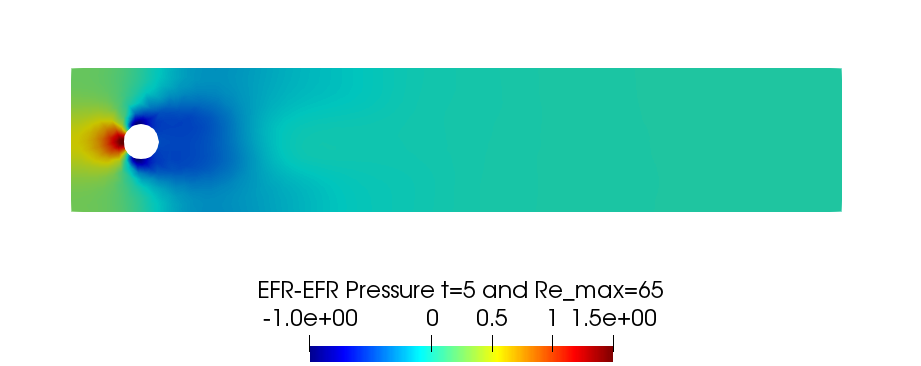}
\end{minipage}
\caption{\scriptsize{(Experiment 4: $\delta=h_{min}$, $\chi = 0.002$ and $r_{\bu}=33, r_p=r_s=3$, $Re_{max}=65$, and $t=5$.) \emph{Top.} Full order EFR pressure field. \emph{Bottom Left.} Reduced {EFR-noEFR} pressure field. \emph{Bottom Right.} Reduced {{EFR-EFR}} pressure field.}
}
\label{fig:4p_65}
\end{figure}

\begin{figure} \centering
        \includegraphics[scale=0.4]{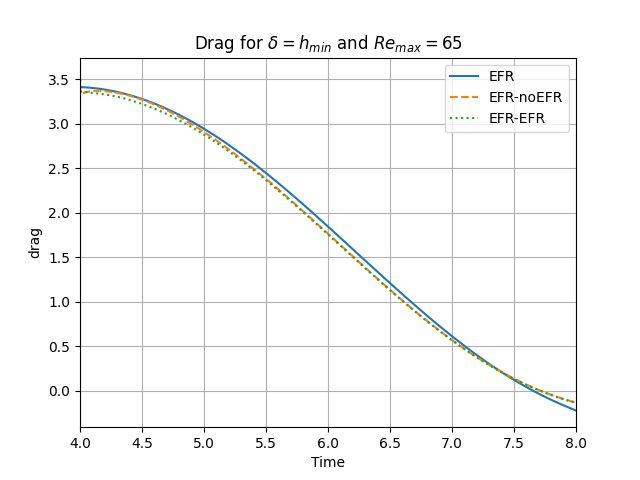}
         \includegraphics[scale=0.4]{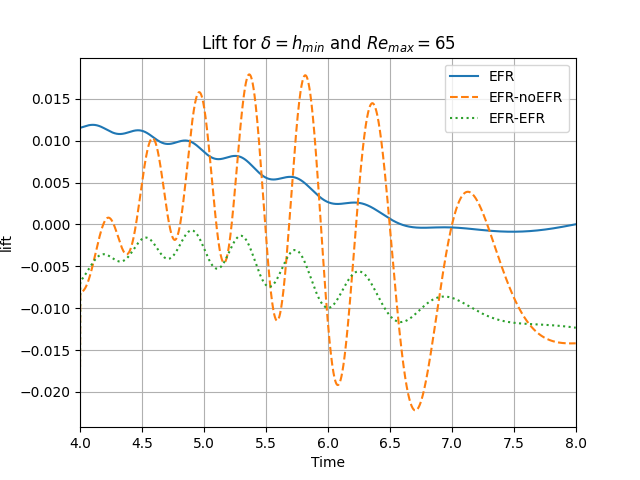}
\caption{\scriptsize{(Experiment 4: $\delta=h_{\text{min}}$, $\chi = 0.002$ and $r_{\bu}=33, r_p=r_s=3$, $Re_{max}=65$. Reconstruction for $t \in [4,8]$.) \emph{Left.}  $C_D(t)$ comparison over time.
\emph{Right.}  $C_L(t)$ comparison over time.}
}
\label{fig:4cd_cl_65}
\end{figure}

The relative log-errors 
plotted in Figure \ref{fig:4err_65} for both the
velocity and the pressure are lower for the {{EFR-EFR}} than for the {EFR-noEFR}.
The {{EFR-EFR}} is also more accurate than the {EFR-noEFR} in approximating the velocity field at $t=8$ (Figure \ref{fig:4u_65}) and the pressure field at $t=5$ (Figure \ref{fig:4p_65}). 
Both algorithms yield  accurate approximations for the drag coefficient, $C_D(t)$ (Figure \ref{fig:4cd_cl_65}, left) and relatively inaccurate approximations for the 
lift coefficient, $C_L(t)$ (Figure \ref{fig:4cd_cl_65}, right).
\A{Table \ref{tab:table65} lists the maximum, minimum, and average error values over time for the velocity and pressure fields, and shows that the {EFR-EFR} is more accurate than the {EFR-noEFR}. However, the EFR-noEFR strategy is slightly more accurate in the reconstruction of the force coefficients in terms of $L^2-$errors: $\overline E_{C_L} = 1.75$, $\hat E_{C_L} = 1.84$, $\overline E_{C_D} = 0.025$, and $\hat E_{C_D} = 0.030$.\\[0.1cm]
Overall, the improvement in the {EFR-EFR} is highlighted by the left Pareto plot in Figure \ref{fig:pareto4}. 
Indeed, fixing $r_p=r_s=3$ and choosing $r_u = 20, 22, 24, 26, 28, 30, 32, 34, 36, 38, 40$ shows that, over this range of $r_u$ values, the {EFR-EFR} performs better than the {EFR-noEFR} with respect to both the velocity and the pressure approximations.
Indeed, the {EFR-EFR} with $r_u=20$ yields a low relative error that the {EFR-noEFR} cannot attain by increasing its $r_u$ value (and, consequently, its relative wall time).
}


\begin{table}[]
\caption{\scriptsize{\A{(Experiment 4: $\delta=h_{min}$, $\chi = 0.002$,  $r_{\bu}=33, r_p=r_s=3$, and $Re_{max}=65$. Reconstruction for $t \in [4,8]$.) Maximum, minimum,  and average relative error over the considered time interval for the  velocity and pressure fields.}}}
\label{tab:table65}
\begin{center}
\begin{tabular}{c|ccc|ccc|}
\cline{2-7}
                                   & \multicolumn{3}{c|}{{EFR-noEFR}}                                                   & \multicolumn{3}{c|}{{EFR-EFR}}                                                \\ \cline{2-7} 
                                   & \multicolumn{1}{c|}{maximum}    & \multicolumn{1}{c|}{minimum}    & average    & \multicolumn{1}{c|}{maximum}    & \multicolumn{1}{c|}{minimum}    & average    \\ \hline
\multicolumn{1}{|c|}{$E_{\bu}(t)$} & \multicolumn{1}{c|}{$1.086e$-1} & \multicolumn{1}{c|}{$1.229e$-3} & $2.636e$-1 & \multicolumn{1}{c|}{$3.635e$-2} & \multicolumn{1}{c|}{$1.230e$-3} & $4.269e$-3 \\ \hline
\multicolumn{1}{|c|}{$E_{p}(t)$}   & \multicolumn{1}{c|}{$1.157e$-1} & \multicolumn{1}{c|}{$1.136e$-2} & $6.507e$-1 & \multicolumn{1}{c|}{$6.316e$-2} & \multicolumn{1}{c|}{$8.833e$-3} & $3.651e$-2 \\ \hline
\end{tabular}%
\end{center}
\end{table}

\subsection{Reynolds number $Re_{max}=100$: \A{reconstructive regime}}
    \label{sec:re100}

\begin{figure} \centering
        \includegraphics[scale=0.4]{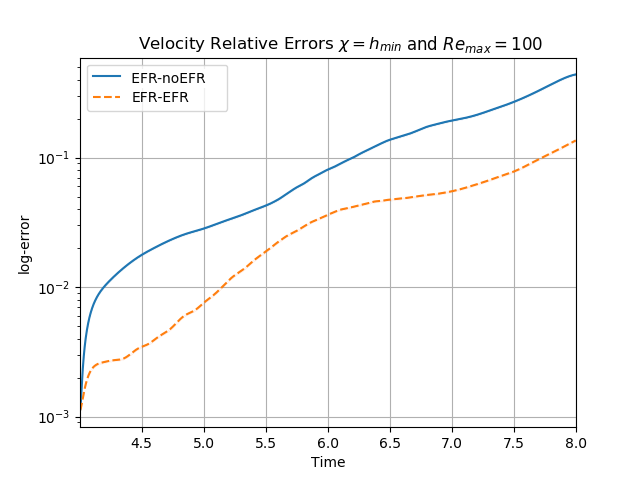}
         \includegraphics[scale=0.4]{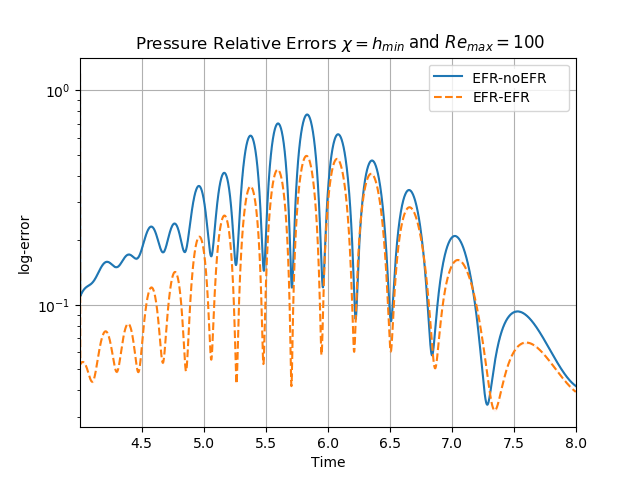}
\caption{\scriptsize{(Experiment 4: $\delta=h_{\text{min}}$, $\chi = 0.002$, $r_{\bu}=33, r_p=r_s=3$, and $Re_{max}=100$. Reconstruction for $t \in [4,8]$.) \emph{Left.}  
Comparison of relative log-errors of the velocity profiles: 
{EFR-noEFR} (solid blue line) and {{EFR-EFR}} (dashed orange line).
\emph{Right.}  
Comparison of relative log-errors of the pressure profiles: 
{EFR-noEFR} (solid blue line) and {{EFR-EFR}} (dashed orange line).
}
}
\label{fig:4err_100}
\end{figure}
\begin{figure}
      \centering  \includegraphics[scale=0.21]{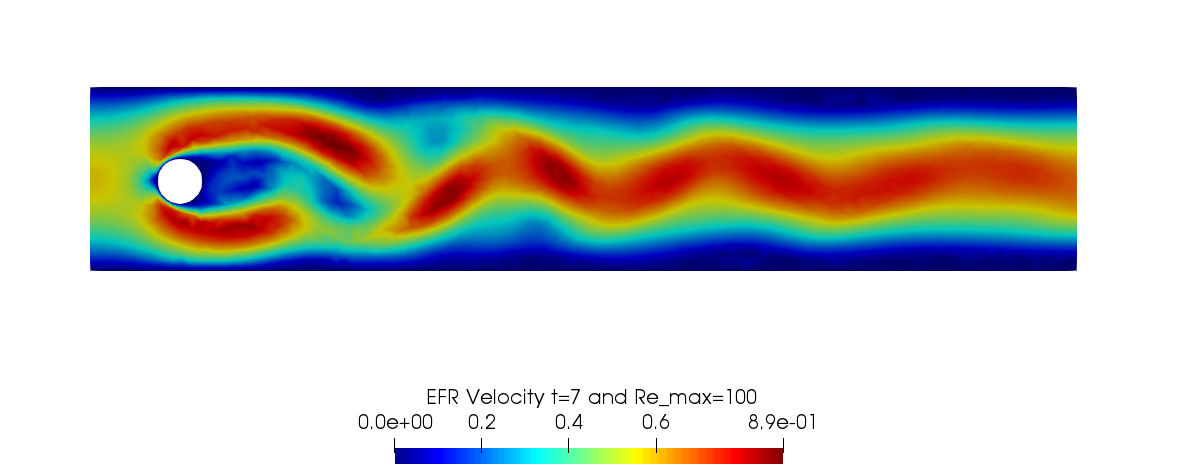}\\
\begin{minipage}{.49\textwidth}
 \centering
         \includegraphics[scale=0.21]{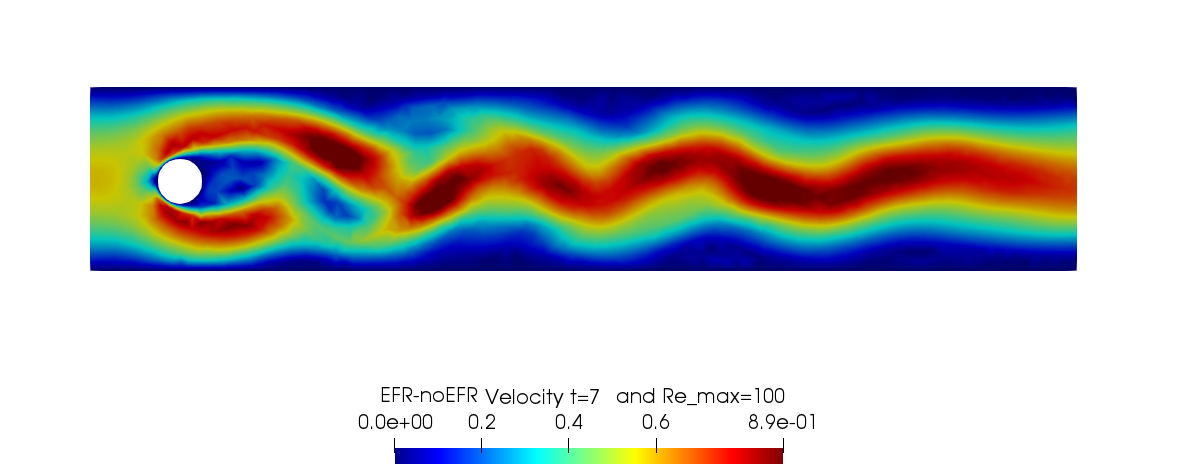}
\end{minipage}
\begin{minipage}{.49\textwidth}
 \centering
         \includegraphics[scale=0.21]{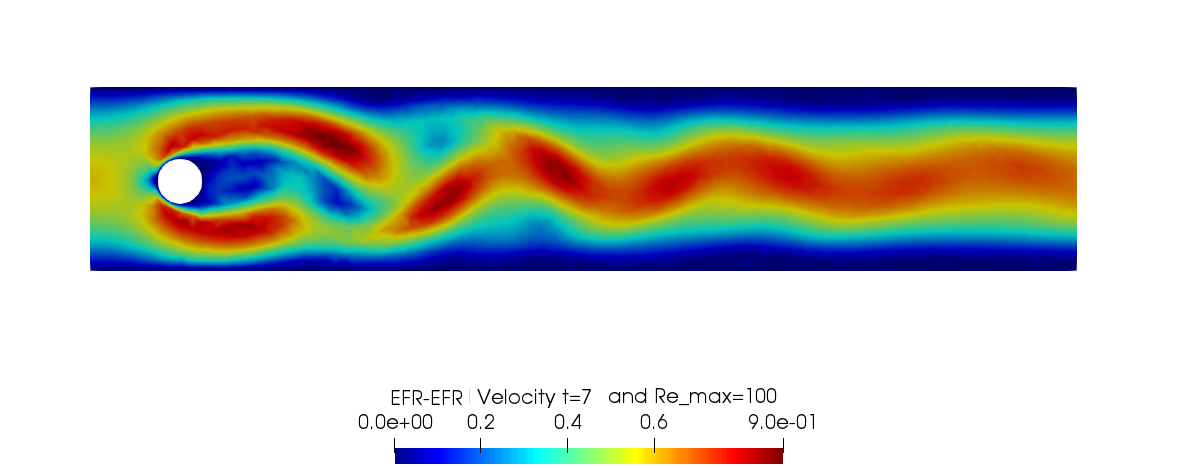}
\end{minipage}
\caption{\scriptsize{(Experiment 4: $\delta=h_{\text{min}}$, $\chi = 0.002$ and $r_{\bu}=33, r_p=r_s=3$, $Re_{max}=100$, and $t=7$.) \emph{Top.} Full order EFR velocity magnitude. \emph{Bottom Left.} Reduced {EFR-noEFR} velocity magnitude. \emph{Bottom Right.} Reduced {{EFR-EFR}} velocity magnitude.}
}
\label{fig:4u_100}
\end{figure}
\begin{figure}
      \centering  \includegraphics[scale=0.21]{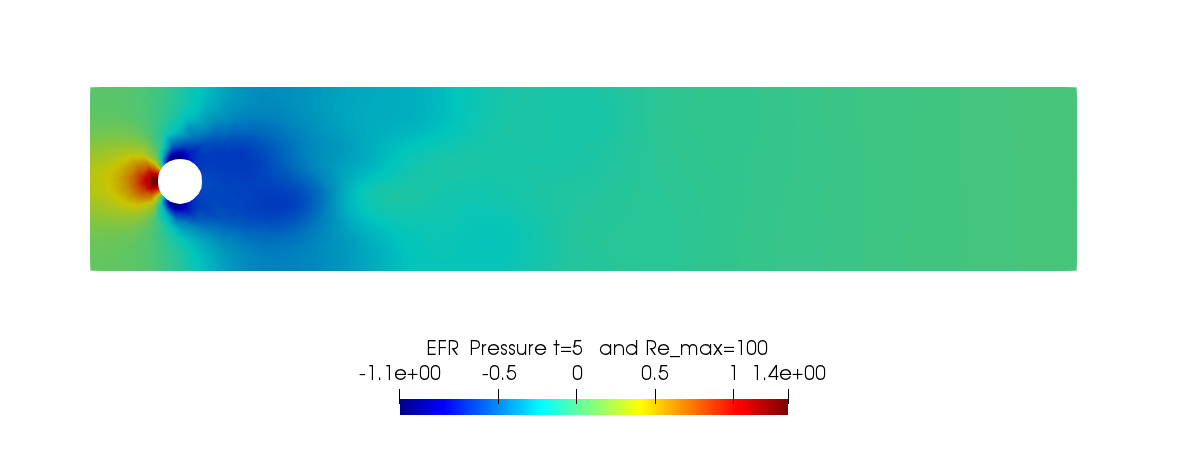}\\
\begin{minipage}{.49\textwidth}
 \centering
         \includegraphics[scale=0.21]{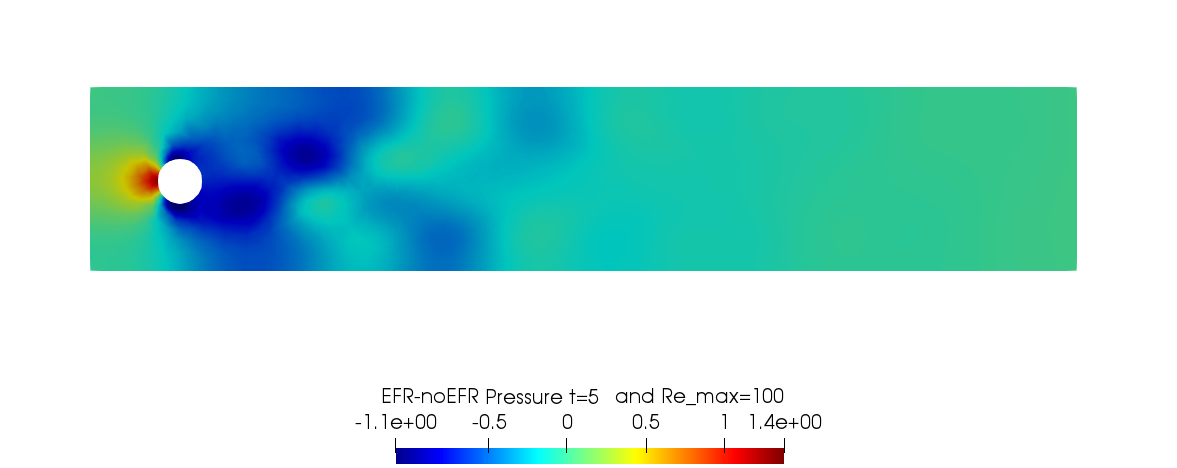}
\end{minipage}
\begin{minipage}{.49\textwidth}
 \centering
         \includegraphics[scale=0.21]{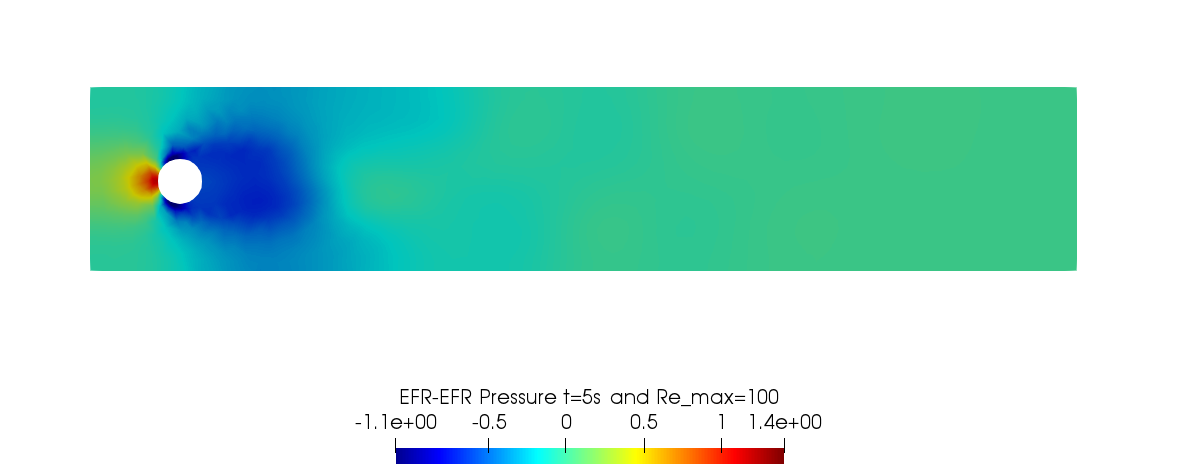}
\end{minipage}
\caption{\scriptsize{(Experiment 4: $\delta=h_{\text{min}}$, $\chi = 0.002$ and $r_{\bu}=33, r_p=r_s=3$, $Re_{max}=100$, and $t=5$.) \emph{Top.} Full order EFR pressure field. \emph{Bottom Left.} Reduced {EFR-noEFR} pressure field. \emph{Bottom Right.} Reduced {{EFR-EFR}} pressure field.}}
\label{fig:4p_100}
\end{figure}
\begin{figure} \centering
        \includegraphics[scale=0.4]{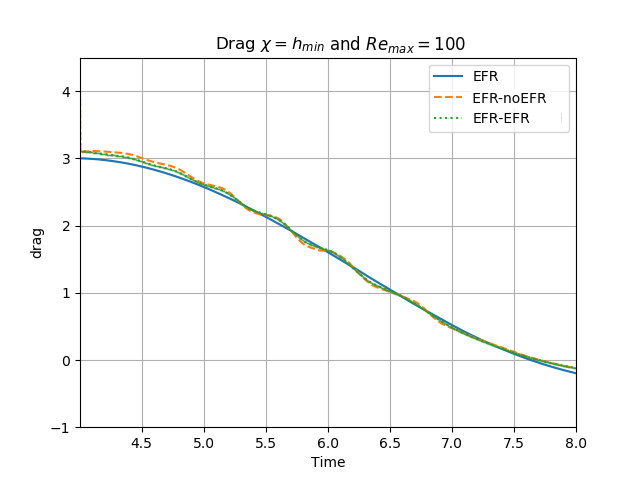}
         \includegraphics[scale=0.4]{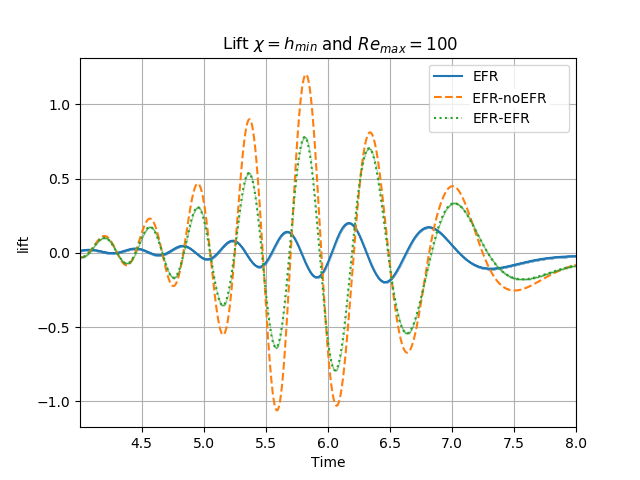}
\caption{\scriptsize{(Experiment 4: $\delta=h_{\text{min}}$, $\chi = 0.002$, $r_{\bu}=33, r_p=r_s=3$, and $Re_{max}=100$. Reconstruction for $t \in [4,8]$.) \emph{Left.}  $C_D(t)$ comparison over time.
\emph{Right.}  $C_L(t)$ comparison over time.}
}
\label{fig:4cd_cl_100}
\end{figure}

The relative log-errors plotted in Figure \ref{fig:4err_100} for both the velocity and the pressure are significantly lower for the {{EFR-EFR}} than for the {EFR-noEFR}.
The {{EFR-EFR}} is also more accurate than the {EFR-noEFR} in approximating the velocity field at $t=7$ (Figure \ref{fig:4u_100}) and the pressure field at $t=5$ (Figure \ref{fig:4p_100}). 
Furthermore, both the {{EFR-EFR}} and the {EFR-noEFR} yield accurate drag coefficients, $C_D(t)$ (Figure \ref{fig:4cd_cl_100}, left).
Although both the {{EFR-EFR}} and the {EFR-noEFR} lift coefficients, $C_L(t)$ (Figure \ref{fig:4cd_cl_100}, right), are relatively inaccurate, the {{EFR-EFR}} approximation is more accurate than the {EFR-noEFR}.



\A{Focusing on Table \ref{tab:table100}, we observe that EFR-EFR and EFR-noEFR present comparable results for the minimum value of the velocity relative error. For the other pointwise error values, EFR-EFR is more accurate than EFR-noEFR, both for velocity and pressure. In terms of $L^2-$errors of the force coefficients, EFR-EFR is always more accurate than EFR-noEFR. Indeed, $\overline E_{C_L} = 5.55$, $\hat E_{C_L} = 4.08$, $\overline E_{C_D} = 0.04$, and $\hat E_{C_D} = 0.02$. 
We also present a Pareto plot for $Re_{max}=100$ in the right panel of Figure \ref{fig:pareto4}. We fix $r_p=r_s=3$ and 
choose $r_{\bu} = 20, 22, 24, 26, 28, 30, 32, 34, 36, 38, 40$. 
Over this range of $r_{\bu}$ values, we observe that, although EFR-noEFR is optimal for smaller $r_{\bu}$ values, the relatively low EFR-EFR error for $r_{\bu}=20$ 
cannot be reached by EFR-noEFR even by  increasing its $r_{\bu}$ value (and, consequently, its relative wall time).}
\begin{table}[]
\caption{\scriptsize{\A{(Experiment 4: $\delta=h_{min}$, $\chi = 0.002$,  $r_{\bu}=33, r_p=r_s=3$, and $Re_{max}=100$. Reconstruction for $t \in [4,8]$.) Maximum, minimum,  and average relative error over the considered time interval for velocity and pressure fields.}}}
\label{tab:table100}
\begin{center}
\begin{tabular}{c|ccc|ccc|}
\cline{2-7}
                                   & \multicolumn{3}{c|}{{EFR-noEFR}}                                                   & \multicolumn{3}{c|}{{EFR-EFR}}                                                \\ \cline{2-7} 
                                   & \multicolumn{1}{c|}{maximum}    & \multicolumn{1}{c|}{minimum}    & average    & \multicolumn{1}{c|}{maximum}    & \multicolumn{1}{c|}{minimum}    & average    \\ \hline
\multicolumn{1}{|c|}{$E_{\bu}(t)$} & \multicolumn{1}{c|}{$4.383e$-1} & \multicolumn{1}{c|}{$1.115e$-3} & $1.220e$-1 & \multicolumn{1}{c|}{$1.354e$-1} & \multicolumn{1}{c|}{$1.117e$-3} & $3.871e$-2 \\ \hline
\multicolumn{1}{|c|}{$E_{p}(t)$}   & \multicolumn{1}{c|}{$7.720e$-1} & \multicolumn{1}{c|}{$3.419e$-2} & $2.426e$-1 & \multicolumn{1}{c|}{$4.949e$-1} & \multicolumn{1}{c|}{$3.215e$-2} & $1.581e$-1 \\ \hline
\end{tabular}%
\end{center}
\end{table}
\A{
The numerical results for $Re_{max}=65$ 
and $Re_{max}=100$ show that, for both Reynolds numbers and for all criteria, the {{EFR-EFR}} is consistently more accurate than the {EFR-noEFR} (although this difference between the EFR-EFR and EFR-noEFR seems to be somewhat lower for $Re_{max}=100$). 
Thus, the numerical investigation in this section suggests that the FOM-ROM consistency is important when the EFR stabilization is used and model reduction is performed both in time and in the parametric domain.
}

\begin{figure}
\centering
        \includegraphics[scale=0.4]{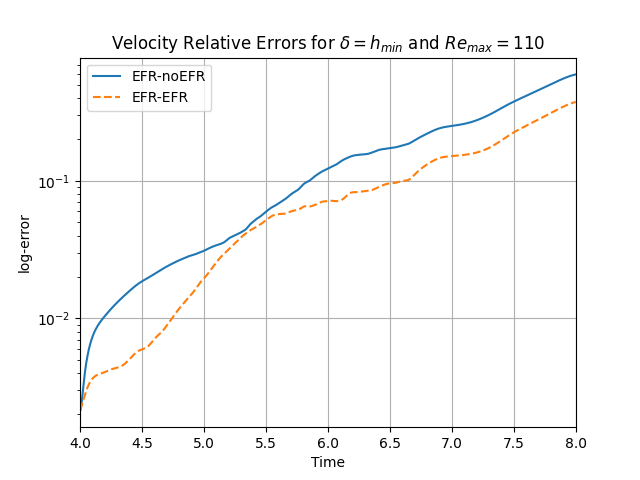}
         \includegraphics[scale=0.4]{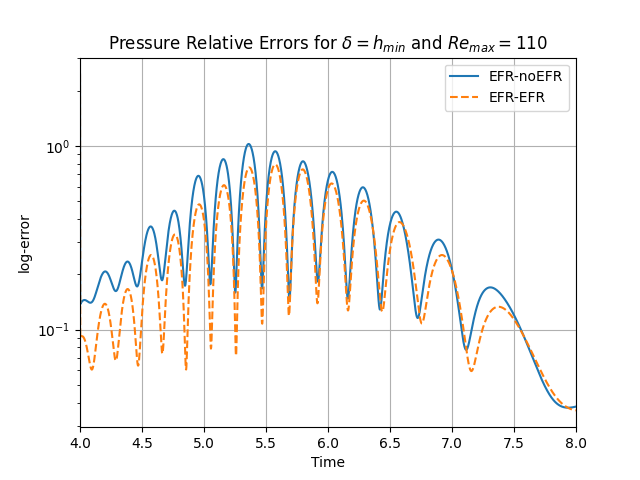}
\caption{\both{\scriptsize{(Experiment 4: $\delta=h_{min}$, $\chi = 0.002$, $r_{\bu}=33$, and $r_p=r_s=3$. Prediction for $Re_{max}=110$ in $[4,8]$.) \emph{Left.}  Comparison of relative log-errors over time of the velocity profiles: EFR full order versus {EFR-noEFR} solutions and EFR full order versus {EFR-EFR} solutions, represented by solid blue and dashed orange lines, respectively.   \emph{Right.}  Analogous representation for the relative log-errors over time of the pressure profiles.}}}

\label{fig:errs_110}
\end{figure}

\begin{figure}
\centering
        \includegraphics[scale=0.4]{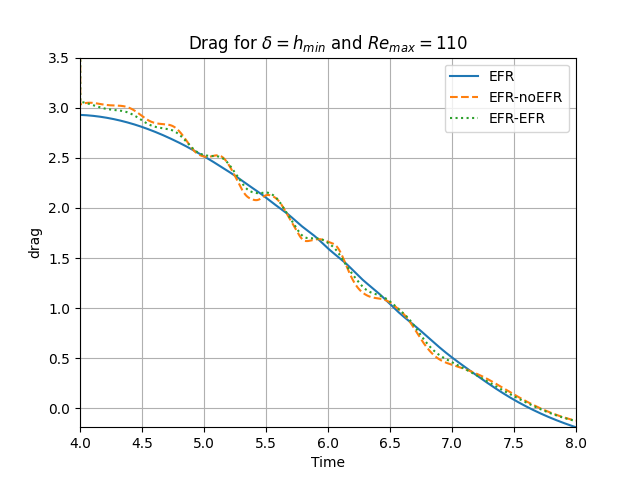}
         \includegraphics[scale=0.4]{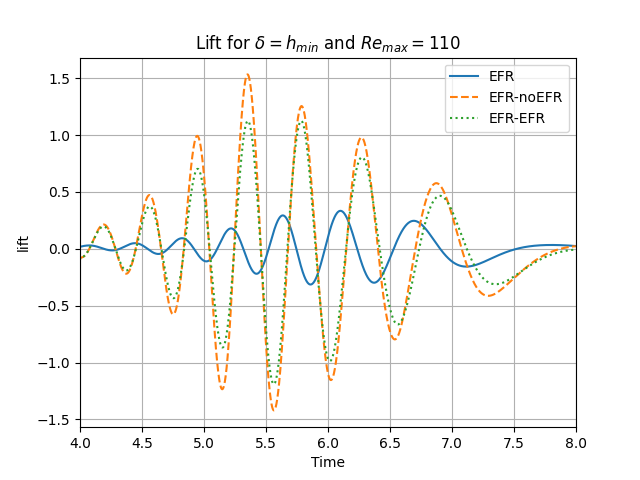}
\caption{\both{\scriptsize{Experiment 4: $\delta=h_{min}$, $\chi = 0.002$, $r_{\bu}=33$, and  $r_p=r_s=3$. Prediction for $Re_{max}=110$ in $[4,8]$.) \emph{Left.}  $C_D(t)$ comparison over time. \emph{Right.}  $C_L(t)$ comparison over time.}}
}
\label{fig:cd_cl110}
\end{figure}

\both{
\subsection{Reynolds number $Re_{max}=110$: predictive regime}
    \label{sec:re110}
In this section, we analyze the predictive capabilities of the 
EFR-EFR and EFR-noEFR algorithms with respect to the Reynolds number. 
We note that we do not investigate the predictive capabilities of the two algorithms with respect to both time and Reynolds number since both algorithms struggled in the predictive regime for Experiment 2 in Section \ref{sec_pred_time}.
We follow the approach used in Section \ref{sec_pred_time} and try to answer 
the following questions: 
(i) Are the the  EFR-noEFR and EFR-EFR algorithms predictive? 
(ii) Which algorithm performs better in the predictive regime with respect to the Reynolds number?

We first  present results for $Re_{max}=110$. 
In our numerical investigation, we use the same computational setting as that used in Experiment 4.
Specifically, we use 
$\delta = h_{min}$ and $\chi = 0.002$. 
We also collect $N_{\bu} = N_p = 2000$ equally spaced snapshots in the time interval $[4,8]$ for $N_{\nu}=20$.
For these snapshots, we choose 
a log-equispaced distribution in  the parametric domain $\nu \in 
[10^{-3},1.575 
\times 10^{-3}]$. 
We employ $r_{\bu}=33$ and $r_s = r_p = 3$ to retain $99.9\%$ of the snapshot energy. 
The relative error plots in Figure \ref{fig:errs_110} show that EFR-EFR is consistently more accurate than the EFR-noEFR in approximating both the velocity and the pressure.
We note, however, that both the EFR-EFR and the EFR-noEFR are relatively inaccurate in approximating the pressure field.
\begin{table}[b]
\caption{\scriptsize{\both{Experiment 4: $\delta=h_{min}$, $\chi = 0.002$, $r_{\bu}=33$, and $r_p=r_s=3$. Prediction for $Re_{max}=110$ in $[4,8]$.) Maximum, minimum, and average relative error over the considered time interval for the velocity and pressure fields.}}}
\label{tab:table110}
\begin{center}
\begin{tabular}{c|ccc|ccc|}
\cline{2-7}
                                   & \multicolumn{3}{c|}{{EFR-noEFR}}                                                   & \multicolumn{3}{c|}{{EFR-EFR}}                                                \\ \cline{2-7} 
                                   & \multicolumn{1}{c|}{maximum}    & \multicolumn{1}{c|}{minimum}    & average    & \multicolumn{1}{c|}{maximum}    & \multicolumn{1}{c|}{minimum}    & average    \\ \hline
\multicolumn{1}{|c|}{$E_{\bu}(t)$} & \multicolumn{1}{c|}{$5.943e$-1} & \multicolumn{1}{c|}{$2.138e$-3} & $1.636e$-1 & \multicolumn{1}{c|}{$3.752e$-1} & \multicolumn{1}{c|}{$2.146e$-3} & $9.869e$-2 \\ \hline
\multicolumn{1}{|c|}{$E_{p}(t)$}   & \multicolumn{1}{c|}{$2.451e$+0} & \multicolumn{1}{c|}{$3.758e$-2} & $3.247e$-1 & \multicolumn{1}{c|}{$2.451e$+0} & \multicolumn{1}{c|}{$3.633e$-2} & $2.558e$-1 \\ \hline
\end{tabular}%
\end{center}
\end{table}
A similar behavior is observed with respect to the force coefficients, which are displayed in Figure \ref{fig:cd_cl110}: EFR-EFR is consistently more accurate than the EFR-noEFR.
We note that the EFR-EFR improvement over the EFR-noEFR is only marginal for the drag coefficient.
Furthermore, both the EFR-EFR and the EFR-noEFR produce relatively inaccurate approximations of the lift coefficient.
The qualitative behavior of the plots in Figures \ref{fig:errs_110} and \ref{fig:cd_cl110} is  supported by the maximum, minimum, and average relative errors listed in Table \ref{tab:table110},  and by the $L^2-$error of the force coefficients: 
$\A{\overline E_{C_L}} = 4.77$, $\A{\hat E_{C_L}}=3.99$, and $\A{\overline E_{C_D}} = 0.048$, $\A{\hat E_{C_D}} = 0.037$.

Overall, these results yield the following conclusions:
(i) The EFR-EFR and EFR-noEFR algorithms are predictive in the approximation of the velocity field and the drag coefficient, but both struggle in the approximation of the pressure field and the lift coefficient.
(ii) The EFR-EFR algorithm is consistently more accurate than the EFR-noEFR algorithm with respect to all criteria, especially in the approximation of the velocity field.
}

\both{
\subsection{Reynolds number $Re_{max}=140$: predictive regime}
    \label{sec:re140}}

\begin{figure}
\centering
        \includegraphics[scale=0.4]{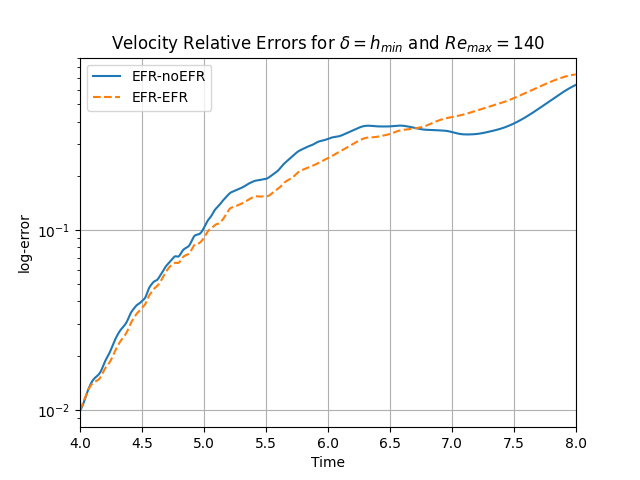}
         \includegraphics[scale=0.4]{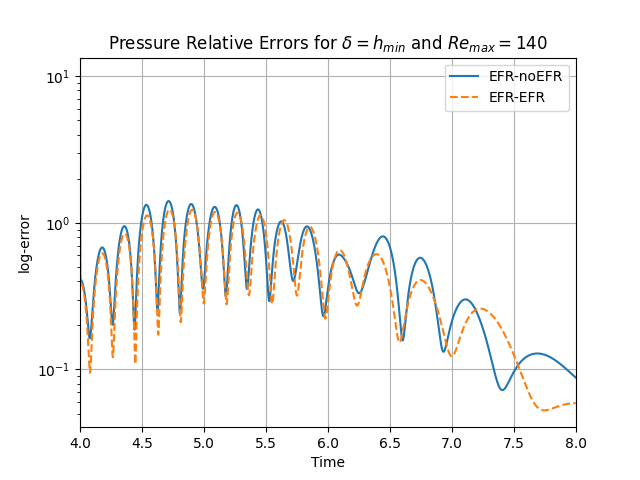}
\caption{\both{\scriptsize{(Experiment 4: $\delta=h_{min}$, $\chi = 0.002$, $r_{\bu}=33$, and $r_p=r_s=3$. Prediction for $Re_{max}=140$ in $[4,8]$.) \emph{Left.}  Comparison of relative log-errors over time of the velocity profiles: EFR full order versus {EFR-noEFR} solutions and EFR full order versus {EFR-EFR} solutions, represented by solid blue  and dashed orange lines, respectively.   \emph{Right.}  Analogous representation for the relative log-errors over time of the pressure profiles.}}}

\label{fig:errs_140}
\end{figure}

\begin{figure}
\centering
        \includegraphics[scale=0.4]{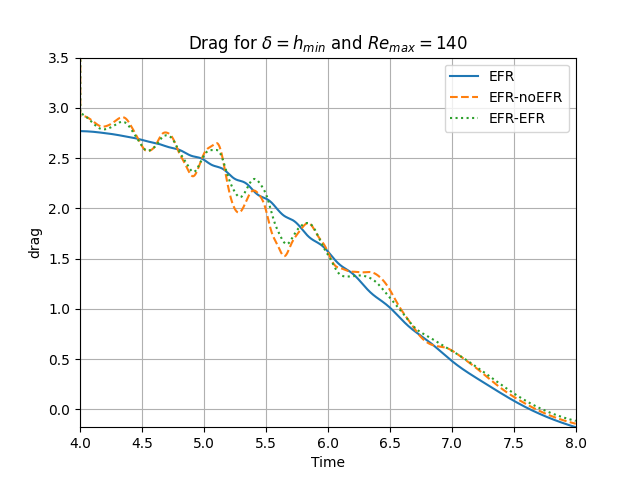}
         \includegraphics[scale=0.4]{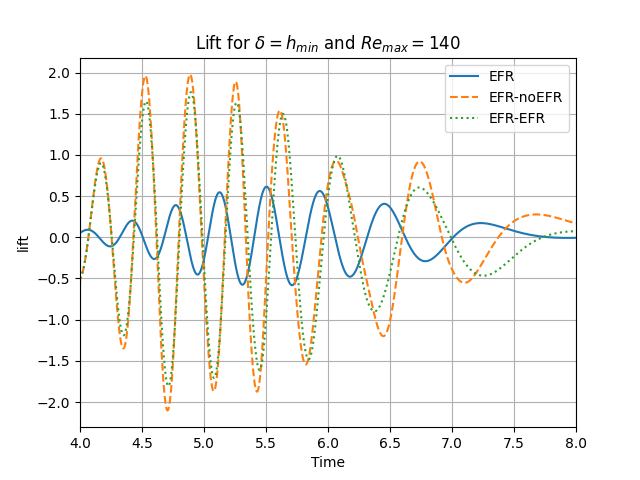}
\caption{\both{\scriptsize{Experiment 4: $\delta=h_{min}$, $\chi = 0.002$, $r_{\bu}=33$, and $r_p=r_s=3$. Prediction for $Re_{max}=140$ in $[4,8]$.) \emph{Left.}  $C_D(t)$ comparison over time. \emph{Right.}  $C_L(t)$ comparison over time.}}
}
\label{fig:cd_cl140}
\end{figure}

\both{
In this numerical investigation, we increase the Reynolds number to $Re_{max} = 140$ and use the same computational setting as that used in Experiment 4.
Specifically, we use $\delta = h_{min}$ and $\chi = 0.002$, and collect $N_{\bu} = N_p = 2000$ equally spaced snapshots in the time interval $[4,8]$ for $N_{\nu}=20$.
For these snapshots, we choose a log-equispaced distribution in  the parametric domain $\nu \in 
[10^{-3},1.575 \times 10^{-3}]$. 
We employ $r_{\bu}=33$ and $r_s = r_p = 3$ to retain $99.9\%$ of the snapshot energy. 
The relative error plots in Figure \ref{fig:errs_140} show that the EFR-EFR and EFR-noEFR algorithms perform similarly:
They predict accurately the velocity field at the beginning, but their accuracy starts to degrade toward the end of the time interval.
Their predictions of the pressure field are inaccurate at the beginning of the simulation, but they become more accurate toward the end of the time interval.
A similar behavior is observed with respect to the force coefficients, which are displayed in Figure \ref{fig:cd_cl140}. 
The EFR-EFR and EFR-noEFR algorithms perform similarly and 
provide relatively accurate approximations of the drag coefficient, but their approximations of the lift coefficient are  inaccurate.
The qualitative behavior of the plots in Figures \ref{fig:errs_140} and \ref{fig:cd_cl140} is  supported by the maximum, minimum, and average relative errors listed in Table \ref{tab:table140},  and by the $L^2-$error of the force coefficients: 
$\A{\overline E_{C_L}} = 3.91$, $\A{\hat E_{C_L}}=3.64$, and $\A{\overline E_{C_D}} = 0.098$, $\A{\hat E_{C_D}} = 0.11$.
Overall, these results yield the following conclusions:
(i) The EFR-EFR and EFR-noEFR algorithms are predictive in the approximation of the velocity field and the drag coefficient, but both struggle in the approximation of the pressure field and the lift coefficient.
(ii) The EFR-EFR and EFR-noEFR algorithms perform similarly with respect to all criteria.
We believe that, to increase the predictive capabilities of the EFR-EFR and EFR-noEFR algorithms, the parameters $\delta$ and $\chi$ should be tuned appropriately.
This, however, goes beyond the scope of the current investigation.
}

\begin{table}[]
\caption{\scriptsize{\both{Experiment 4: $\delta=h_{min}$, $\chi = 0.002$, $r_{\bu}=33$, and $r_p=r_s=3$. Prediction for $Re_{max}=140$ in $[4,8]$.) Maximum, minimum, and average relative error over the considered time interval for velocity and pressure fields.}}}
\label{tab:table140}
\begin{center}
\begin{tabular}{c|ccc|ccc|}
\cline{2-7}
                                   & \multicolumn{3}{c|}{{EFR-noEFR}}                                                   & \multicolumn{3}{c|}{{EFR-EFR}}                                                \\ \cline{2-7} 
                                   & \multicolumn{1}{c|}{maximum}    & \multicolumn{1}{c|}{minimum}    & average    & \multicolumn{1}{c|}{maximum}    & \multicolumn{1}{c|}{minimum}    & average    \\ \hline
\multicolumn{1}{|c|}{$E_{\bu}(t)$} & \multicolumn{1}{c|}{$6.387e$-1} & \multicolumn{1}{c|}{$9.922e$-3} & $2.617e$-1 & \multicolumn{1}{c|}{$7.310e$-1} & \multicolumn{1}{c|}{$9.930e$-3} & $2.768e$-1 \\ \hline
\multicolumn{1}{|c|}{$E_{p}(t)$}   & \multicolumn{1}{c|}{$1.029e$+1} & \multicolumn{1}{c|}{$7.233e$-2} & $5.289e$-1 & \multicolumn{1}{c|}{$1.029e$+1} & \multicolumn{1}{c|}{$5.260e$-2} & $4.744e$-1 \\ \hline
\end{tabular}%
\end{center}
\end{table}
\section{Conclusions}
\label{sec:conc}

In this paper, we took a step in the study of FOM-ROM consistency when the EFR algorithm is used as numerical stabilization in convection-dominated, marginally-resolved  flows. 
To this end, as a mathematical model we considered the incompressible Navier-Stokes equations. 
We used moderate Reynolds numbers, which yielded a convection-dominated regime.
We performed FOM and ROM simulations in the marginally-resolved regime, i.e., when the number of degrees of freedom is barely capable of capturing the main features of the underlying flow.
To tackle the inaccuracies of the FOM and ROM simulations in the marginally-resolved regime, we employed the EFR algorithm, which leverages spatial filtering to alleviate the spurious oscillations.

To investigate the FOM-ROM consistency, we considered two models:
\begin{itemize}
    \item[$\circ$] the \A{EFR-noEFR}, in which the EFR regularization is used at a FOM level, but not at a ROM level;
    \item[$\circ$] the \A{EFR-EFR}, in which the EFR regularization is used both at a FOM and at a ROM level.
\end {itemize}
We investigated the \A{EFR-noEFR } and \A{EFR-EFR } in the numerical simulation of a 2D flow past a circular cylinder at time-dependent Reynolds numbers with a maximum value $Re=100$.
As criteria for our comparison, we used the relative velocity error, the relative pressure error, and the lift and drag coefficients.
We also considered two types of model reduction:
(i) model reduction in time, for which we used the POD algorithm; and 
(ii) model reduction in time and in the parameter space, for which we used the nested-POD algorithm.
In all our tests, for both types of model reduction, and for all three criteria, the \A{EFR-EFR } was more accurate than the \A{EFR-noEFR}.
These results suggest that FOM-ROM consistency is beneficial for the EFR regularization in a convection-dominated,  marginally-resolved regime.

These first steps in the study of the FOM-ROM consistency of regularized models are encouraging.
There are, however, other research directions that should be investigated for a deeper understanding of this important topic.
Probably the most important investigation should focus on the FOM-ROM consistency for regularized models in the {\it under-resolved} regime, which is important in many realistic settings (e.g., turbulent flows) where the ROM dimension is significantly lower than the number of degrees of freedom needed to accurately represent the complex dynamics of the underlying system.
\A{Related to this investigation, higher Reynolds number flows should be considered. }
Another important research direction is the investigation of FOM-ROM consistency when {\it different} regularized models are used at the FOM and ROM levels (e.g., the Leray model is used at the FOM level and the EFR model is used at the ROM level).
Related to this, one could also investigate the {\it parameter} FOM-ROM consistency, which is complementary to the {\it model} FOM-ROM consistency investigated in this paper.
Specifically, one could consider the same regularized model at the FOM and ROM levels, but use different parameters (e.g., different $\delta$ values) in these regularized models.
\A{
Using different $\delta$ or $\chi$ values at the FOM and ROM levels (i.e., $\delta^{FOM} \neq \delta^{ROM}$ or $\chi^{FOM} \neq \chi^{ROM}$) could yield more accurate ROM solutions
in several settings, such as hyper-reduction. 
}\\
\C{  
Finally, we emphasize that most of the existing studies (including this paper) on the FOM-ROM consistency have been numerical investigations.
Although theoretical investigations could support the existing numerical investigations and shed new light on the FOM-ROM consistency, these studies are relatively scarce (for notable examples, see the numerical analysis performed in~\cite{giere2015supg,pacciarini2014stabilized} for FOM-ROM consistency of the streamline upwind Petrov-Galerkin (SUPG) stabilization, and~\cite{ingimarson2021full} for FOM-ROM consistency with respect to the discretization of the nonlinearity of the Navier-Stokes equations).
In a future study, we plan to perform the numerical analysis of the FOM-ROM consistency with respect to the EFR regularization, and investigate whether the theoretical results support the numerical findings in the numerical investigation in this paper.
}
\color{black}{
These numerical and theoretical investigations of various types of FOM-ROM consistency could provide a new impetus for the development of ROMs that are consistent with their corresponding FOMs.
}


\section*{Acknowledgements}
We acknowledge the support by European Union Funding for Research and Innovation -- Horizon 2020 Program -- in the framework of European Research Council Executive Agency: Consolidator Grant H2020 ERC CoG 2015 AROMA-CFD project 681447 ``Advanced Reduced Order Methods with Applications in Computational Fluid Dynamics.'' We also acknowledge the PRIN 2017  ``Numerical Analysis for Full and Reduced Order Methods for the efficient and accurate solution of complex systems governed by Partial Differential Equations'' (NA-FROM-PDEs) and the INDAM-GNCS project ``Tecniche Numeriche Avanzate per Applicazioni Industriali.''
The fourth author acknowledges support through National Science Foundation Grant No. DMS-2012253.
The computations in this work have been performed with RBniCS \cite{rbnics} library, 
which is an implementation in FEniCS \cite{fenics} of several reduced order modeling techniques, and is developed at SISSA mathLab; we acknowledge developers and contributors to both libraries.

\bibliographystyle{abbrv}
\bibliography{main}

\end{document}